\documentclass{mn2e}

\usepackage{graphicx,lscape,amssymb,amsmath,url,placeins}

\title[A Deep ALMA image of the HUDF]{A deep ALMA image of the Hubble Ultra Deep Field}
\author[J.S. Dunlop et al.]{J.S.\,Dunlop$^{1}$\thanks{E-mail: jsd@roe.ac.uk},
R.J.\,McLure$^{1}$, A.D.\,Biggs$^2$, J.E.\,Geach$^{3}$, M.J.\,Micha{\l}owski$^{1}$, R.J.\,Ivison$^{1,2}$,
\and W.\,Rujopakarn$^{4}$,  E.\,van\,Kampen$^{2}$, A.\,Kirkpatrick$^{5}$, A.\,Pope$^{5}$, D.\,Scott$^{6}$, A.M.\,Swinbank$^{7}$,
\and T.A.\,Targett$^{8}$, I.\,Aretxaga$^{9}$, J.E.\,Austermann$^{10}$, P.N.\,Best$^{1}$, V.A.\,Bruce$^{1}$, E.L.\,Chapin$^{11}$,
\and S.\,Charlot$^{12}$, M.\,Cirasuolo$^{2}$, K.\,Coppin$^{3}$, R.S.\,Ellis$^{2}$, S.L.\,Finkelstein$^{13}$, C.C.\,Hayward$^{14}$,
\and D.H.\,Hughes$^{9}$, E.\,Ibar$^{15}$, P.\,Jagannathan$^{16}$, S.\,Khochfar$^{1}$, M.P.\,Koprowski$^{3}$,
\and D.\,Narayanan$^{17}$, K.\,Nyland$^{18}$, C.\,Papovich$^{19}$, J.A.\,Peacock$^{1}$, G.H.\,Rieke$^{20}$,
\and B.\,Robertson$^{21}$, T.\,Vernstrom$^{22}$, P.P.\,van\,der\,Werf$^{23}$,
G.W.\,Wilson$^{5}$, M.\,Yun$^{5}$\\
$^{1}$Institute for Astronomy, University of Edinburgh, Royal Observatory, Edinburgh, EH9 3HJ\\
$^{2}$European Southern Observatory, Karl-Schwarzchild-Str. 2, 95748 Garching b. Munchen, Germany\\
$^{3}$Centre for Astrophysics Research, Science \& Technology Research Institute, University of Hertfordshire, College Lane, Hatfield, AL10 9AB\\
$^{4}$Department of Physics, Faculty of Science, Chulalongkorn University, 254 Phayathai Road, Pathumwan, Bangkok 10330, Thailand\\
$^{5}$Department of Astronomy, University of Massachusetts, Amherst, MA 01002, USA\\
$^{6}$Department of Physics and Astronomy, University of British Columbia, Vancouver, BC V6T1Z1, Canada\\
$^{7}$Centre for Extragalactic Astronomy, Department of Physics, Durham University, South Road, Durham DH1 3LE\\
$^{8}$Department of Physics and Astronomy, Sonoma State University, 1801 East Cotati Avenue, Rohnert Park, CA 94928-3609, USA\\
$^{9}$Instituto Nacional de Astrofisica, Optica y Electronica (INAOE), Aptdo. Postal 51 y 216, 72000 Puebla, Pue., Mexico\\
$^{10}$NIST Quantum Devices Group, 325 Broadway Mailcode 817.03, Boulder, CO 80305, USA\\
$^{11}$Herzberg Astronomy and Astrophysics, National Research Council Canada, 5071 West Saanich Road, Victoria, BC V9E 2E7, Canada\\
$^{12}$Sorbonne Universites, UPMC-CNRS, UMR7095, Institut d’Astrophysique de Paris, F-75014 Paris, France\\
$^{13}$Department of Astronomy, The University of Texas at Austin, Austin, TX 78712, USA\\
$^{14}$Center for Computational Astrophysics,
Flatiron Institute, 162 Fifth Avenue, New York, NY 10010, USA\\
$^{15}$Instituto de Fisica y Astronomia, Universidad de Valparaiso, Avda. Gran Bretana 1111, Valparaiso, Chile\\
$^{16}$National Radio Astronomy Observatory, 1003 Lopezville Road, Socorro, NM 87801, USA\\
$^{17}$Department  of Physics and Astronomy, Haverford College, Haverford, PA 19041, USA\\
$^{18}$National Radio Astronomy Observatory, Charlottesville, VA 22903, USA\\
$^{19}$Department of Physics and Astronomy, Texas A\,\&\,M University, College Station, TX 77843-4242, USA\\
$^{20}$Steward Observatory, University of Arizona, Tucson, AZ 85721, USA\\
$^{21}$Department of Astronomy and Astrophysics, University of California,
Santa Cruz, 1156 High Street, Santa Cruz, CA 95064, USA\\
$^{22}$Dunlap Institute for Astronomy and Astrophysics, University of Toronto, Toronto, ON M5S 3H4, Canada\\
$^{23}$Leiden Observatory, Leiden University, PO Box 9513, 2300 RA Leiden, the Netherlands
\vspace*{-0.5cm}}

\begin{document}

\date{
\vspace*{-0.9cm}
}

\pagerange{\pageref{firstpage}--\pageref{lastpage}} \pubyear{???}

\maketitle

\label{firstpage}

\vspace{-1.5cm}

\begin{abstract}
We present the results of the first, deep ALMA imaging covering the
full $\simeq~4.5$\,arcmin$^2$ of the HUDF imaged with
WFC3/IR on {\it~HST}. Using a 45-pointing mosaic, we have obtained a
homogeneous 1.3-mm image reaching $\sigma_{1.3}~\simeq~35$\,${\rm \mu Jy}$,
at a resolution of
$\simeq~0.7$\,arcsec. From an initial list of $\simeq~50>3.5\sigma$ peaks,
a rigorous analysis confirms 16 sources with $S_{1.3}>120$\,${\rm \mu Jy}$.
All of these have secure galaxy counterparts
with robust redshifts ($\langle~z~\rangle~=~2.15$).
Due to the unparalleled supporting data, the physical properties of the ALMA
sources are well constrained, including
their stellar masses ($M_*$) and UV+FIR star-formation rates (SFR).
Our results show that stellar mass is the best predictor of
SFR in the high-redshift Universe; indeed at $z~\geq~2$
our ALMA sample contains 7 of the 9 galaxies in the HUDF with
$M_*~\geq~2~\times~10^{10}\,{\rm M_{\odot}}$, and we detect only one
galaxy at $z~>~3.5$, reflecting the rapid drop-off of high-mass galaxies
with increasing redshift.
The detections, coupled with stacking, allow us to probe the redshift/mass
distribution of the 1.3-mm background down to $S_{1.3}~\simeq~10\,{\rm \mu Jy}$.
We find strong evidence for a steep star-forming `main sequence'
at $z~\simeq~2$, with SFR~$\propto~M_*$ and a mean specific
SFR~$\simeq~2.2$\,Gyr$^{-1}$.
Moreover, we find that $\simeq~85$\% of total star formation at $z~\simeq~2$
is enshrouded in dust, with $\simeq~65$\% of all star formation at
this epoch occurring in high-mass galaxies
($M_*~>~2~\times~10^{10}\,{\rm M_{\odot}}$), for which the
average obscured:unobscured SF ratio is $\simeq~200$.
Finally, we revisit the cosmic evolution of SFR density; we find
this peaks at $z~\simeq~2.5$, and that
the star-forming Universe transits from primarily unobscured to
primarily obscured at $z~\simeq~4$. 
\end{abstract}

\begin{keywords}
galaxies: high-redshift, evolution, starburst - cosmology: observations - submillimetre: galaxies
\vspace*{-2.3cm}
\end{keywords}

\section{Introduction}
\label{sec:intro}

A complete understanding of cosmic star-formation history, and the physical 
mechanisms that drive galaxy formation and evolution, requires that we connect
our UV/optical and infrared/mm views of the Universe (e.g. Hopkins \& Beacom 2006; Dunlop 2011;
Burgarella et al. 2013; Madau \& Dickinson 2014). 
Until the advent of ALMA, these two views have been largely disconnected, for both technical and 
physical reasons. Benefiting from low background and high angular resolution,
deep UV/optical surveys have proved extremely effective at completing our 
inventory of {\it unobscured} star formation which, certainly at high redshift,
is dominated by large numbers of low-mass galaxies with individual
star-formation rates ${\rm SFR \simeq 1\,M_{\odot} yr^{-1}}$ (e.g. McLure et al. 2013; Bouwens et al. 2015;
Bowler et al. 2015; Finkelstein et al. 2015; McLeod et al. 2015). 

However, UV/optical observations are unable to uncover the most extreme star-forming 
galaxies, which, following the breakthroughs in far-IR/sub-mm astronomy 
at the end of the last century, are known to be enshrouded in dust (Smail, Ivison \& Blain 1997;
Hughes et al. 1998; Barger et al. 1998; Eales et al. 1999). 
Such objects have now been uncovered in significant numbers through 
far-IR/mm surveys with the JCMT (e.g. Scott et al. 2002, 2006; Coppin et al. 2006; Austermann et al. 2010; Geach et al. 2013),
IRAM (Dannerbauer et al. 2004; Greve et al. 2004; Lindner et al. 2011), 
APEX (Weiss et al. 2009; Smolcic et al. 2012),
ASTE (Scott et al. 2008, 2010; Hatsukade et al. 2010, 2011),
BLAST (Devlin et al. 2009; Dunlop et al. 2010) 
and {\it Herschel} (Elbaz et al. 2011; Lutz et al. 2011;
Eales et al. 2010; Oliver et al. 2010), with `sub-mm galaxies' now known out to redshifts $z > 6$ (Reichers et al. 2013).

While sub-mm surveys for high-redshift galaxies benefit from a strong negative K-correction (provided by a modified 
blackbody spectral energy distribution (SED); Blain \& Longair 1993; Hughes et al. 1993; Dunlop et al. 1994),
the high background and/or the relatively poor angular resolution provided by single-dish telescopes at these wavelengths means 
that they are only really effective at uncovering rare, extreme star-forming
galaxies with ${\rm SFR > 300\,M_{\odot} yr^{-1}}$ (albeit reaching down to ${\rm SFR > 100\,M_{\odot} yr^{-1}}$
in the very deepest SCUBA-2 450/850\,$\mu$m imaging; Geach et al. 2013; Roseboom et al. 2013; Koprowski et al. 2016).
The existence of such objects presents an interesting and 
important challenge to theoretical models of galaxy formation (e.g. Baugh et al. 2005; Khochfar \& Silk 2009; 
Dav\'e et al. 2010; Hayward et al. 2011; Narayanan et al. 2015),
but they provide only $\simeq 10-15$\% of the known far-infrared/mm background
(Fixsen et al. 1998; Scott et al. 2012; Geach et al. 2013),
and attempts to complete our inventory of obscured star formation have had to rely on
stacking experiments (e.g. Peacock et al. 2000; Marsden et al. 2009; Geach et al. 2013; Coppin et al. 2015).

A key goal, therefore, of deep surveys with ALMA is to close the depth/resolution gap between
UV/optical and far-infrared/mm studies of the high-redshift Universe, and hence
enable a complete study of visible+obscured star formation within the 
overall galaxy population. Over the last two years, ALMA has begun to make 
important contributions in this area. Most early 
ALMA programmes have focused (sensibly) on pointed observations 
of known objects (e.g. Hodge et al. 2013; Karim et al. 2013;
Ouchi et al. 2013; Bussmann et al. 2015; Capak et al. 2015;
Maiolino et al. 2015; Simpson et al. 2015a, 2015b; Scoville et al. 2016),
including gravitationally lensed sources (e.g. Weiss et al. 2013; Watson et al. 2015; B\'ethermin et al. 2016; Knudsen et al. 2016;
Spilker et al. 2016). However, strenuous efforts have been made to exploit the resulting
combined `blank-field' survey by-product to improve our 
understanding of the deep mm source counts
(e.g. Ono et al. 2014; Carniani et al. 2015, Fujimoto et al. 2016; Oteo et al. 2016) albeit with interestingly 
different results. More recently, time has been awarded to 
programmes that aim to deliver contiguous ALMA mosaic imaging of small regions
of sky with excellent multi-wavelength supporting data (e.g. Hatsukade et al. 2015;
Umehata et al. 2015). Such programmes offer
not only further improvements in our knowledge of the sub-mm/mm source 
counts, but also the ability to determine the nature and physical properties 
(redshifts, stellar masses, star-formation rates) of the ALMA-detected galaxies. For example, ALMA 1.1-mm
imaging of 1.5\,arcmin$^2$ within the CANDELS/UDS field (PI: Kohno) has provided new results
on the 1.1-mm counts, and enabled the study of several  ALMA-detected galaxies 
(Hatsukade et al. 2016; Tadaki et al. 2015).

However, to date, no homogeneous ALMA imaging has been
undertaken within the best-studied region of deep `blank-field' sky, the Hubble Ultra Deep Field (HUDF).
On scales of a few arcmin$^2$, the HUDF remains unarguably the key ultra-deep extragalactic survey field
and, lying within the GOODS-South field at RA 03$^h$, Dec $-$28$^{\circ}$, is ideally located for deep ALMA observations.
While four of the six Hubble Frontier
Fields\footnote{http://www.stsci.edu/hst/campaigns/frontier-fields/}
provide alternative target fields for deep ALMA
observations, the quality of the optical--near-infrared data in these
fields will never seriously rival 
that which has already been achieved in the HUDF. In part this is due 
to the huge investment in {\it HST} optical imaging in this field made 
prior to the degradation of the ACS camera (Beckwith et al. 2006).
However, it is also a result of the more recent investment in 
imaging with WFC3/IR on the {\it HST} since 2009. Specifically,
the combination of the
UDF09 campaign (Oesch et al. 2010; Bouwens et al. 2010;
Illingworth et al. 2013)
followed by the UDF12 programme (Ellis et al. 2013; Koekemoer et al. 2013; Dunlop et al. 2013),
has delivered the deepest near-infrared imaging ever achieved
(reaching 30 AB mag, 5-$\sigma$) over an area of 4.5\,arcmin$^2$. 
As a result of coupling this multi-band {\it HST} imaging with the recently augmented ultra-deep {\it Spitzer} 
data (Labb\'e et al. 2015), accurate photometric redshifts,
stellar masses and UV star-formation rates are now known for $\simeq 3000$
galaxies in this field (e.g. Parsa et al. 2016). 
For a field of this size, the HUDF is also uniquely rich in optical/infrared 
{\it spectroscopic} information, with a combination of ground-based optical spectroscopy
and {\it HST} WFC3/IR near-infrared grism spectroscopy delivering redshifts
and emission-line strengths for over 300 galaxies (see Section 2.2).
Finally, the
HUDF lies in the centre of the {\it Chandra} Deep Field South (CDFS) 4-Ms X-ray imaging (Xue et al. 2014),
and has recently been the focus of a new programme of ultra-deep radio imaging
with the JVLA (PI: Rujopakarn).

The aim of the work presented here was to exploit this unique database by using ALMA
to construct the deepest homogeneous mm-wavelength image obtained to date on the relevant scales.
As described in detail in the next Section, $\simeq 20$\,hr of ALMA observations were approved
in Cycle 1, and rolled over into Cycle 2, to enable us to complete a 1.3-mm mosaic
covering the full 4.5\,arcmin$^2$ imaged with WFC3/IR, seeking to reach an rms
depth of $\sigma_{1.3} \simeq 30\,{\rm \mu Jy}$ (PI: Dunlop). We chose to undertake this first deep
ALMA image of the HUDF at 1.3\,mm (rather than at shorter wavelengths)
for three reasons. First, in practice it maximises sensitivity to higher redshift
dusty star-forming galaxies at $z > 3$. Second, it is at these longer wavelengths that the resolution
of single dish surveys is undoubtedly poorest, and hence the imaging most confused.
Third, this decision aided the feasibility of the observations in early ALMA cycles,
with only 45 pointings required to complete the mosaic, and both nightime and daytime observations being acceptable.
Astrophysically, we sought to reach detections 4--5 times deeper than can be achieved
with the deepest single-dish surveys (corresponding to star-formation rates $\simeq 25\,{\rm M_{\odot} yr^{-1}}$
out to the very highest redshifts), and to exploit the uniquely complete HUDF galaxy database in deep
stacking experiments.

Data taking for this project commenced in 2014, and was completed in summer 2015, and in this paper
we present the first results. We present and discuss the properties of the ALMA map, the sources uncovered within it,
and the implications for our understanding of cosmic star formation and galaxy evolution.
The remainder of this paper is structured as follows. In Section 2 we describe
the ALMA observations, explain
how the data were reduced, and provide a summary of
all the key multi-wavelength supporting data in the field. In Section 3
we explain how sources were extracted from the ALMA map, and then, in Section
4, describe how cross-identifications with the {\it HST} sources in the field
enabled us to clean the source list to a final sample of 16 robust ALMA
sources. In Section 5 we consider the implications of the
number of sources we have detected, aided by the results of source injection
and retrieval simulations, and compare our results to other recent
estimates of deep mm number counts.
In Section 6 we derive the physical properties of the
sources we have detected, and explore the implications for star formation
in galaxies at $z = 1 -3$. Then, in Section 7 we present the results
of stacking the 1.3-mm signal on the positions of known galaxy
populations in the HUDF, and consider the consequences for the mm-wavelength
background and for the ratio of obscured/unobscured star formation over
cosmic history. Finally we discuss the astrophysical implications
of our findings in Section 8, and summarize our conclusions in Section
9. Throughout, all magnitudes  are  quoted  in the  AB  system
(Oke 1974; Oke \& Gunn 1983), and all cosmological calculations assume a flat
cold dark matter cosmology
with $\Omega_{\rm M} = 0.3$, $\Omega_{\rm \Lambda} = 0.7$, and
$H_0 = 70\,{\rm km s^{-1} Mpc^{-1}}$.

\begin{table}
  \caption{Summary of the ALMA observations of the HUDF. The date of each Execution Block (EB)
    is given along with the approximate maximum baseline length and the average amount of precipitable water vapour (PWV).}
\label{tab:obs}
\centering
\begin{tabular}{lcc} \\ \hline
Observing date & Maximum baseline / m & PWV / mm \\ \hline
2014 July 18 & \phantom{1}650 & 0.43 \\
2014 July 29 & \phantom{1}820 & 1.04 \\ 
2014 Aug 17  &           1100 & 0.94 \\
2014 Aug 18  &           1250 & 1.51 \\
2014 Aug 18  &           1250 & 1.45 \\
2014 Aug 27  &           1100 & 1.35 \\
2014 Aug 28  &           1100 & 1.20 \\
2014 Aug 28  &           1100 & 1.25 \\
2014 Sep 1   &           1100 & 1.08 \\
2015 May 16  & \phantom{1}550 & 0.65 \\
2015 May 16  & \phantom{1}550 & 0.80 \\
2015 May 17  & \phantom{1}550 & 1.00 \\
2015 May 17  & \phantom{1}550 & 1.80 \\ \hline
\end{tabular}
\end{table}

\begin{figure*}
\begin{center}
\includegraphics[scale=0.72]{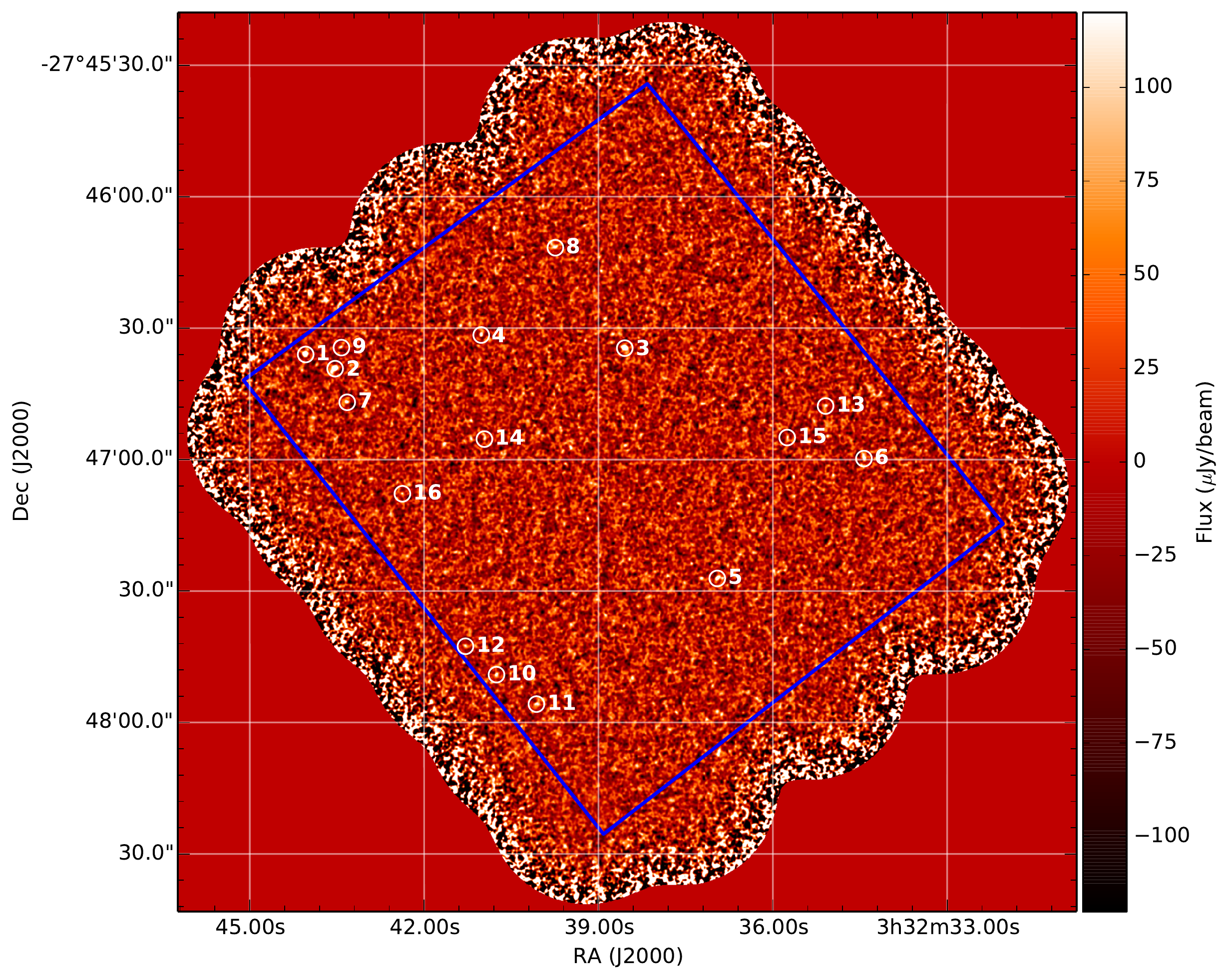}
\end{center}
\caption{The ALMA 1.3-mm map of the HUDF, with the
  positions of the 16 sources listed in Table\,2 marked by 3.6-arcsec diameter
  circles. The border of the homogenously deep region of near-infrared WFC3/IR
  imaging obtained through the UDF09 and UDF12 {\it HST} programmes is indicated
  by the dark-blue rectangle. The ALMA image, constructed from a mosaic of
  45 individual pointings, provides homogeneous 1.3-mm coverage of this region, with
  a typical noise per beam of $\sigma_{1.3} \simeq 35\,{\rm \mu Jy}$.}
\label{fig:alma_image}
\end{figure*}

\section{Data}
\label{sec:data}

\subsection{ALMA observations and data reduction}
\label{sec:alma}

The ALMA observations of the HUDF were taken during two separate observing seasons --
the first nine Execution Blocks (EBs) in July to September of 2014 and the remaining
four in May 2015. As the primary goal of these observations was to produce a continuum map of the
HUDF, the correlator was configured to process the maximum 7.5-GHz bandwidth in the form of
four 1875~MHz-wide spectral windows, each with 128 dual-polarization channels. However,
the velocity resolution of 40--45\,${\rm km s^{-1}}$ is still sufficient to resolve spectral lines that are
typically observed in high-redshift star-forming galaxies. The correlator averaging time was
2~sec per sample.

The HUDF was observed using a 45-pointing mosaic, with each pointing separated by 0.8 times
the antenna beamsize. This mosaic pattern was observed twice per EB, except
for one which terminated after only 20 pointings of the first mosaic pass. However, no problems were
found with these data and they were included in the final map. The amplitude and bandpass calibrator
for each EB was the unresolved quasar J0334$-$401, this also serving as the phase calibrator during
the second observing season. Although relatively far (12.4~degrees) from the HUDF, the phase
solutions varied smoothly over the course of each EB, and maps made from these data demonstrated
that phase referencing had indeed been successful. In the first season, the phase calibrator was
J0348$-$2749 which is only 3.5~degrees from the target. The array configuration varied greatly
during the observations, with the first season generally using baselines twice as long as
required to achieve the requested angular resolution. A summary of the observations is 
given in Table\,1.

All data reduction was carried out using {\sc CASA} and followed standard procedures.
Firstly, the data from the second season needed to be corrected for incorrect
antenna positions which had been used during correlation. Other {\em a priori}
calibrations included application of system temperature tables and water vapour
radiometer phase corrections. The latter were particularly large and time variable
for the second season, presumably as a result of observing during the day.
Very little data needed to be flagged, a notable exception being the outer four
channels of each spectral window which have very poor sensitivity. After the
removal of the frequency response of each antenna using the bandpass calibrator,
amplitude and phase corrections were calculated as a function of time for the
flux and phase calibrators. The flux scale was then set with reference to the
regularly-monitored flux density of J0334$-$401 and the gain solutions
interpolated onto the HUDF scans.

A continuum mosaiced image of the calibrated data was produced using the task
{\sc clean}. To enhance mapping speed, the data were first averaged in both frequency and
time to produce a dataset with 10 frequency channels per spectral window and a time sampling of 10~s.
The data were naturally weighted for maximum sensitivity, but the relatively large array
configurations still produced a synthesized beam ($589 \times 503$~mas$^2$) that was
significantly smaller than the circular 0.7-arcsec beam that had been requested.
As this would potentially lead to problems with detecting resolved sources,
we experimented with various $u,v$ tapers in order to find the best combination
of angular resolution and mosaic sensitivity.
A $\simeq 220 \times 180\,k\lambda$ taper,
with PA oriented to circularize the beam as much as
possible, produced a beam close to that requested
($707 \times 672$~mas$^2$) and a final mosaic sensitivity as
measured over a large central area of the map of 34~$\mu$Jy\,beam$^{-1}$.
As the detected source flux densities were very weak,
and the synthesized beam sidelobes very low, no deconvolution (cleaning)
was performed. The resulting image is shown in Fig.\,1.
Finally, to aid checks on data quality, and source reality, we also constructed three alternative 50:50 splits of
the ALMA 1.3-mm image, splitting the data in half by observing date, sideband, and polarization.

\subsection{Supporting multi-frequency data}
\label{sec:auxdata}

\subsubsection{Optical/near-infrared imaging}
The key dataset which defined the area that we aimed to cover with the ALMA 1.3-mm mosaic
is the ultra-deep near-infrared imaging of the HUDF obtained with WFC3/IR on {\it HST}
via the UDF09 (e.g. Bouwens et al. 2010; McLure et al. 2010; Oesch et al. 2010; Finkelstein et al. 2010, 2012; Bunker et al. 2010)
and UDF12 (e.g. Ellis et al. 2013; McLure et al. 2013; Dunlop et al. 2013; Illingworth et al. 2013; Finkelstein et al. 2015) programmes.
As described in Koekemoer et al. (2013), the final UDF12 WFC3/IR imaging reaches
a 5-$\sigma$ detection depth of $29.7$\,mag in the $Y_{105}$ filter, and $29.2$\,mag  in the $J_{125}$, $J_{140}$, and
$H_{160}$ filters (total magnitudes, as derived from small-aperture magnitudes assuming
point-source corrections). These unparalleled near-infrared data, covering an area $\simeq 4.5$\,arcmin$^2$,
are complemented by what remains the deepest ever optical imaging obtained with ACS on {\it HST}
(Beckwith et al. 2006). This provides imaging in the $B_{435}$, $V_{606}$, $i_{775}$ and $z_{850}$ filters, reaching 5-$\sigma$ detection
depths of 29.7, 30.2, 29.9, and 29.8\,mag respectively. More recently, the CANDELS programme (Grogin et al. 2011) has provided 
deep $i_{814}$ data across the HUDF (reaching 29.8\,mag, 5-$\sigma$) as part of the CANDELS-DEEP imaging of GOODS-South
(Koekemoer et al. 2011; see also Guo et al. 2013).

The core {\it HST} imaging dataset is extended to shorter wavelengths by the inclusion of deep VLT VIMOS imaging
in the $U$-band (reaching 28\,mag, 5-$\sigma$; Nonino et al. 2009), and to longer wavelengths by the deepest ever
$K_s$-band imaging obtained with HAWK-I on the VLT through the HUGS survey
(Fontana et al. 2014), which reaches $K_s = 26.5$\,mag (5-$\sigma$).
Imaging longward of 2.2\,$\mu$m has been obtained with {\it Spitzer},
with new ultra-deep IRAC imaging of the HUDF at $3.6\,\mu$m  and $4.5\,\mu$m being provided by our own
stack of the available public data described by Labb\'{e} et al. (2015) (see also Ashby et al. 2013, 2015).
This reaches deconfused 5-$\sigma$ detection depths of $\simeq 26.5$\,mag at $3.6\,\mu$m and $\simeq 26.3$\,mag at $4.5\,\mu$m.

Galaxy detection and photometry in the deep {\it HST} imaging
dataset was undertaken using {\sc sextractor} v2.8.6 (Bertin \& Arnouts 1996) in dual image mode with $H_{160}$
as the detection image, and the {\it HST} photometry homogenised through appropriately scaled apertures at
shorter wavelengths.
The ground-based ($U$ and $K_s$) and {\it Spitzer} photometry was extracted by deconfusing the data
using {\it HST} positional priors both using the method described in McLure et al. (2011, 2013), and
independently using {\sc TPHOT} (Merlin et al. 2015).

The resulting optical--near-infrared catalogue contains $\simeq 2900$ objects with 12-band photometry
(see, for example, Parsa et al. 2016).

\subsubsection{Mid/far-infrared/sub-mm imaging}

Longward of 4.5\,$\mu$m, the original GOODS {\it Spitzer} imaging (PID 104; PI Dickinson)
provides the deepest available data at 5.6\,$\mu$m, 8.0\,$\mu$m (from IRAC: Fazio et al. 2004)
and 24\,$\mu$m (from MIPS). The 24\,$\mu$m imaging has been augmented and incorporated within the
{\it Spitzer} Far-Infrared Deep Extragalactic Legacy (FIDEL)\footnote{PI M. Dickinson, see 
  http://www.noao.edu/noao/fidel/} survey (Magnelli et al. 2009) and reaches 
a 5-$\sigma$ detection limit of $S_{24} \simeq 30\,{\rm \mu Jy}$.

Data at longer far-infrared wavelengths is provided by {\it Herschel} (Pilbratt et al. 2010),
and we ultilise here the final public image products
from three major guaranteed-time surveys.
PACS (Poglitsch et al. 2010) images at 100\,$\mu$m and 160\,$\mu$m, reaching rms depths
of 0.17 and 0.42\,mJy respectively are provided by a combination of the data
obtained through the GOODS-{\it Herschel} (Elbaz et al. 2011) and the PACS Evolutionary Probe
(PEP; Lutz et al. 2011) surveys,
while SPIRE (Griffin et al. 2010) images at 250\,$\mu$m, 350\,$\mu$m and 500\,$\mu$m, reaching
5.86, 6.34 and 6.88\,mJy respectively (including confusion noise) are provided by
the {\it Herschel} Multi-tiered Extragalactic Survey (HerMES; Oliver et al. 2010, 2012).

Because the {\it Herschel} (and especially the SPIRE) imaging has such low angular resolution
compared to the ALMA imaging, care must be taken to attempt to deconfuse the {\it Herschel}
images in order to avoid obtaining biased, or artifically accurate far-infrared photometry
for the ALMA sources (see Appendix A, Fig.\,A1).
We therefore fitted the {\it Herschel} maps of the HUDF region with appropriate beams
centred at the positional priors of all the ALMA and $24\,{\rm \mu m}$ sources in the field.
The best-fitting beam normalizations, and associated covariance errors then allowed us
to extract {\it Herschel} fluxes/non-detections.

\subsubsection{Radio imaging}
Until recently, the deepest available radio imaging in the HUDF was provided by the 1.4\,GHz
observations of the Extended CDFS as described by Kellermann et al. (2008) and Miller et al. (2008). This produced imaging
with a 2.8\,$\times$\,1.6\,arcsec beam reaching a typical rms sensitivity of $\sigma_{1.4} \simeq 7.5\,{\rm \mu Jy}$.
However, recently (March 2014 -- September 2015) a new, ultra-deep, JVLA
4--8\,GHz survey has been
undertaken within GOODS-South, with a single pointing (7.2\,arcmin primary beam
at 6\,GHz) centred on the HUDF
(at RA 03$^h$ 32$^m$ 38.6$^s$, Dec $-$27$^{\circ}$46$^{\prime}$59.83$^{\prime \prime}$). This new imaging, which we utilise here and
in a companion paper on mm/radio source sizes within the HUDF (Rujopakarn et al. 2016),
comprises 149, 17 and 11 hours of imaging in the A, B, and C configurations respectively. The result
is an image with a synthesized beam of 0.31\,$\times$\,0.61 arcsec (PA = $-$3.6\,deg), reaching an 
rms sensitivity at 6\,GHz of $\sigma_{6} \simeq 0.32\,{\rm \mu Jy}$ per beam at the phase centre, and
$\sigma_{6} \simeq 0.35\,{\rm \mu Jy}$ per beam at the edge of the HUDF.
This imaging, which in effect (for a power-law radio spectral slope of $\alpha = 0.8$, $f_{\nu} \propto \nu^{-\alpha}$)
is $\simeq 10$ times deeper than the pre-existing 1.4\,GHz radio map, reveals 27 radio sources
with peak S/N $>$ 5 within the 4.5\,arcmin$^{2}$ area of the HUDF marked in
Fig.\,1.

\subsubsection{X-ray imaging}
The deepest X-ray imaging in the HUDF is provided by the 4\,Msec imaging 
with {\it Chandra} of the Chandra Deep Field South (CDFS)
which reaches an on-axis flux-density detection limit of
$\simeq 3 \times 10^{-17}\,{\rm erg\,cm^{-2}\,s^{-1}}$ across the full soft+hard 
band (0.5--8\,keV) (Xue et al. 2011). Various authors have studied
the galaxy counterparts of the X-ray sources
within the wider GOODS-South field (e.g. Rangel et al. 2013; Hsu et al. 2014; Giallongo
et al. 2016) but for the present study focussed on the
HUDF we work with the original X-ray positions, and establish
our own galaxy identifications and redshift information as required.

\subsubsection{Optical/near-infrared spectroscopy}
New spectroscopic observations of the HUDF were taken with MUSE as part of the
guaranteed-time programme between 2014 September and 2015 January.
The MUSE IFU provides full spectral coverage spanning 4770$-$9300\AA, and a
contiguous field of view of 60\,$\times$\,60\,arcsec, with a spatial sampling of
0.2\,arcsec\,/\,pixel, and a spectral resolution of $R$\,=\,3500 at $\lambda$\,=\,7000\AA.
The publicly-available MUSE data in the HUDF comprises a 3\,$\times$\,3 mosaic
of $\simeq$\,18.2\,ksec integrations, plus a single deep $\simeq$\,65\,ksec exposure
in the centre of the field.

We downloaded
the public dataset and reduced it using the {\sc esorex} pipeline.
This pipeline identifies the location of the data on the CCD using the flat-field
image, and then extracts the flat-field, arc and science data. It
then wavelength calibrates and flat-fields each slice and constructs the data cube.
Each science exposure was interspersed with a flat field to
improve the slice-by-slice flat field (illumination).
Residual slice-to-slice variations were then modelled and removed using a set of custom
routines which attempt to model the (wavelength-dependent) offsets.
Sky subtraction was performed on each sub-exposure by identifying and subtracting
the sky emission using blank areas of sky at each wavelength slice (after masking
continuum sources), and the final mosaics were constructed using an average with
a 3$\sigma$-clip to reject cosmic rays, using point sources in each (wavelength collapsed)
image to register the cubes.  The final cube was then registered to
the \emph{HST}\,/\,WFC3/IR $J_{125}$ image using point sources in both
frames.  Flux calibration was carried out using observations of known standard stars
at similar airmass which were taken immediately before or after the science observations
(and in each case we confirmed the flux calibration by measuring the flux density
of stars with known photometry in the MUSE science field).
More details on the MUSE HUDF project will be provided in Bacon et al. (in preparation).

To search for redshifts from each ALMA-identified source, we extracted one- and two-dimensional
spectra from a 1.5\,$\times$\,1.5\,arcsec region centred at the ALMA position and
searched for emission and absorption lines.  This yielded spectroscopic redshifts for
6 of the 16 sources listed in Table\,2, of which 4 are new, with the other 2 confirming previous
ground-based redshifts derived using FORS2 on the VLT (Kurk et al. 2013; Vanzella et al. 2008).

The new redshifts being provided by MUSE add to an already impressive database of
spectroscopic redshifts in the GOODS-South field, and in the HUDF in particular. The
various pre-existing ground-based spectroscopic campaigns are summarized in Parsa et al. (2016),
but in recent years {\it HST} WFC3/IR near-infrared grism spectroscopy has also
made an important contribution, with a combination of the 3D-HST programme, and CANDELS supernovae
grism follow-up observations delivering $\simeq 1000$ redshifts
in the GOODS-South field (Brammer et al. 2012; Skelton et al. 2014;
Morris et al. 2015; Momcheva et al. 2016). While many of the {\it HST} redshifts simply provide
(useful) confirmation of the results of earlier ground-based spectroscopic observations,
this grism work has been particularly helpful in filling in the traditional `redshift desert'
between $z \simeq 1.2$ and $z \simeq 2$, where relatively few strong emission lines are
accessible in the optical regime.

In total, these multiple efforts (extending over the last $\simeq 15$ years) have yielded
spectroscopic redshifts for nearly 3000 galaxies in the GOODS-South field, with over 200
robust spectrocopic redshifts now available within the sub-region defined by the HUDF.
As a consequence of this uniquely rich/dense spectroscopic database, we are able to provide
spectroscopic redshifts for 13 of the 16 galaxies in the final ALMA-selected sample (the selection of
which is described below in Sections 3 and 4). These redshifts, along with the
appropriate references, are given in the final column of Table\,2.

The redshift of one ALMA-identified source (UDF3) is confirmed independently
from our ALMA observations, via the detection of 3 spectral lines from CO,
CI and H$_2$O in our ALMA datacube (see later, and Ivison et al.,
in preparation).

\begin{table*}
\begin{normalsize}
\begin{center}
  \caption{Details of the final sample of 16 ALMA-detected sources in the HUDF, selected and refined as
    described in Sections 3 and 4. Column 1 gives source numbers as also used in Fig.\,1, while
    columns 2 and 3 give the positions of the ALMA sources as determined from the 1.3-mm map.
    Estimated total flux densities (see Section 4.3 for details on corrections to
    point-source flux densities) and peak S/N at 1.3-mm are given in columns 4 and 5. Then in columns 6 and 7 we
    give the co-ordinates of the adopted galaxy counterpart as determined from the $H_{160}$
    WFC3/IR {\it HST} image of the HUDF. Columns 8 and 9 give the positional offsets between the
    ALMA and {\it HST} positions, before ($\Delta_1$) and after ($\Delta_2$)
    moving the {\it HST} positions south by 0.25\,arcsec
    (see Section 4.2 for a discussion of this astrometric shift, and its calculation/motivation; the cumulative
    distributions produced by these two alternative sets of positional offsets are shown in Fig.\,2).
    Column 10 gives the total $H_{160}$ magnitude of each {\it HST} galaxy
    counterpart, while column 11 lists the redshift for each source. The 13 spectroscopic redshifts are given
    to three decimal places, with the three photometric redshifts given to two decimal places.
    The sources of the spectroscopic redshifts are indicated by the flag in column 12, and are as follows:
    1) Brammer (private communication); 2) MUSE, this work; 3) Momcheva et al. (2016); 4) Kurk et al. (2013);
    5) Hathi, Malhotra \& Rhoads (2008); 6) Vanzella et al. (2008).}
\label{tab:Sources}
\setlength{\tabcolsep}{1.43 mm} 
\begin{tabular}{lclcrclcccll}
\hline
ID & RA ({\small ALMA}) &  Dec ({\small ALMA}) & $S_{\rm 1.3mm}$     &  S/N\phantom{1}            & RA ({\small {\it HST}}) & Dec ({\small {\it HST}}) & $\Delta_1$ & $\Delta_2$ & $H_{160}$& \phantom{0}$z$ &Ref\\
   & /deg      &  \phantom{Dec}/deg       & /${\rm \mu Jy}$  &  {\small 1.3mm}  & /deg     & \phantom{Dec}/deg  & /arcsec  & /arcsec  & /ABmag & &\\
\hline
UDF1  & 53.18348 & $-$27.77667 &  924 $\pm$ 76 &  18.37  & 53.18345 & $-$27.77658 & 0.33 & 0.13 & 24.75 &    3.00  &  \\    
UDF2  & 53.18137 & $-$27.77757 &  996 $\pm$ 87 &  16.82  & 53.18140 & $-$27.77746 & 0.38 & 0.15 & 24.70 &    2.794 & \phantom{R}1\\    
UDF3  & 53.16062 & $-$27.77627 &  863 $\pm$ 84 &  13.99  & 53.16060 & $-$27.77613 & 0.51 & 0.27 & 23.41 &    2.541 & \phantom{R}2\\    
UDF4  & 53.17090 & $-$27.77544 &  303 $\pm$ 46 &   6.63  & 53.17090 & $-$27.77539 & 0.18 & 0.06 & 24.85 &    2.43  &  \\    
UDF5  & 53.15398 & $-$27.79087 &  311 $\pm$ 49 &   6.33  & 53.15405 & $-$27.79091 & 0.24 & 0.42 & 23.30 &    1.759 & \phantom{R}3\\    
UDF6  & 53.14347 & $-$27.78327 &  239 $\pm$ 49 &   4.93  & 53.14347 & $-$27.78321 & 0.22 & 0.03 & 22.27 &    1.411 & \phantom{R}2\\    
UDF7  & 53.18051 & $-$27.77970 &  231 $\pm$ 48 &   4.92  & 53.18052 & $-$27.77965 & 0.21 & 0.06 & 24.17 &    2.59  &  \\    
UDF8  & 53.16559 & $-$27.76990 &  208 $\pm$ 46 &   4.50  & 53.16555 & $-$27.76979 & 0.43 & 0.22 & 21.75 &    1.552 & \phantom{R}4\\    
UDF9  & 53.18092 & $-$27.77624 &  198 $\pm$ 39 &   4.26  & 53.18105 & $-$27.77617 & 0.46 & 0.40 & 21.41 &    0.667 & \phantom{R}2\\
UDF10 & 53.16981 & $-$27.79697 &  184 $\pm$ 46 &   4.02  & 53.16969 & $-$27.79702 & 0.42 & 0.56 & 23.32 &    2.086 & \phantom{R}3\\
UDF11 & 53.16695 & $-$27.79884 &  186 $\pm$ 46 &   4.02  & 53.16690 & $-$27.79869 & 0.54 & 0.31 & 21.62 &    1.996 & \phantom{R}2,\,4\\
UDF12 & 53.17203 & $-$27.79517 &  154 $\pm$ 40 &   3.86  & 53.17212 & $-$27.79509 & 0.39 & 0.28 & 27.00 &    5.000 & \phantom{R}5\\
UDF13 & 53.14622 & $-$27.77994 &  174 $\pm$ 45 &   3.85  & 53.14615 & $-$27.77988 & 0.31 & 0.24 & 23.27 &    2.497 & \phantom{R}3\\
UDF14 & 53.17067 & $-$27.78204 &  160 $\pm$ 44 &   3.67  & 53.17069 & $-$27.78197 & 0.24 & 0.06 & 22.76 &    0.769 & \phantom{R}2\\
UDF15 & 53.14897 & $-$27.78194 &  166 $\pm$ 46 &   3.56  & 53.14902 & $-$27.78196 & 0.18 & 0.36 & 23.37 &    1.721 & \phantom{R}3\\
UDF16 & 53.17655 & $-$27.78550 &  155 $\pm$ 44 &   3.51  & 53.17658 & $-$27.78545 & 0.22 & 0.09 & 21.42 &    1.314 & \phantom{R}2,\,6\\
\hline
\end{tabular}
\end{center}
\end{normalsize}
\end{table*}

\section{ALMA source extraction}
\label{sec:sources}

To detect sources in the ALMA image we first constructed a noise map which
provides an estimate of the local pixel-to-pixel variance on scales
comparable to the beam. For every pixel we evaluated the standard
deviation of flux-density values within a window of size $10\times\theta$
where $\theta = \surd(1.331\times a\times b)$ where $a$ and $b$ are
the semi-major and semi-minor axes of the synthesized beam. To mitigate 
the contribution from bright sources we applied local 4$\sigma$ clipping before
evaluating the standard deviation. This noise map then allowed us to
construct a signal-to-noise map which we used as the detection image. A
simple peak-finding algorithm was adopted: first we identified 
significant ($>5\sigma$) peaks, and co-added these to construct a model
point spread function (PSF). This PSF was then used to subtract sources
from the map as they were identified, starting from the most
significant peak and moving down until a threshold floor significance
was reached.

We limited source detection to map regions with $\sigma_{1.3}~<~40\,\mu$Jy, which yielded
an effective survey area of 4.4\,arcmin$^2$. Within this area we detected
47 candidate sources with peak S/N~$>$~3.5 and a point-source flux-density
$S_{1.3}~\ge~120\,{\rm \mu Jy}$. However, running an identical source extraction
on the negative map (i.e. the real map multiplied by $-1$) yielded
29 sources with S/N~$>$~3.5 and $S_{1.3}~\ge~120\,{\rm \mu Jy}$.

It is interesting to consider whether this is as expected. Adopting a beam angular
radius of 0.35\,arcsec, the map contains $\simeq 42000$ beams, and thus, based
on Gaussian statistics, we would expect $\simeq 10$ spurious peaks with S/N $>$ 3.5.
However, if, as pointed out by Condon (1997) and Condon et
al. (1998), there are effectively twice as many statistically independent noise samples
as naively expected, then these numbers rise to $\simeq 20$ spurious peaks with S/N $>$ 3.5, in much better accord with what is actually found from
source extraction on the negative image. The noise level only then needs to be altered by $< 5$\%
to bring the numbers into essentially exact agreement. This suggests that there is no serious
issue with the noise in the map, and indeed a full simulation of the image involving
beam filtering of white noise confirms that the numbers and S/N distribution of the
spurious sources as derived from the negative map are as expected (Peacock et al., in preparation).

The implication is that only $\simeq 15-20$ of the `sources' extracted from the positive
image are real, and the challenge is to identify which these are.

\section{Galaxy counterparts and Source List Refinement}
\label{sec:galid}

\subsection{Galaxy identifications}

Refining the source list is not as straightforward as, for example, confining attention
to sources with S/N $>$ 4, given that there are 7 such `sources' in the negative map.
A clean source list can be produced by limiting the selection to S/N~$>$~6, but this leaves only
5 sources, and clearly does not make optimal use of the new ALMA data. Fortunately we are able
to use the excellent positional accuracy of the ALMA sources, along with the wealth of
supporting multi-frequency data, to identify which of the $>$3.5-$\sigma$ peaks extracted as
described in the previous section correspond to real ALMA sources.

Firstly, it was very evident that the brightest sources in the ALMA source list had obvious
galaxy counterparts in the {\it HST} imaging, with positions coincident to within
$< 0.5$\,arcsec. Excellent positional correspondence is certainly expected since,
even for a 3.5-$\sigma$ source, the predicted 1-$\sigma$ uncertainty in RA and Dec given a beam-size of
0.7\,arcsec (FWHM) is 0.085\,arcsec, and the corresponding conservative 3-$\sigma$ search radius
is $\simeq 0.25$\,arcsec (see Ivison et al. 2007). This level
of positional accuracy is approached by the positional offsets between the ALMA
and radio sources (albeit increased by a factor $\simeq 2$ by image pixelisation for a
10-$\sigma$ source), but ambiguity over the the true centroid of some of the
{\it HST} counterparts, astrometric uncertainties, and potentially even optical-mm
physical offsets combine to make the positional correspondence between the ALMA sources
and their {\it HST} counterparts not quite as precise as theoretically predicted.
Nonetheless, for the obviously secure galaxy identifications confirmed by radio detections
we found that $\sigma_{pos} = 0.2$\,arcsec, and so adopted a   
search radius of 0.6\,arcsec. This very small search radius makes chance
ALMA--{\it HST} coincidences very unlikely for all
but the very faintest galaxies, and indeed, applied to the negative
ALMA source list, yields only 3 random galaxy identifications. 

Applied to the positive source sample, searching for near-infrared galaxy candidates
within a radius of 0.6\,arcsec (which obviously assumes that real
ALMA sources have an {\it HST} counterpart in the UDF09+UDF12 imaging; see below) reduced the potential
source sample to 21 sources, 12 of which are independently confirmed as real sources
in the new ultra-deep JVLA 6\,GHz imaging (see Rujopakarn et al. 2016).

\begin{figure}
\begin{center}
\includegraphics[scale=0.55]{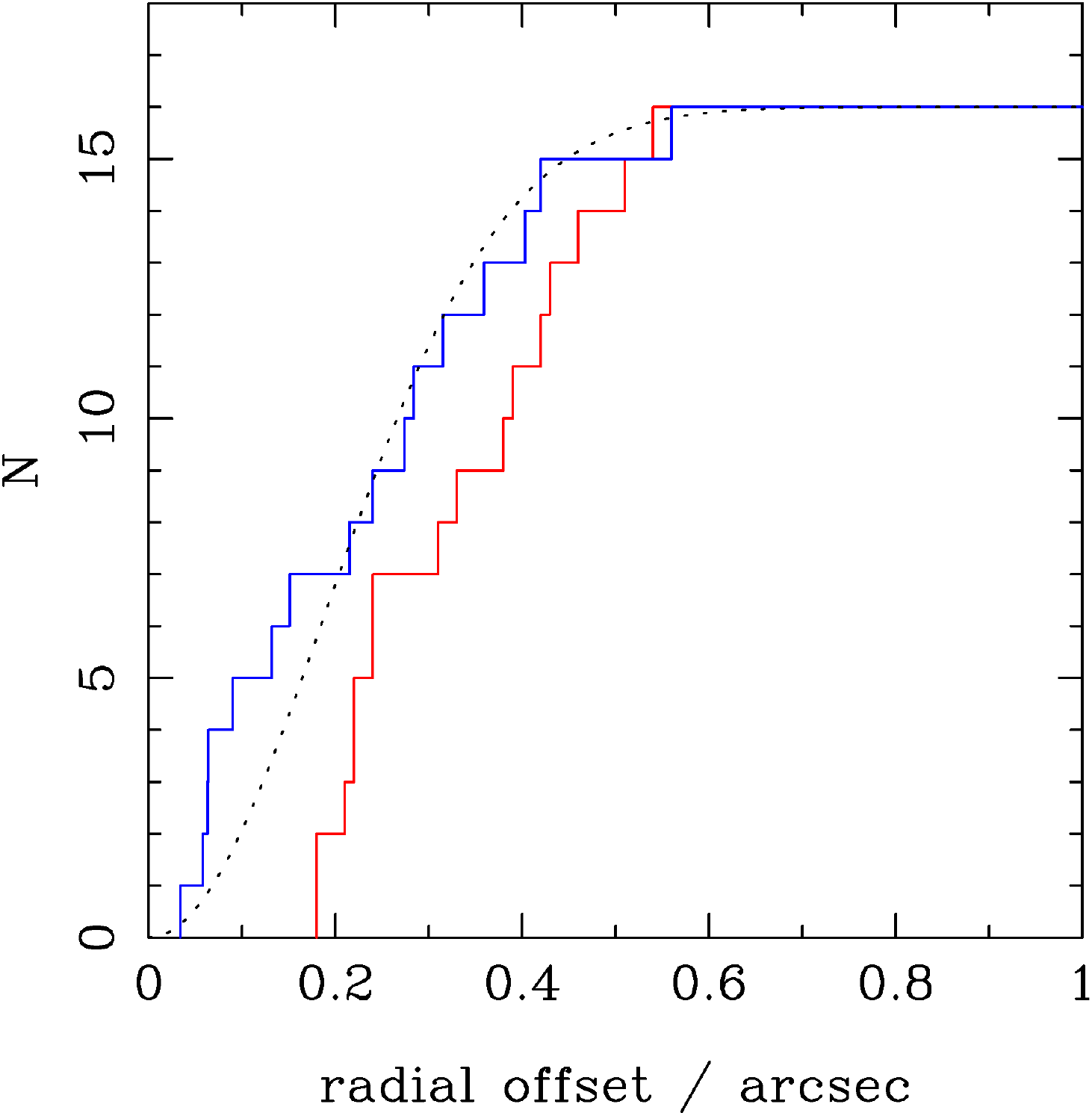}
\end{center}
\caption{The cumulative distribution of radial separations between
  the 16 ALMA 1.3-mm sources and their adopted {\it HST} near-infrared
  galaxy counterparts, as detailed in Table 1. The red line shows the distribution prior to moving the
  {\it HST} astrometric reference frame, while the blue line shows the significantly
  tighter distribution which arises after moving all {\it HST} positions south by 0.25
  arcsec (see Section 4.2). The dashed black line shows the cumulative distribution expected
  assuming a Gaussian distribution of positional errors, with $\sigma_{\rm RA} = \sigma_{\rm Dec} = 0.2$\,arcsec (see Section 4 for
  further details).}
\label{fig:alma_image}
\end{figure}

\subsection{Astrometry issues}
While performing the galaxy counterpart identification
we noticed a systematic positional offset, primarily in declination, between the ALMA positions
and the positions of their (often obvious) galaxy counterparts. We checked this using the stacking
results described below (see Section 7), and deduced that the ALMA positions lie
systematically two ALMA pixels (i.e. $\simeq 0.24$\,arcsec) south of the {\it HST} positions.

At the same time we found that, for the 12 radio-detected ALMA sources, there was
no significant systematic offset in position between the JVLA and ALMA positions, with the mean offset
being $< 35$ mas in both RA and Dec (despite the use of different phase calibrators).

We explored this further, using the radio data which extends over somewhat more of GOODS-South than
just the HUDF region, and found that the {\it HST} positions (based on the
$H_{160}$ HUDF/GOODS-South imaging) are systematically offset from the radio positions
by +0.279\,arcsec in Dec, and $-$0.076\,arcsec in RA, and 
are systematically offset from the 2MASS positions by 
by +0.247\,arcsec in Dec, and +0.035\,arcsec in RA (see Rujopakarn et al. 2016).

Given the apparent consistency of the offset in Dec, we experimented with simply shifting
all {\it HST} positions south by 0.25\,arcsec, and repeating the identification process.
The impact of this change is documented in Table\,2
(which gives the ALMA-{\it HST} offset in arcsec
both before and after modifying the {\it HST} co-ordinate system), and
in Fig.\,2, which shows the tightening of the distribution of positional offsets
after applying this shift.

One might reasonably ask which coordinate system is correct? However, the agreement between the
JVLA, ALMA and 2MASS positions strongly suggests that it is the {\it HST} co-ordinate system
that is wrong. In fact, there is good evidence that this is the case. The HUDF and CANDELS astrometry
has been tied to the GOODS ACS astrometric solution. Referring back
to the documentation accompanying the GOODS 2008 data release, it transpires that, for v2.0 of the GOODS {\it HST} data, the GOODS team
decided to shift the GOODS-North coordinate system south by 0.3\,arcsec in Dec. However, a similar shift
was not applied to the GOODS-South v2.0 image
mosaic\footnote{https://archive.stsci.edu/pub/hlsp/goods/v2/h\_goods\_v2.0\_rdm.html\#4.0}. The stated rationale was the lack of available comparison
data of the necessarily quality in the GOODS-South field
at the time, although it was claimed that
`{\it an analysis of Chandra Deep Field South astrometry by the MUSYC team using
the Yale Southern Observatory Double Astrograph telescope suggests that the
mean GOODS-S world coordinate system is absolutely accurate at a level
better than 0.1 arcseconds}'. It seems clear, now, in the light of the new ALMA and JVLA data, that this is not
the case, and that the GOODS-South world co-ordinate system should, as was done for GOODS-North v2.0, be
moved south by $\simeq 0.25 - 0.3$\,arcsec. This could be implemented for future CANDELS/HUDF releases, but for now
we continue to give {\it HST} co-ordinates in Table\,2 in the existing GOODS-South/CANDELS/HUDF system
(to ease object identification in existing {\it HST}-based catalogues), and simply note the
improved positional correspondence achieved for our galaxy identifications when this astrometric
shift is systematically applied.

\subsection{Final ALMA HUDF source sample}
Application of this astrometric shift, as well as tightening the positional agreement
for solid identifications, also led to the rejection of three others, and finally we also
rejected the two sources for which the only available galaxy counterpart had
$H_{160} > 28.5$. This latter decision was made on continuity grounds (no other remaining
ALMA source has $H_{160} > 27$), and because, as evidenced from searching for galaxy counterparts
to the negative pseudo-sources, the {\it HST} source density at these extreme depths
is expected to yield $\simeq 2$ chance coincidences within a search radius $r < 0.55$\,arcsec.

This leaves the final sample of 16 sources whose positions in the ALMA map are indicated
in Fig.\,1. Accurate positions (both for the ALMA sources and their {\it HST} counterparts) and flux densities
are given in Table\,2 (along with ALMA-{\it HST} positional offsets before and after the afore-mentioned
astrometric shift and, for completeness and ease of reference, redshifts; see Section 5 below).
Reassuringly, this final 16-source sample, culled on the basis of the search for near-infrared
counterparts in the deep {\it HST} imaging, still contains all 12 radio-detected sources
from the original $\simeq 50$-source sample of 1.3-mm peaks (see Table\,3).

Our final sample of 16 sources is thus very similar in size to what would be expected
on the basis of comparing the numbers of positive and negative $>$3.5-$\sigma$ peaks as explained
above in Section 3. Nevertheless, one might be concerned that, by culling the ALMA
source list on the basis of secure galaxy identifications, we are effectively excluding the
possibility that the ALMA map might reveal sources that are not visible in the deep {\it HST}
imaging. In fact, we believe this is not a concern for three reasons. First, it must be remembered
that, as a result of the UDF12 programme (Ellis et al. 2013; Koekemoer et al. 2013; McLure et al. 2013; Dunlop et al. 2013), the near-infrared
imaging in this field is the deepest ever obtained, and we completed our search for galaxy
counterparts in a stack of $Y_{105}$+$J_{125}$+$J_{140}$+$H_{160}$ imaging reaching a detection threshold
of $>30$\,mag. Second, continuity arguments imply no significant number of
near-infrared non-detections of the ALMA sources in our sample; as can be seen from Table\,2, even though the sample has been culled of
objects that lack galaxy counterparts at $H_{160} < 28.5$, in practice
all the objects have $H_{160} \leq 27$, and indeed 15/16 have $H_{160} < 25$. Thus, in the context
of the extremely deep WFC3/IR imaging available here, the galaxy counterparts of the secure ALMA sources
are relatively bright, and it is extremely hard to argue that only slightly fainter
ALMA sources should suddenly have galaxy counterparts that
are two orders-of-magnitude fainter
in the near-infrared. Third, we initially uncovered one reasonably significant source (S/N= 4.9, originally
source number 9 in the master sample) for which we could not find any galaxy counterpart down to
the limit of the WFC3/IR imaging. Notwithstanding the knowledge that there are two $>4.5$-$\sigma$
pseudo-sources in the negative image,
we still explored this source in detail,
in case it represented an extremely unusual
(perhaps very distant) dusty object. As part of this exploration we
interrogated the data splits described
earlier in Section 2.1, and found that this source featured at $\simeq 6$\,$\sigma$ in one half of the
time-stream, but at less that 3\,$\sigma$ in the other half. This is not the behaviour expected for a genuine
5-$\sigma$ source, and confirmed our suspicion that this was indeed our brightest false single-band detection.

\begin{figure*}
\begin{center}
\includegraphics[scale=0.5]{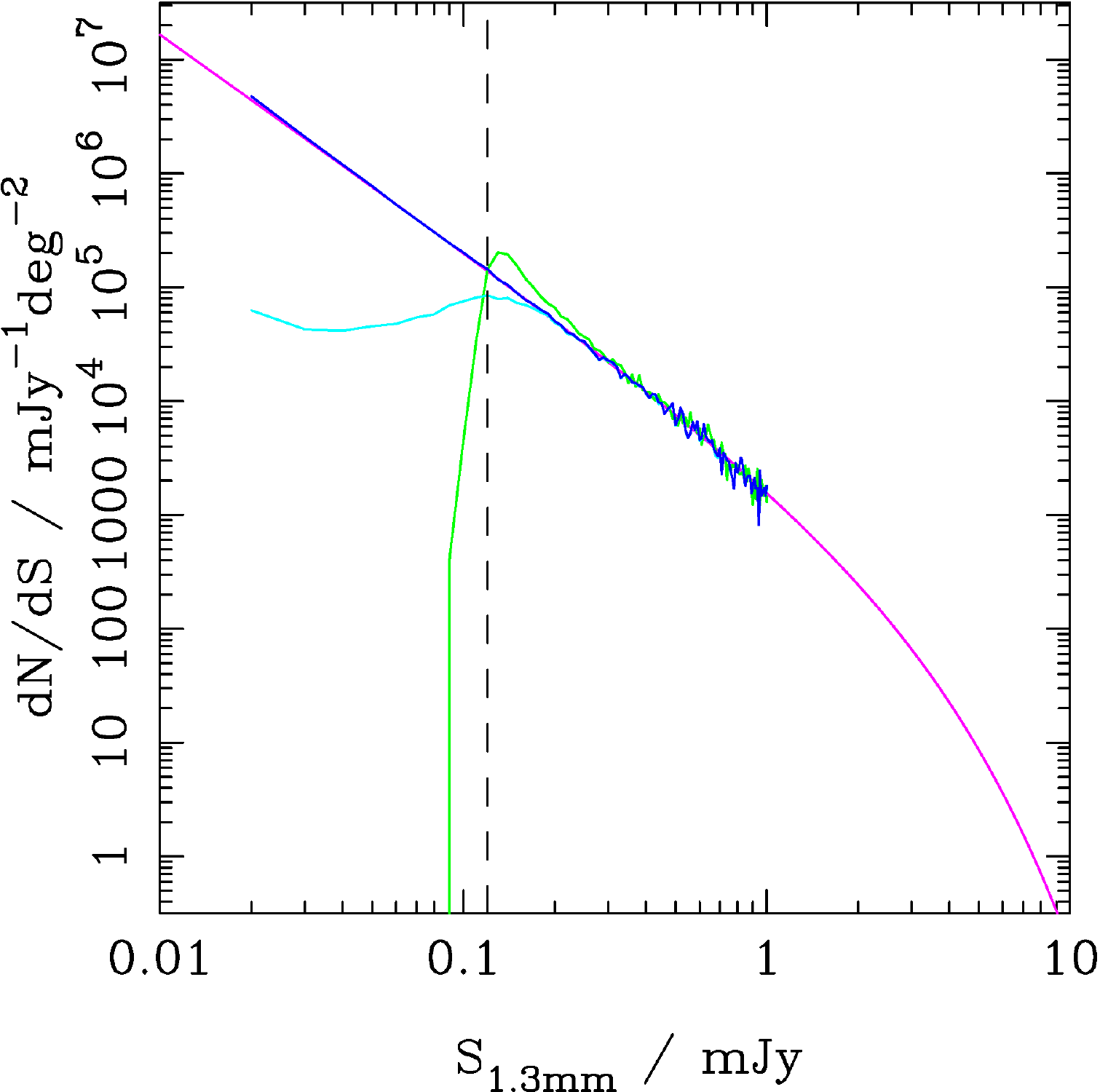}
\hspace*{2cm}
\includegraphics[scale=0.5]{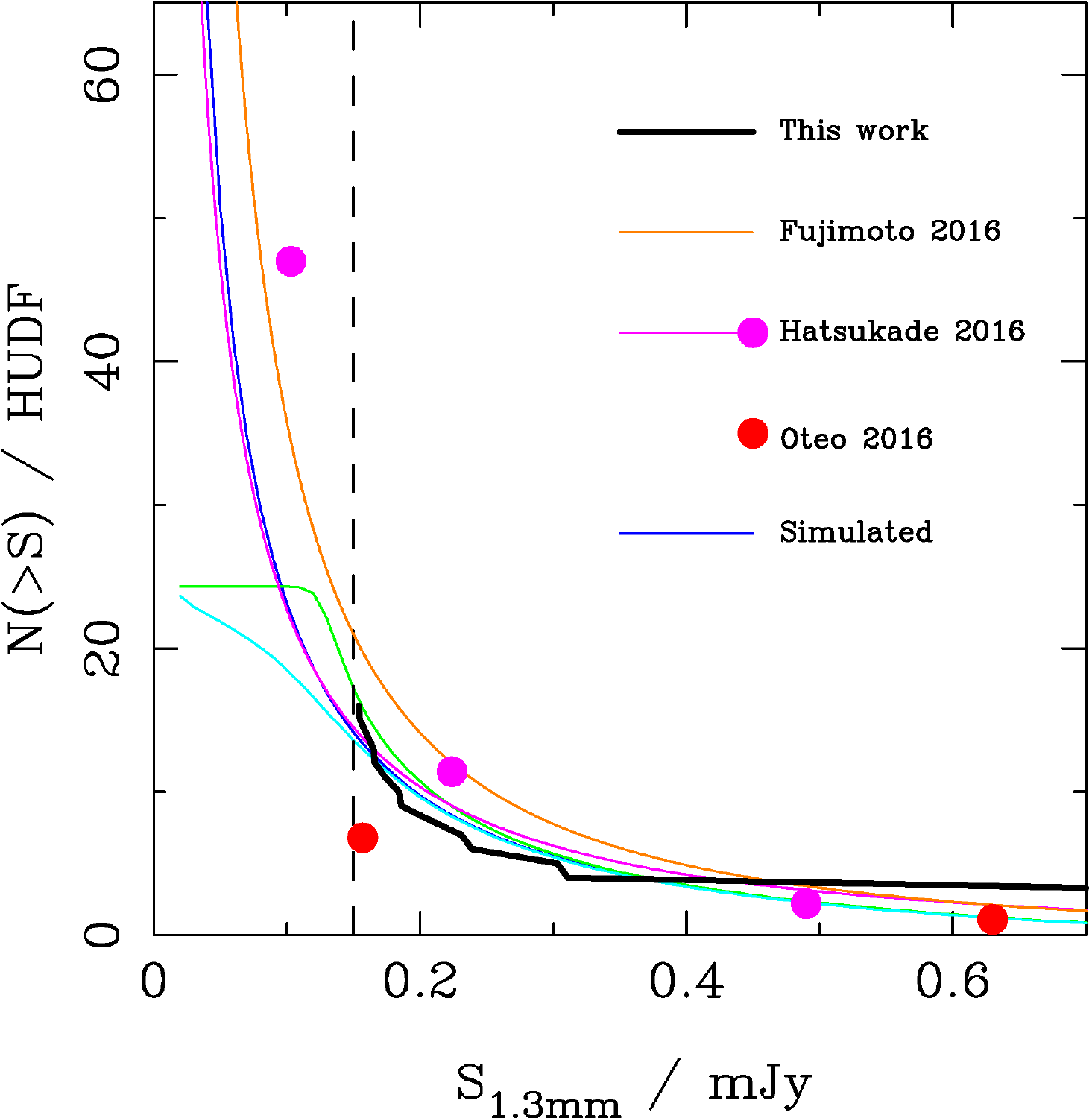}
\end{center}
\caption{{\bf Left:} Completeness and flux boosting in the HUDF ALMA 1.3-mm image. The magenta
  curve shows the differential number counts as given by the Schechter-function fit of Hatsukade
  et al. (2016), shifted from 1.1-mm to 1.3-mm by scaling flux density down by a factor of 1.65
  (Fujimoto et al. 2016). The navy blue line shows the results of randomly drawing 1000 HUDF-size samples
  from this number-count model, and so represents the input for our source-injection simulations.
  The green curve shows the number counts as retrieved from the 1000 ALMA HUDF images using the same
  source extraction technique as applied to uncover the real sources, and insisting on S/N $>$ 3.5.
  The cyan curve shows the same retrieved sources, but at their input rather than retrieved flux densities.
  The vertical dashed line indicates our chosen point-source detection threshold of $S_{1.3} = 0.120$\,mJy, where it can
  be seen that flux boosting almost perfectly offsets the effect of incompleteness.
  {\bf Right:} Cumulative number counts scaled to the size of the HUDF. The results from the
  present study are plotted as the thick black line
  (after scaling by a factor of 1.25 to account for average resolved flux density), and compared with the results of other
  recent studies (scaled to 1.3-mm) as indicated in the legend.
  We found $\simeq 1.5$ times sources than anticipated given the number counts from Fujimoto
  et al. (2016) (orange curve), and also found fewer sources than indicated by the binned
  cumulative number counts reported by Hatsukade et al. (2016)
  (magenta points). However, our ALMA HUDF counts are in good agreement with integration of the
  Schechter fit to the differential counts provided by Hatsukade et al. (2016)
  (magenta line), and lie above the recent number-count
  results reported by Oteo et al. (2016) (red points), which were based on higher significance cuts, and arguably
  less biased pointings than the afore-mentioned studies. As in the left-hand panel,
  our simulated number counts are indicated by the navy blue line, and the impact of flux boosting+incompleteness
  is indicated by the green curve, with incompleteness indicated in cyan. Extrapolation of this model
  to still fainter flux densities suggests that to uncover $\simeq 100$ 1.3-mm sources in the HUDF would
  require reaching a detection limit of $S_{1.3} = 30\,{\rm \mu Jy}$ (i.e. $\simeq 4$ times deeper than achieved here).}
\end{figure*}

We conclude that, to the best of our ability (i.e. using all available supporting information,
utilising the negative ALMA `sample' as a control, and examining carefully various 50:50 splits of
the ALMA data) that the final sample presented in Table\,2 and Fig.\,1 represents all the robust
ALMA sources detected in our map with peak S/N $>$ 3.5, and point-source flux density
$S_{1.3} > 120\,{\rm \mu Jy}$.

While the sources listed in Table\,2 were all selected on the basis
of peak S/N $>$ 3.5, and point-source flux density
$S_{1.3} > 120\,{\rm \mu Jy}$, subsequent fitting to the images showed that at least
the first three sources are clearly resolved. For UDF1, UDF2 and UDF3, the ratio of
total to point-source flux density was found to be 1.26, 1.56, and 1.50 respectively, and
it is the total integrated flux density that is given in Table\,2. Thereafter, however,
we found that the fainter sources in the robust list were too faint for accurate total fluxes
to be estimated by individual source fitting. While it may well be the case that the fainter sources
are smaller, we decided it was unreasonable to assume they were simply point sources, an assumption
which would clearly bias there estimated fluxes systematically low (albeit a subset will be flux-boosted).

We therefore decided to make a systematic correction to the point-source
flux densities of the remaining 13 sources,
to provide a best estimate of their true total 1.3-mm flux densities. We created a stack of the brightest 5 sources,
and found that fitting to this yielded a total to point-source flux density ratio of $\simeq 1.3$. We therefore
decided to make a conservative, systematic correction to the point-source flux densities of
sources UDF4 through UDF16, by simply multiplying their point-source flux densities (and associated errors)
by a factor 1.25. It is these estimated total flux densities that are tabulated in column 4 of Table\,2, but
in column 5 we also give the original peak S/N ratio for each source (as derived at the detection stage).

Finally, we note that, for source UDF3, we have detected molecular
emission lines from H$_2$O, CO and CI which, as well as confirming its spectroscopic redshift at mm wavelengths,
also in this case make a significant contribution to the total flux density given in Table\,2.
Our best estimate is that removal of the line contribution
reduces the flux density of UDF3 from $S_{1.3} = 863 \pm 84\,{\rm \mu Jy}$ to $S_{1.3} = 717 \pm 134\,{\rm \mu Jy}$,
and we use this latter value as approriate for star-formation rate estimates later in this paper.
Sources UDF8 and UDF11 also appear to have emission lines within our sampled band-pass, but not at a level
that seriously impacts on the estimated continuum flux (see Ivison et al., in preparation).

\begin{table}
\begin{normalsize}
\begin{center}
  \caption{The radio (6\,GHz, JVLA) and X-ray (0.5--8\,keV) detections of the 16 ALMA
    sources in the HUDF. Radio flux densities and associated uncertainties are
    from the new ultra-deep JVLA image of the HUDF region obtained by
    Rujopakarn et al. (2016). We do not report radio-source positions here, simply
    because they are coincident with the ALMA positions within $50$\,milli-arcsec.
    The X-ray flux densities, and derived luminosities are the total
    (i.e. soft+hard) values derived from the
    {\it Chandra} 4\,Msec imaging (Xue et al. 2011; Hsu et al. 2014). The X-ray positions for all five detected sources differ
    from the ALMA positions by $< 0.5$\,arcsec.}
  \label{tab:RadXray}
\setlength{\tabcolsep}{1.45 mm} 
\begin{tabular}{lcrr}
\hline
ID & $S_{\rm 6GHz}$ & $S_{\rm X}$/$10^{-17}$ & $L_{\rm X}/10^{42}$  \\
   & /${\rm \mu Jy}$  & /${\rm erg\,cm^{-2}\,s^{-1}}$  &/${\rm erg\,s^{-1}}$ \\
\hline
UDF1  & \phantom{1}9.02 $\pm$ 0.57 &  150 $\pm$ \phantom{1}7\phantom{11}            &    25.1 $\pm$ 1.2 \\    
UDF2  & \phantom{1}6.21 $\pm$ 0.57 &                 &         \\    
UDF3  &           12.65 $\pm$ 0.55 &  \phantom{15}6 $\pm$ \phantom{1}3\phantom{11}  &    \phantom{1}0.8 $\pm$ 0.4\\    
UDF4  & \phantom{1}3.11 $\pm$ 0.62 &                 &         \\    
UDF5  & \phantom{1}6.25 $\pm$ 0.46 &                 &         \\    
UDF6  & \phantom{1}8.22 $\pm$ 0.51 &                 &         \\    
UDF7  &           18.69 $\pm$ 0.60 &  \phantom{15}8 $\pm$ \phantom{1}3\phantom{11}  &   \phantom{1}1.0 $\pm$ 0.4 \\    
UDF8 & \phantom{1}7.21 $\pm$ 0.47 &  330 $\pm$ 15\phantom{11}            &    20.0 $\pm$ 0.9 \\    
UDF9 & \phantom{1}2.92 $\pm$ 0.58 &                 &         \\
UDF10 & $< 0.70$                   &                 &         \\
UDF11 & \phantom{1}9.34 $\pm$ 0.74 &  \phantom{1}11 $\pm$ \phantom{1}4\phantom{11}  &    \phantom{1}0.8 $\pm$ 0.4 \\
UDF12 & $< 0.70$                   &         &         \\
UDF13 & \phantom{1}4.67 $\pm$ 0.53 &         &         \\
UDF14 & $< 0.68$                   &         &         \\
UDF15 & $< 0.68$                   &         &         \\
UDF16 & \phantom{1}5.49 $\pm$ 0.46 &         &         \\
\hline
\end{tabular}
\end{center}
\end{normalsize}
\end{table}

\section{Number Counts}

\subsection{Simulations, completeness, and flux boosting}

To quantify incompleteness and the impact of flux boosting, we performed a series of
source injection and retrieval simulations. To make this as realistic as possible,
we drew random samples from current best estimates of the
source counts at the depths of interest. The results shown in Fig.\,3a were based on 1000
realisations of an HUDF-size image, with the sources drawn randomly from the
Schechter-function fits to the 1.1-mm source counts given by
Hatsukade et al. (2016), after scaling the 1.1-mm flux densities to 1.3-mm values by
dividing by 1.65 (Fujimoto et al. 2016).
The scaled Hatsukade et al. (2016) differential number-count model fit
is plotted as a magenta line in Fig.\,3a,
with the resulting input to our simulations shown in navy blue
(1000 HUDF samples are not sufficient to sample the 1.3-mm number
counts brighter than $\simeq 1$\, mJy, but this is not important here).

We created a fake sky map by randomly placing single-pixel
point sources into an equivalent pixel grid as the real map
(with no clustering), convolved this with the ALMA PSF, and added this model
to the real map. These simulated sources were
then located and extracted from the image in exactly the
same way as for the real sources (see Section 3). We are then
able to quantify incompleteness by plotting the number of
sources retrieved (at S/N\,$>3.5 \sigma$) as a function of their
input flux density (cyan curve), and
the combined impact of incompleteness and flux boosting by
plotting the number of sources retrieved (at S/N\,$>3.5 \sigma$)
as a function of output flux density (green curve).

From this plot it can be seen that the differential number counts at our chosen
point-source selection limit of S/N\,$> 3.5 \sigma$, $S_{1.3} > 0.12$\,mJy should
be basically correct, with the effect of incompleteness
almost exactly compensated by the flux-boosting of intrinsically fainter sources.

\begin{figure*}
\begin{center}
\begin{tabular}{cccc}
\includegraphics[scale=0.228]{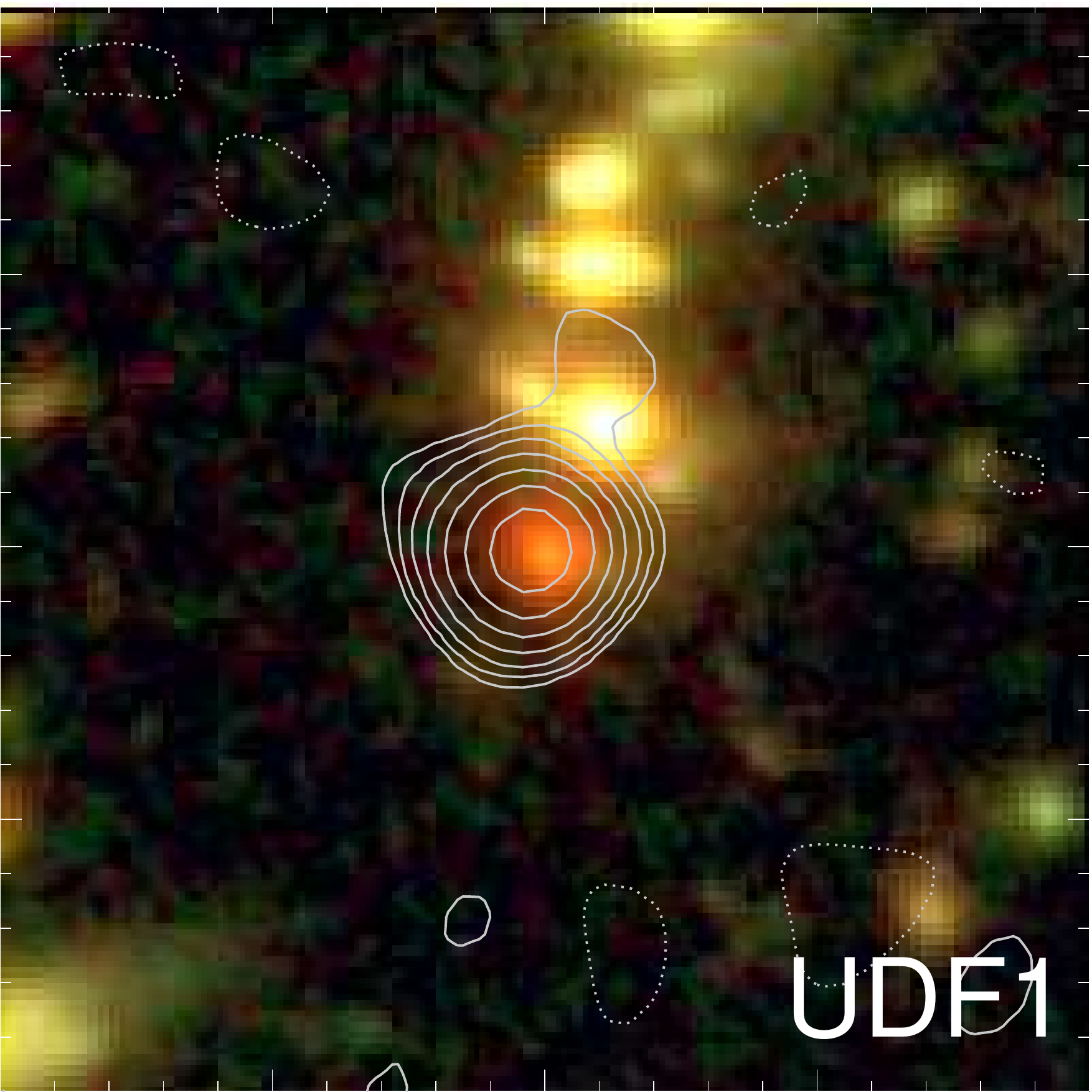}&
\includegraphics[scale=0.228]{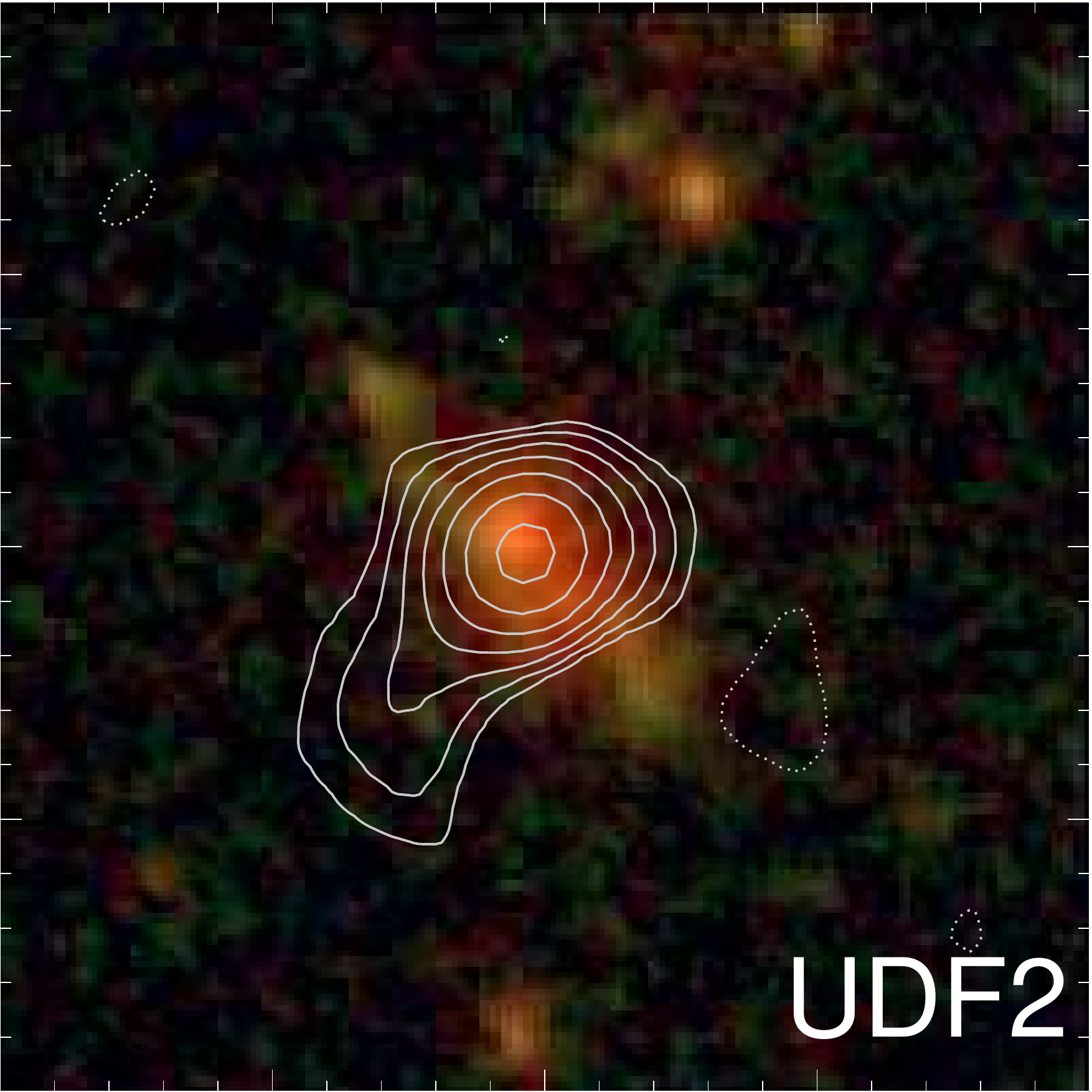}&
\includegraphics[scale=0.228]{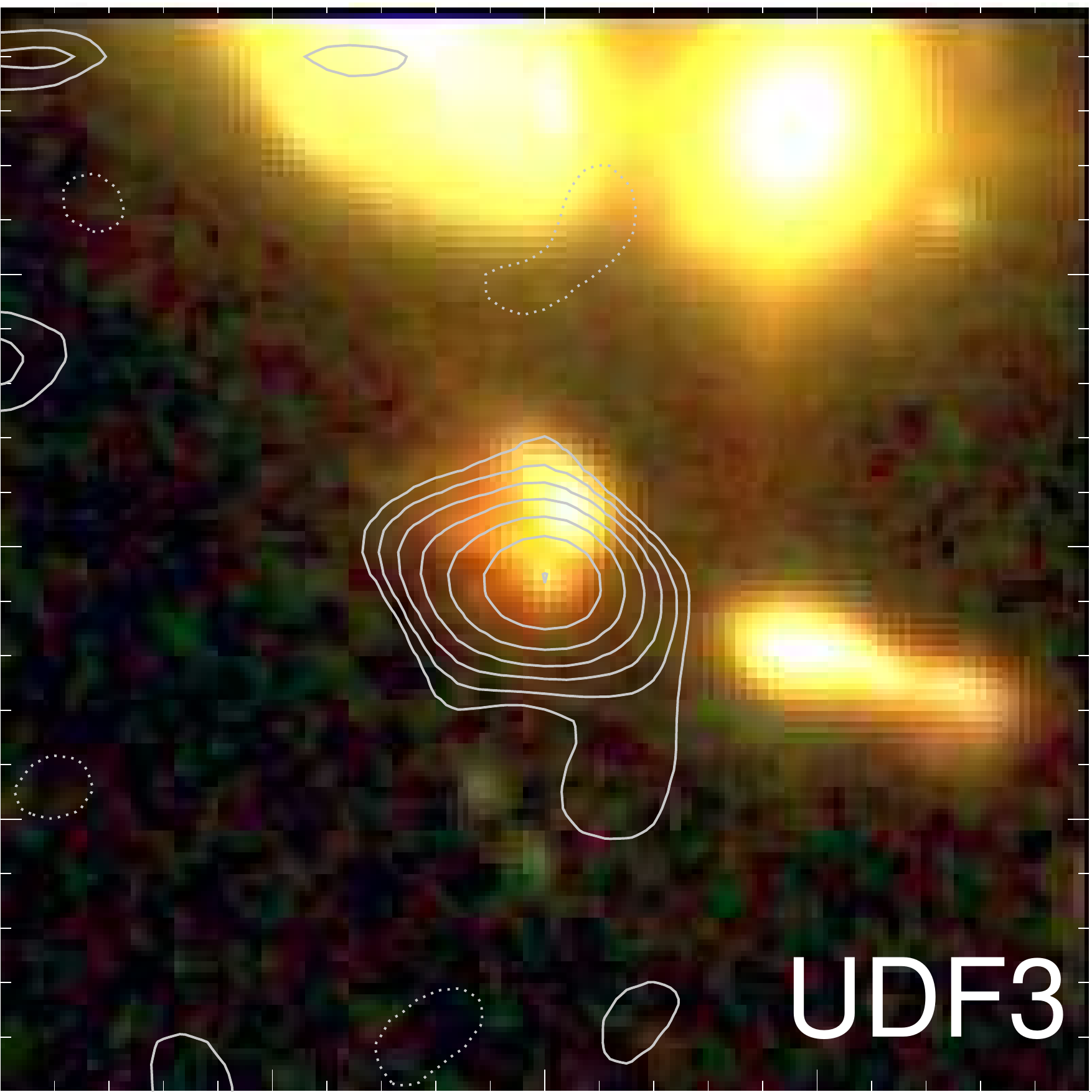}&
\includegraphics[scale=0.228]{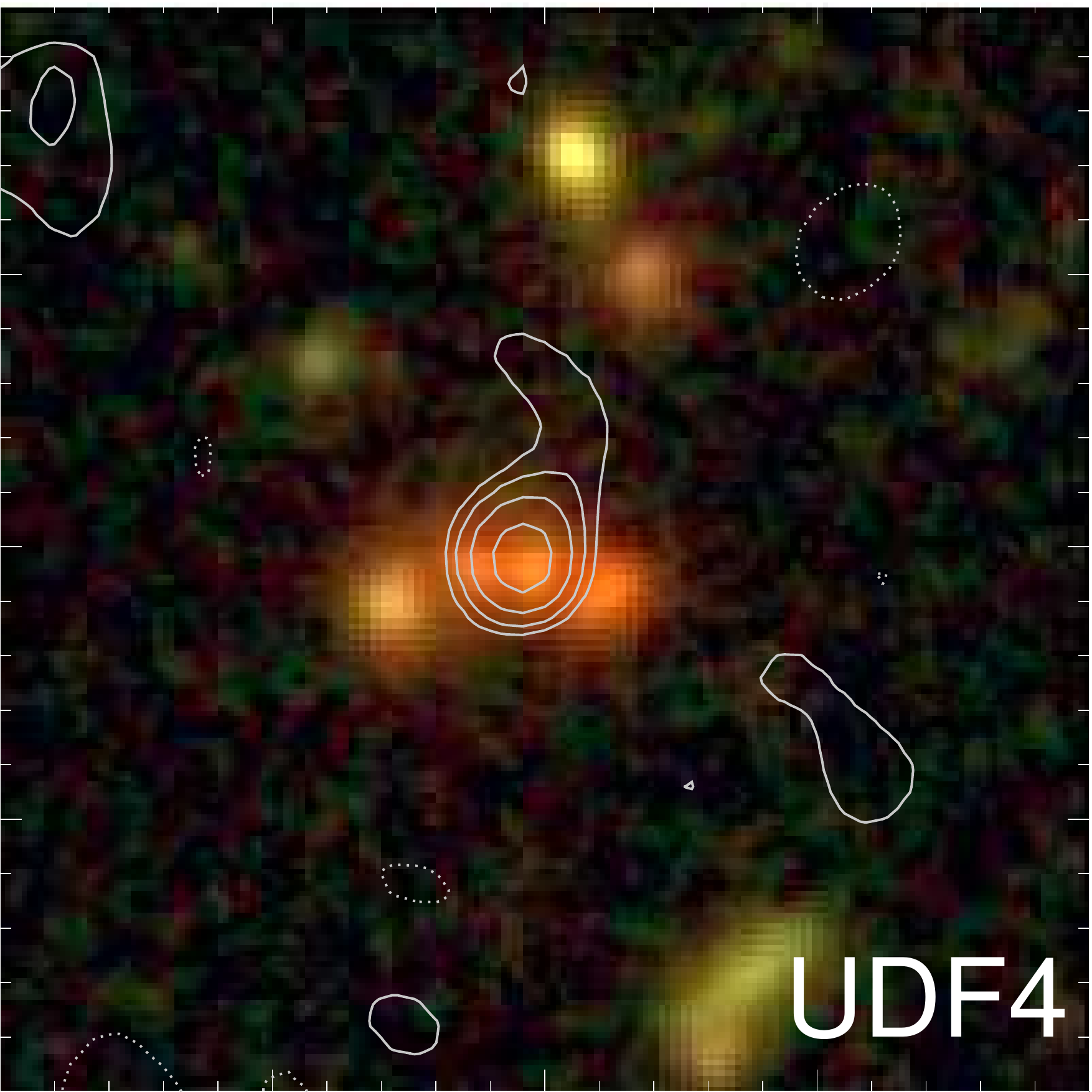}\\
\\
\includegraphics[scale=0.228]{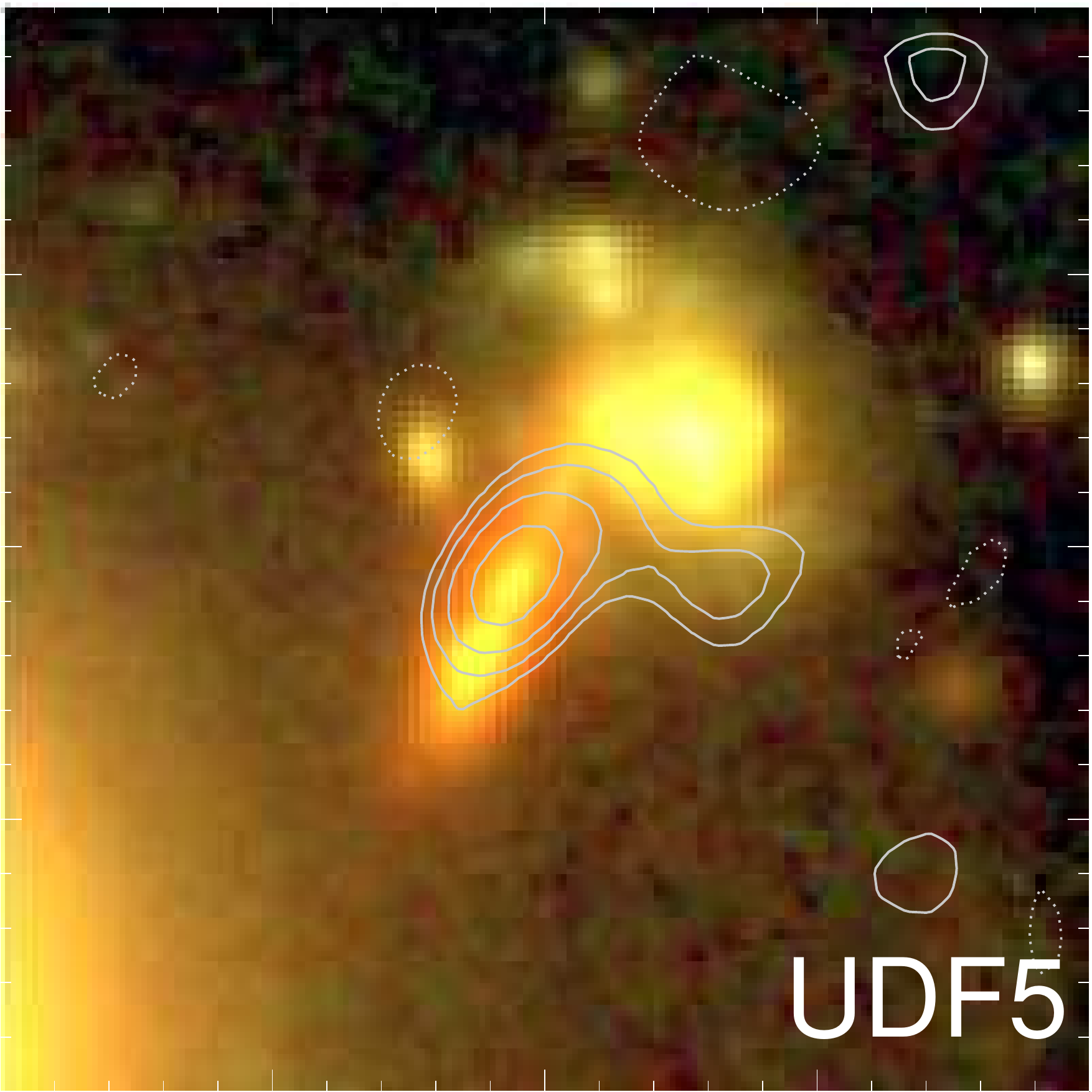}&
\includegraphics[scale=0.228]{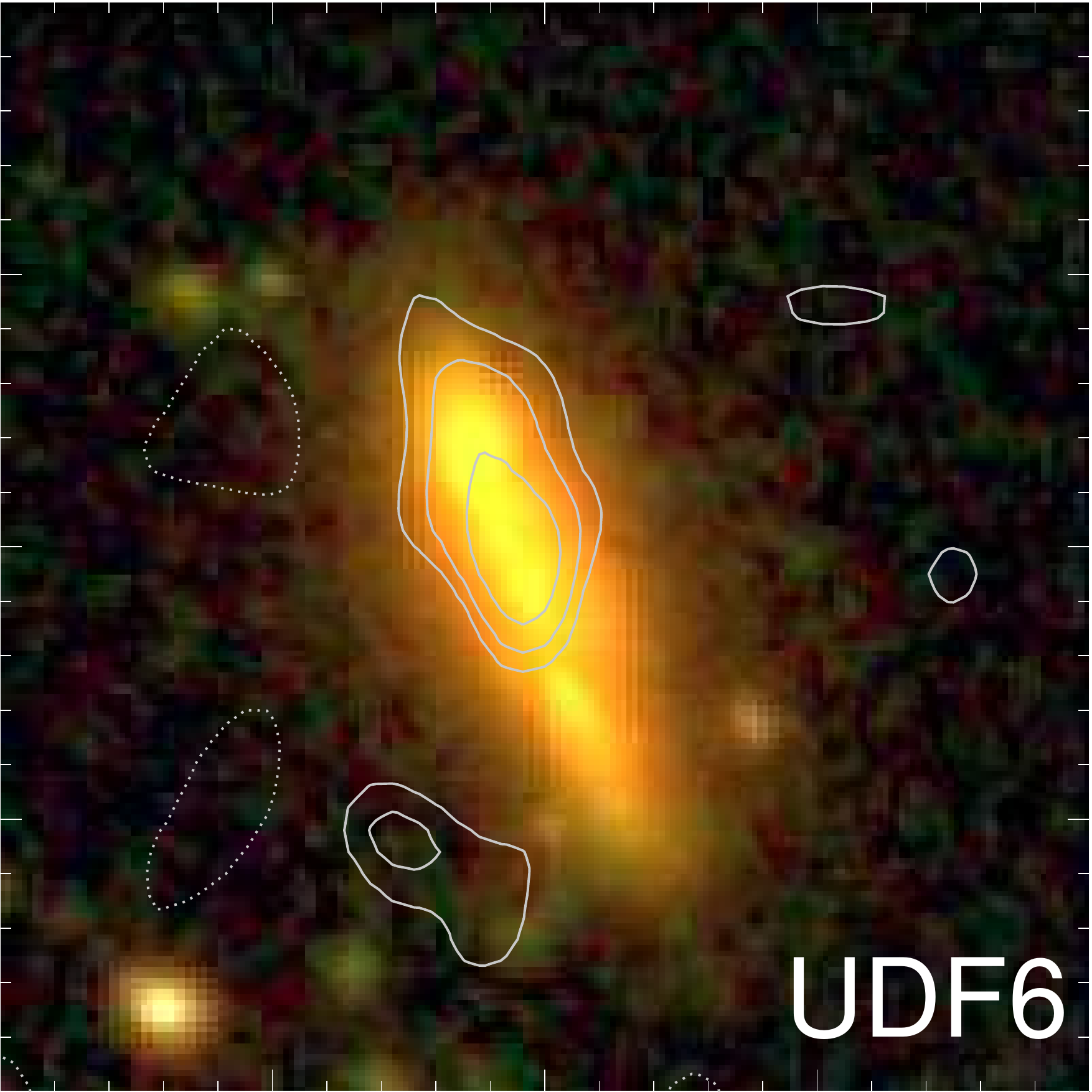}&
\includegraphics[scale=0.228]{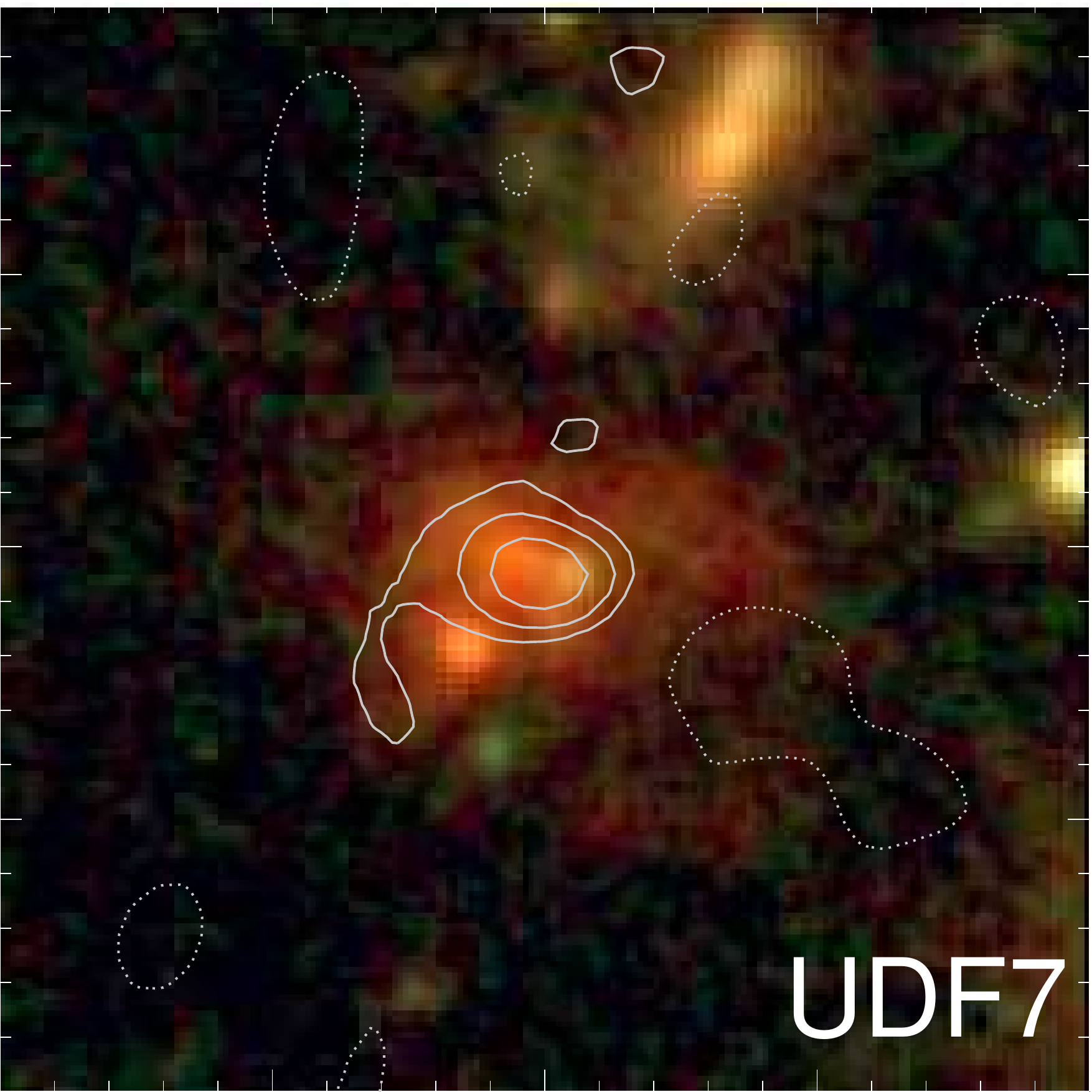}&
\includegraphics[scale=0.228]{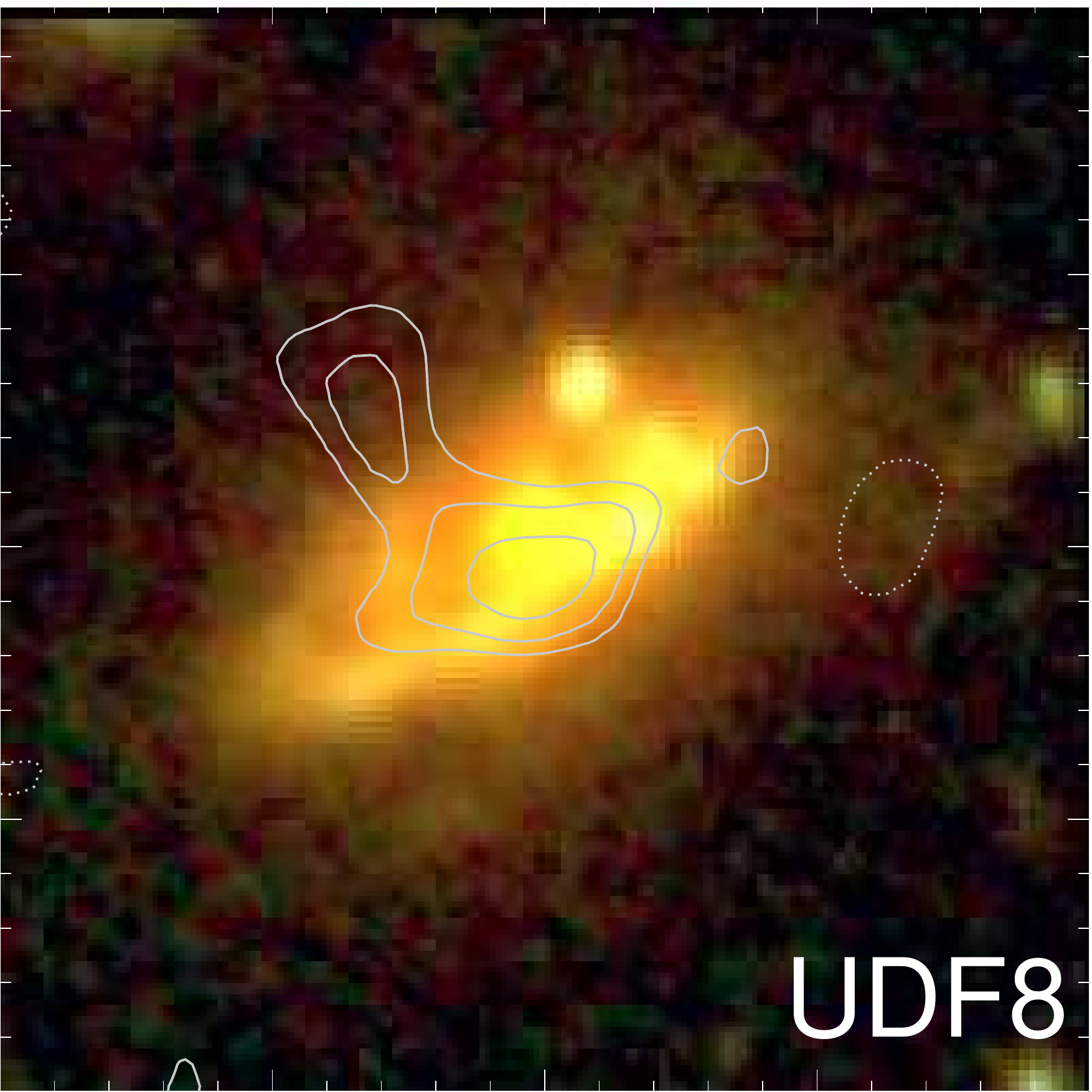}\\
\\
\includegraphics[scale=0.228]{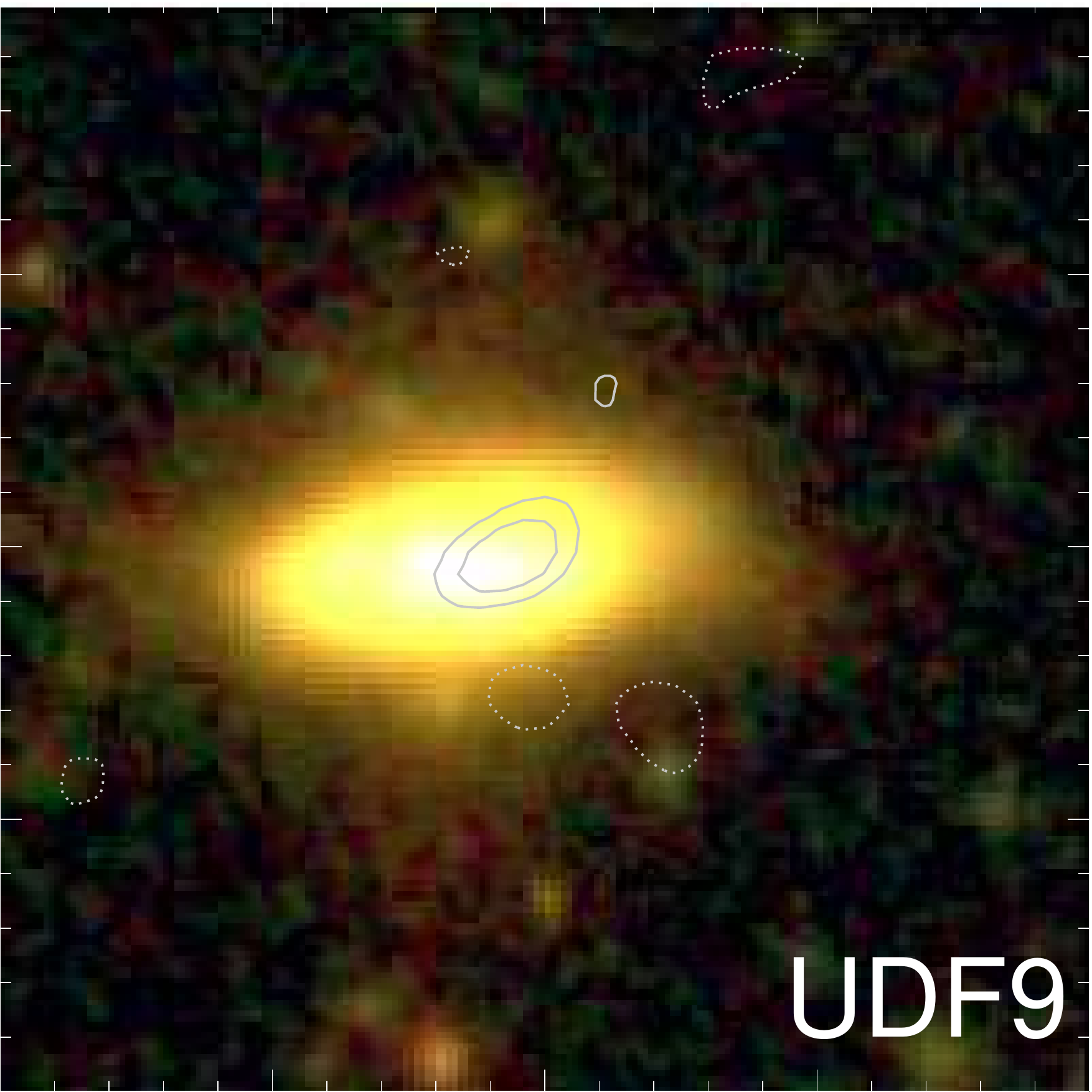}&
\includegraphics[scale=0.228]{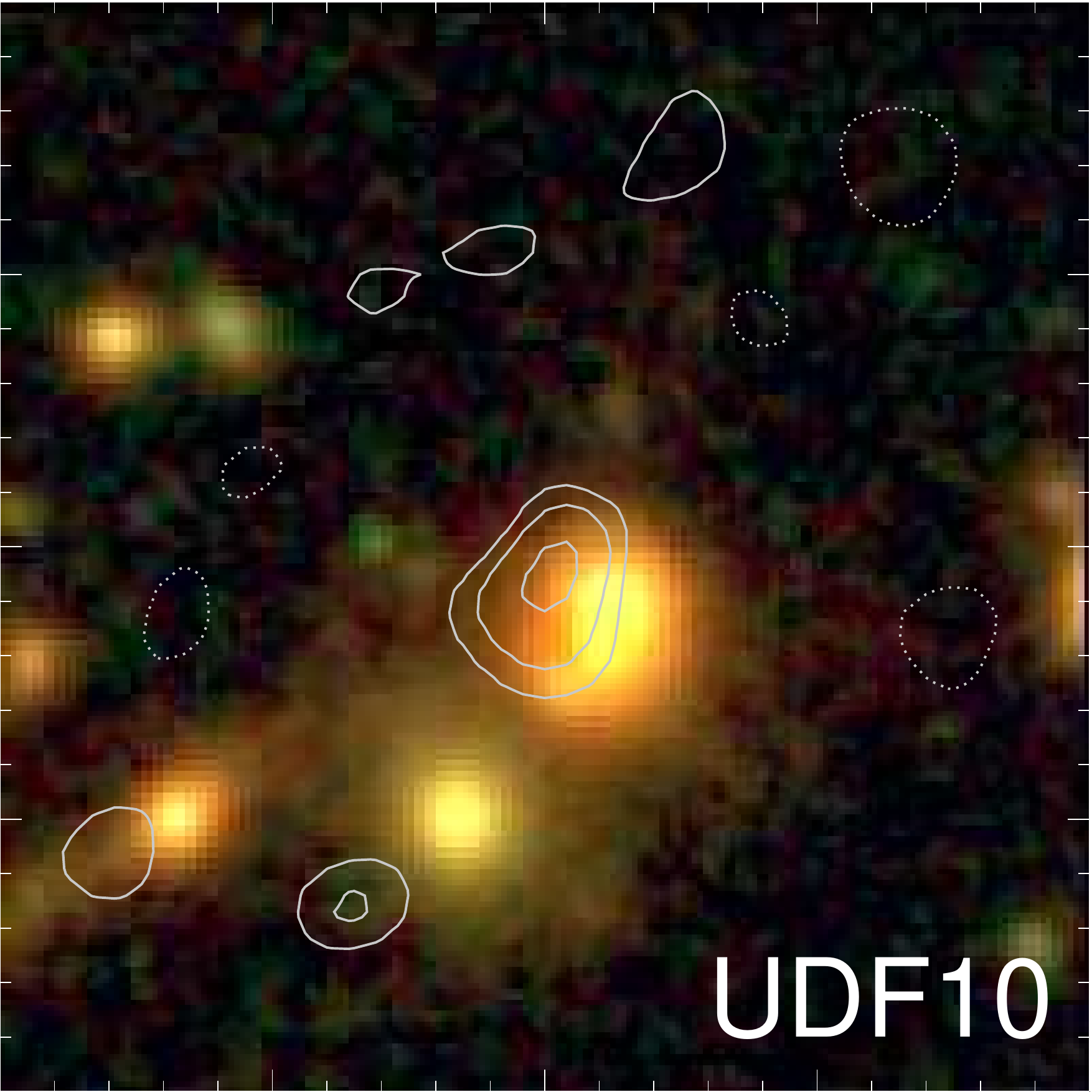}&
\includegraphics[scale=0.228]{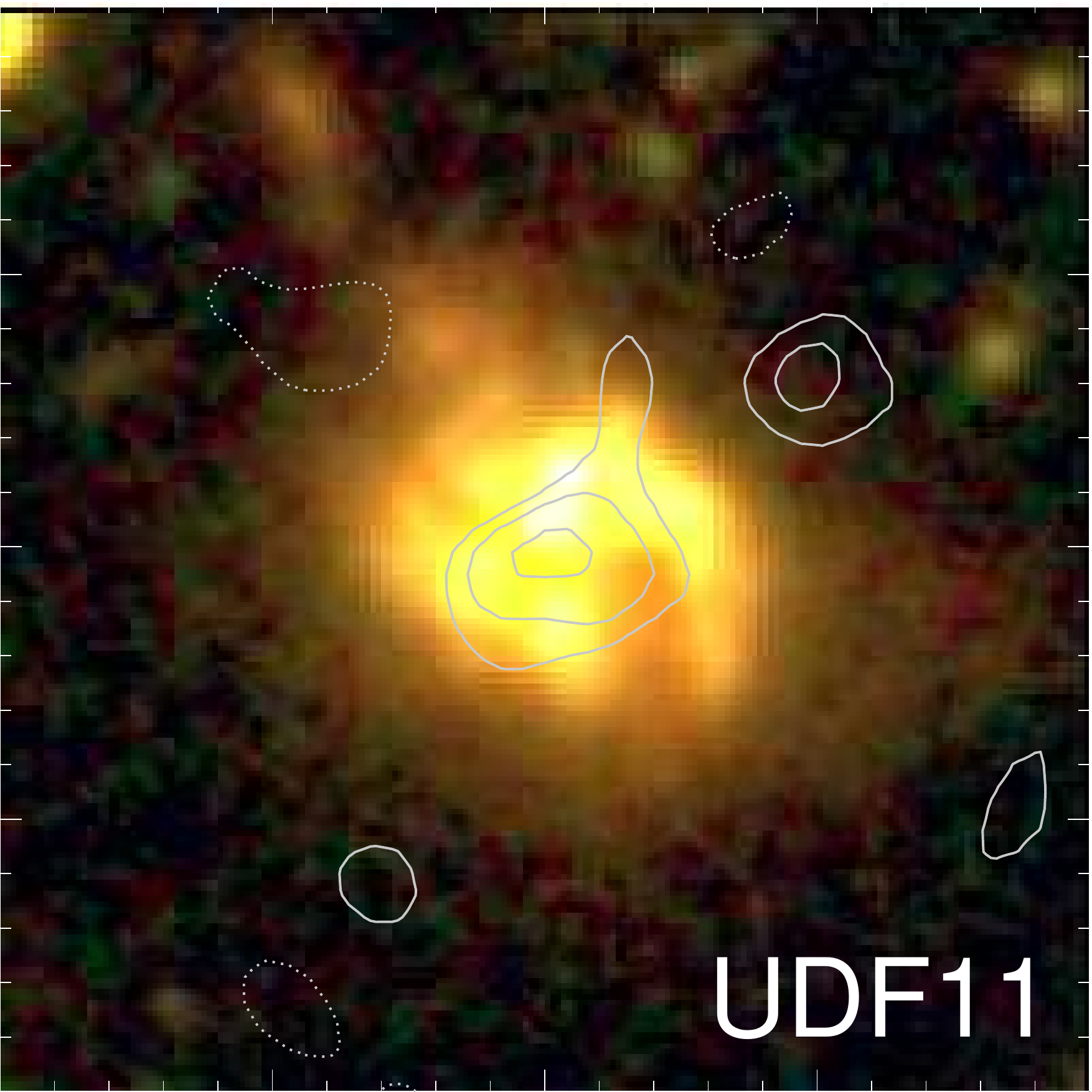}&
\includegraphics[scale=0.228]{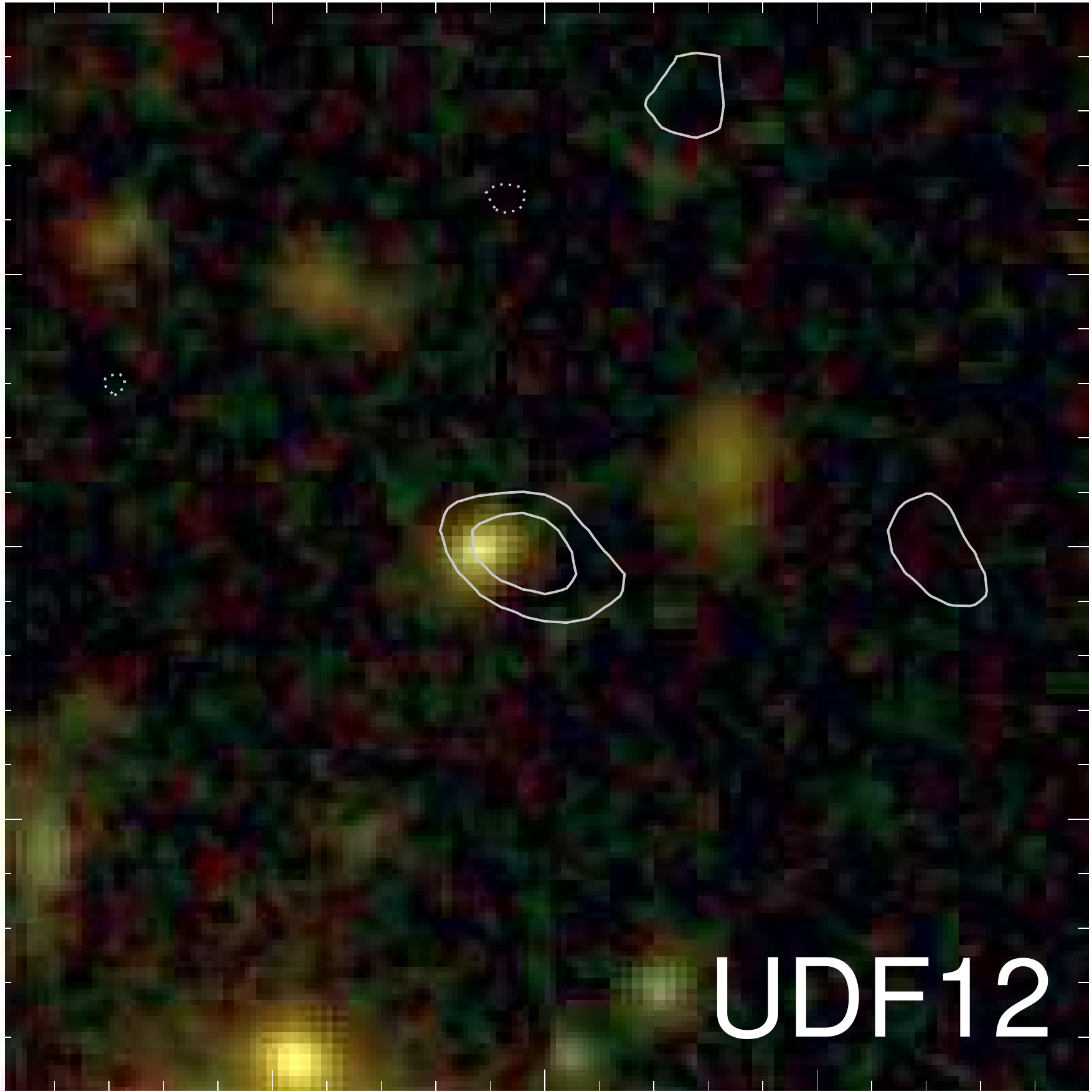}\\
\\
\includegraphics[scale=0.228]{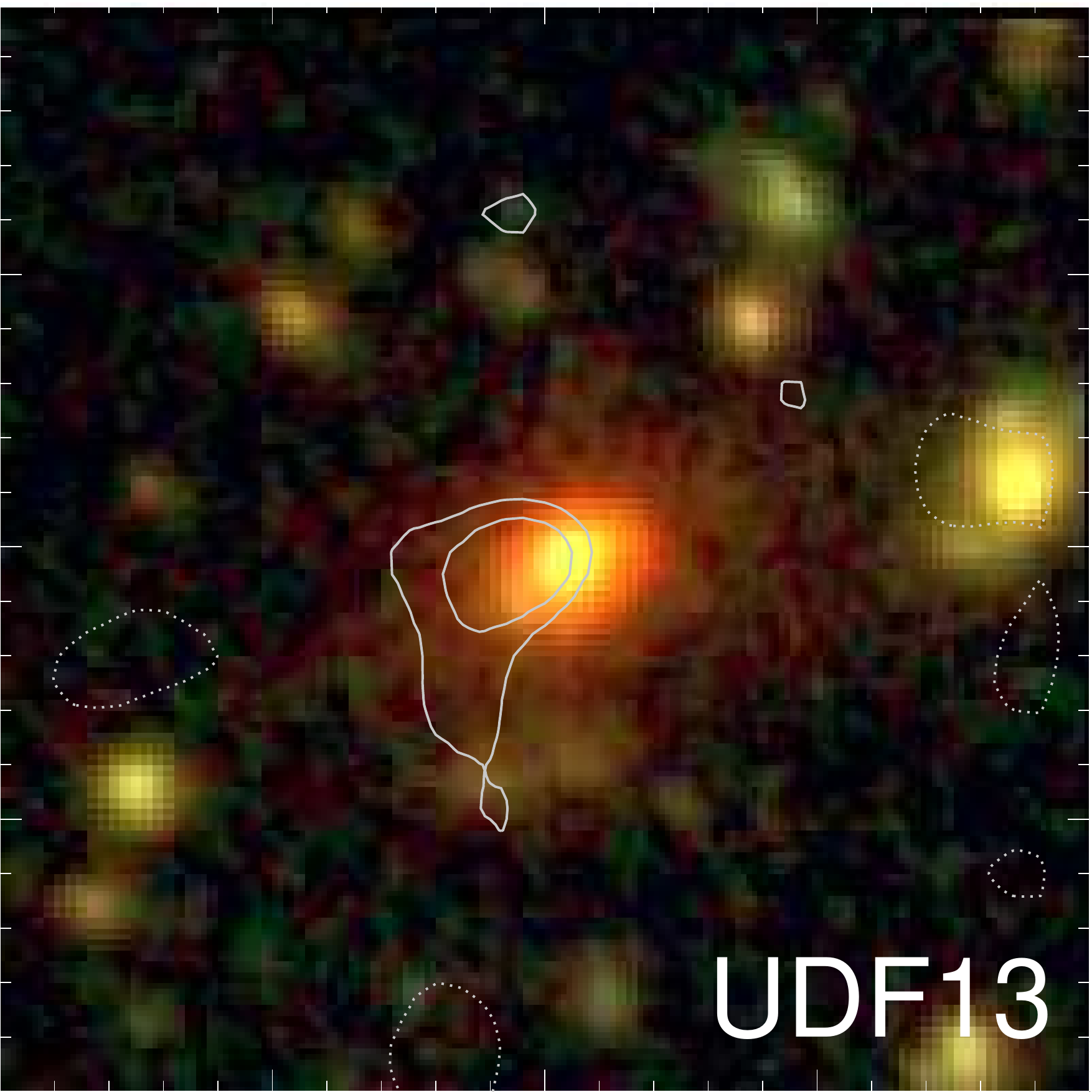}&
\includegraphics[scale=0.228]{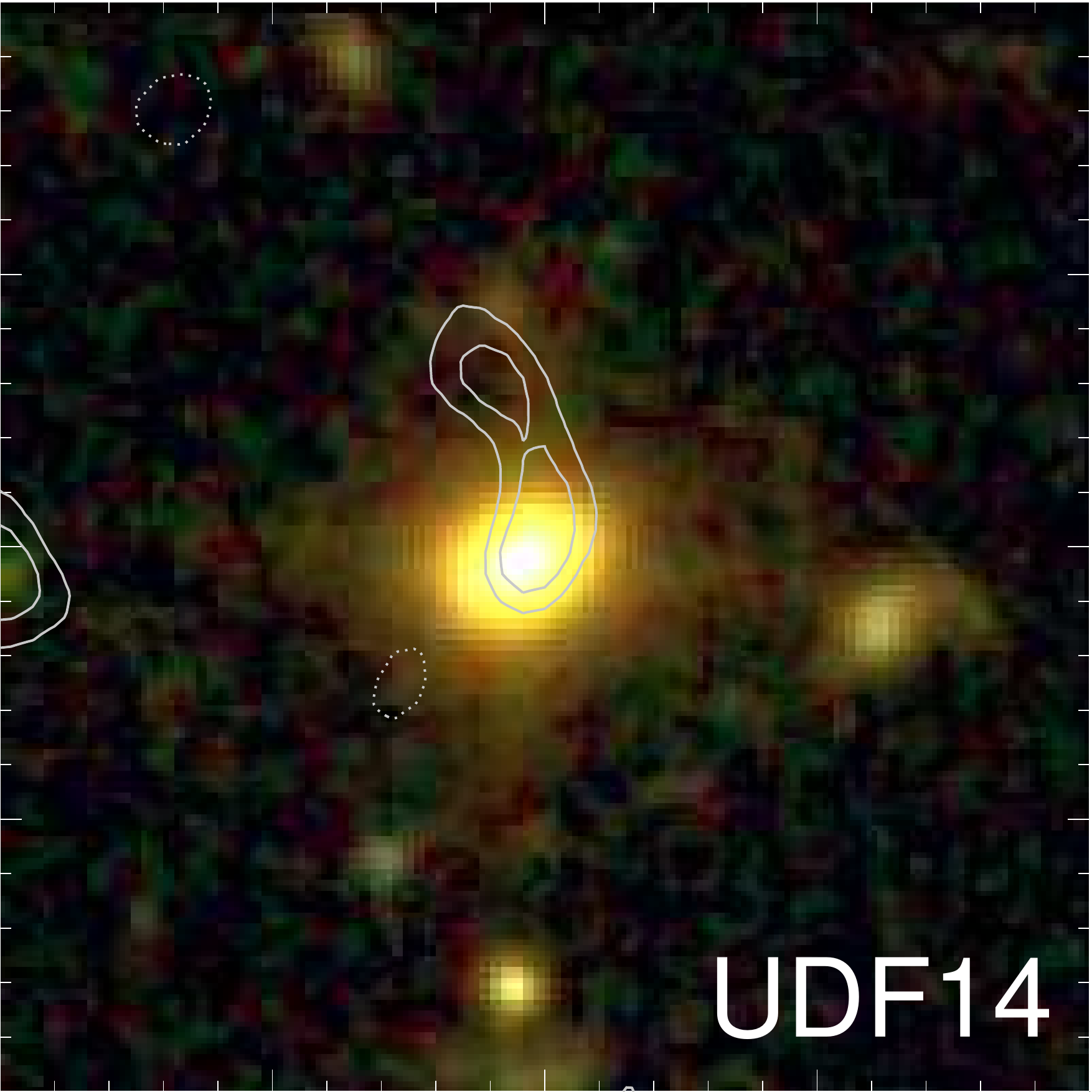}&
\includegraphics[scale=0.228]{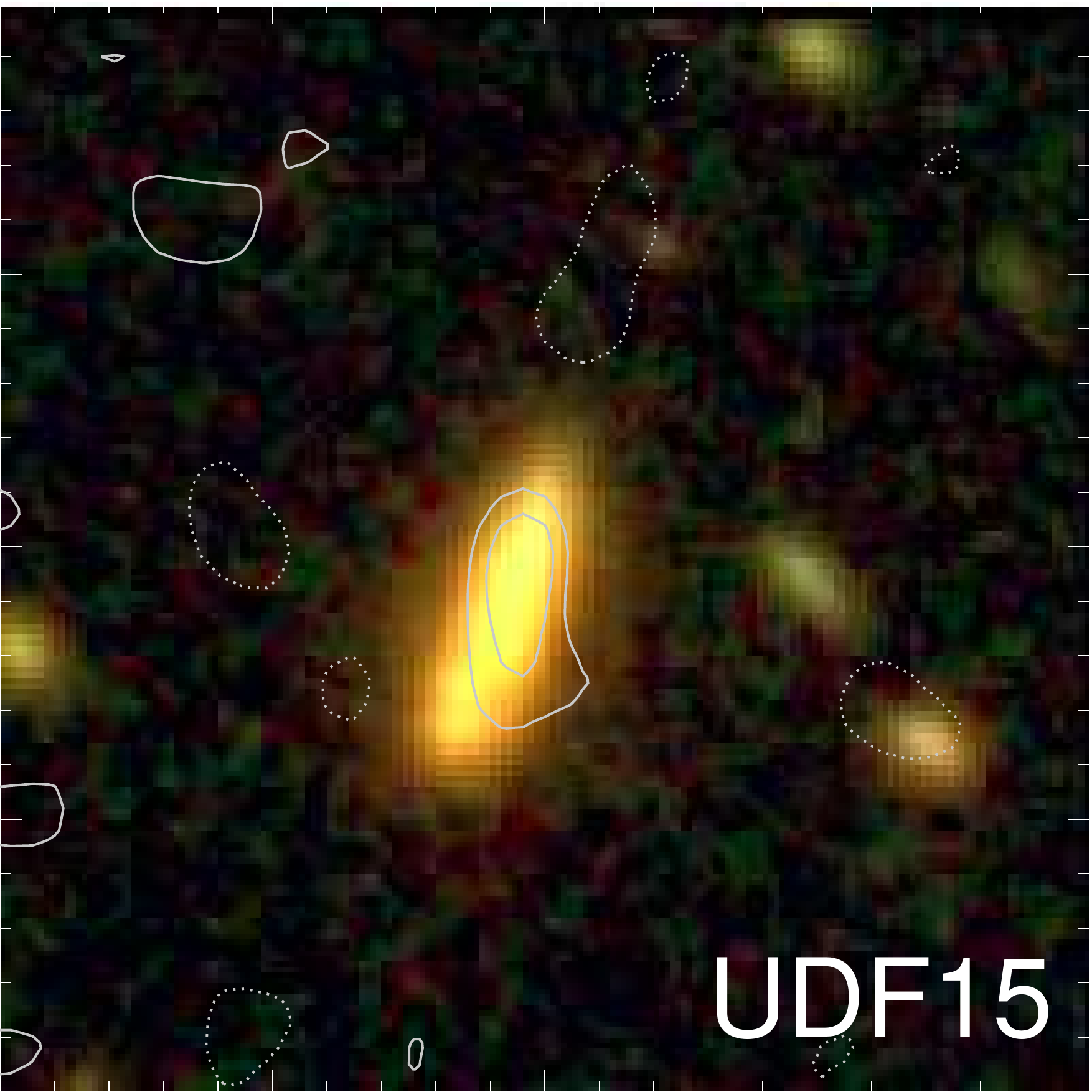}&
\includegraphics[scale=0.228]{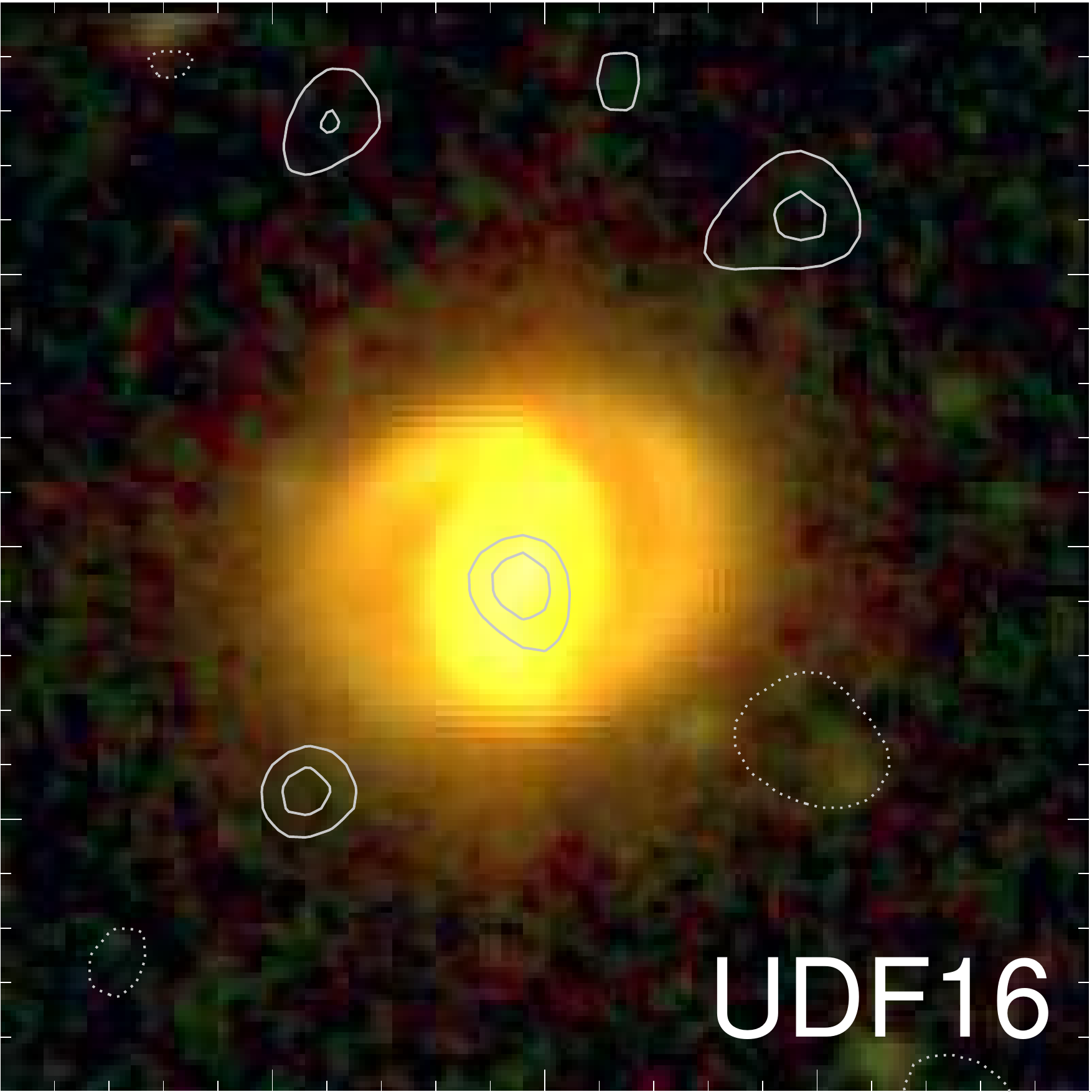}\\
\\
\end{tabular}
\end{center}
\caption{Colour ($i_{775}$+$Y_{105}$+$H_{160}$) {\it HST} postage-stamp
  images of the 16 ALMA-detected galaxies listed in Table 1, with the contours
  from the ALMA 1.3-mm image overlaid (in pale grey). Each stamp is $6 \times 6$\,arcsec
  in size, with North to the top, and East to the left.
  ALMA contours are at $-$2$\sigma$, 2$\sigma$, 2$^{1.5}\sigma$, 2$^{2.0}\sigma$ and 2$^{2.5}\sigma$.}
\end{figure*}

\subsection{Observed and predicted number counts}

Investigating the 1.3-mm number counts is not the main focus of this study, as the area of
sky imaged here is small, and the flux-density range
limited. Indeed, several recent
studies have explored the 1.1--1.3-mm number counts at comparable depths by combining
existing ALMA surveys and single pointings that together sample a significantly larger
sky area, and include subregions of imaging reaching significantly
deeper than achieved here (e.g. Ono et al. 2014; Carniano et al. 2015;
Fujimoto et al. 2015; Oteo et al. 2016). Moreover, such studies,
by including multiple sightlines, can potentially
mitigate the impact of cosmic variance. Nonetheless, given the homogeneity of our
dataset, and the unbiased nature of the HUDF (as compared to pointings centred
on known extragalactic objects of specific interest for ALMA follow-up), it is
of interest to check how the number of sources detected here compares with
expectations based on recent number-count studies.

This comparison is presented in Fig.\,3b, which shows our observed cumulative counts,
compared with appropriately scaled results from several of the aforementioned recent
studies (integrated Schechter functions, and also binned counts).
This shows that, down to a flux density $S_{1.3}~=~0.15$\,mJy (our effective flux-density
limit after scaling the flux densities by a factor $\simeq 1.25$),
the 1.3-mm number counts in the HUDF are lower by a factor of $\simeq 1.5$ than
predicted by the Schechter-function source-count fit provided by Fujimoto et al. (2016),
and also lower than the {\it binned} cumulative counts
given by Hatsukade et al. (2016). However, our number counts are in very good agreement with
integration of the 
Schechter-function fit to the differential counts
provided by Hatsukade et al. (2016) (plotted in Fig.\,3a as the magenta line),
and somewhat higher than the binned cumulative
counts published by Oteo et al. (2016) (who noted 
that their derived counts
are a factor $\simeq 2$ lower than those reported in most other recent studies).

In summary, Fig.\,3b shows that significant work still needs to be done to clarify the faint-end of the
mm number counts with ALMA, and that our results from the HUDF
are consistent with current contraints.

Our ALMA image of the HUDF can also be used to explore the number counts significantly
faintward of $S_{1.3}~=~0.15$\,mJy both via a $P(D)$ analysis
(which will be presented elsewhere) and by stacking on the positions of the known galaxies
in the HUDF at all redshifts. The results of such stacking experiments are presented in
Section 7, along with the implications for the 1.3-mm background.

\begin{table*}
\begin{normalsize}
\begin{center}
  \caption{Stellar masses, estimated dust extinction, unobscured and obscured star-formation rates, and specific star-formation rates
    for the 16 ALMA-detected galaxies in the HUDF.
    Column 1 gives source numbers as also used in Figs\,1 and 3, and Tables 2 and 3.
    Column 2 gives the stellar mass of each galaxy, determined as described in Section 6.2,
    while column 3 then gives the raw UV SFR (SFR$_{\rm UV}$) based on the uncorrected UV absolute magnitude.
    Column 4 then gives the best-estimate of extinction, $A_V$, as derived from the optical-infrared
    SED fitting (see Section 6.2).
    Then, in columns 5 to 8, we give alternative measures of star-formation rates derived
    as described in Section 6.3, namely: i) the dust-corrected SFR derived from the optical-infrared SED fitting undertaken to
    determine the stellar mass (SFR$_{\rm SED}$); ii) the SFR derived from fitting the
    long-wavelength {\it Spitzer}+{\it Herschel}+ALMA (24${\rm \mu m} - 1.3$\,mm) photometry (see Appendix A, Fig.\,A1)
    with the star-formation template plotted in Fig.\,8
    (SFR$_{\rm FIR1}$);
    iii) the SFR derived from the same long-wavelength photometry, but adopting the best-fitting
    of three alternative long-wavelength star-forming SEDs (SFR$_{\rm FIR2}$);  
    iv) the SFR inferred from the radio detections (SFR$_{\rm Rad}$).
    For the reasons described in Section 6.3, we adopt SFR$_{\rm FIR1}$ as the best/simplest estimate of true star-formation rate, and use this to calculate the
    ratio of obscured:unobscured SFR given in Column 9, and finally the estimates of specific star-formation rates 
    given in column 10. All values given here assume a Chabrier (2003) IMF. Finally we note that, for source UDF3, marked by a $\dagger$, the derived values
    of SFR$_{\rm FIR1}$ and  SFR$_{\rm FIR2}$ (and hence also the ratio of obscured:unobscured SFR and sSFR) are given after correcting
    for the unusually large contribution made to the 1.3-mm flux density in this object by molecular line emission (see Section 4.3 for details, and Ivison et al.,
    in preparation).}
\label{tab:Sources}
\setlength{\tabcolsep}{1.43 mm} 
\begin{tabular}{lccccrrrcr}
\hline
ID & $\log_{10}(M_*/{\rm M_{\odot}})$  & SFR$_{\rm UV}$  & $A_{\rm V}$  & SFR$_{\rm SED}$   & SFR$_{\rm FIR1}$        & SFR$_{\rm FIR2}$         & SFR$_{\rm Rad}$& SFR$_{\rm obs}$/ & sSFR\phantom{1}\\
   &             &/${\rm M_{\odot} yr^{-1}}$&/mag&/${\rm M_{\odot} yr^{-1}}$&/${\rm M_{\odot} yr^{-1}}$&/${\rm M_{\odot} yr^{-1}}$&/${\rm M_{\odot} yr^{-1}}$& SFR$_{\rm UV}$      &/Gyr$^{-1}$\\
\hline
UDF1  & 10.7$\pm$0.10  & 0.31$\pm$0.05 & 3.1 &           399.4 & 326$\pm$83\phantom{1}           & 364$\pm$82            & 439$\pm$28          &  1052$\pm$317          &  6.50$\pm$2.24  \\
UDF2  & 11.1$\pm$0.15  & 0.32$\pm$0.10 & 2.2 & \phantom{1}50.2 & 247$\pm$76\phantom{1}           & 194$\pm$64            & 242$\pm$22          &  \phantom{1}772$\pm$339           &  1.96$\pm$0.92  \\ 
UDF3$^{\dagger}$  & 10.3$\pm$0.15  & 4.70$\pm$0.30 & 0.9 & \phantom{1}42.0 & 195$\pm$69\phantom{1} & 173$\pm$1             & 400$\pm$17          &  \phantom{11}41$\pm$\phantom{1}15 &  9.77$\pm$4.88  \\ 
UDF4  & 10.5$\pm$0.15  & 0.43$\pm$0.20 & 1.6 & \phantom{1}20.0 &  94$\pm$\phantom{1}4\phantom{1} &  58$\pm$\phantom{1}5  & 89$\pm$17           &  \phantom{1}219$\pm$102           &  2.97$\pm$1.05  \\ 
UDF5  & 10.4$\pm$0.15  & 0.20$\pm$0.05 & 2.4 & \phantom{1}36.1 &  102$\pm$\phantom{1}7\phantom{1} & 67$\pm$25            & 86$\pm$\phantom{1}6 &  \phantom{1}510$\pm$132           &  4.06$\pm$1.46  \\ 
UDF6  & 10.5$\pm$0.10  & 0.10$\pm$0.02 & 2.8 & \phantom{1}78.0 &  87$\pm$11\phantom{1}           &  66$\pm$\phantom{1}5  & 68$\pm$\phantom{1}5 &  \phantom{1}870$\pm$205           &  2.75$\pm$0.73  \\ 
UDF7  & 10.6$\pm$0.10  & 0.50$\pm$0.03 & 1.5 & \phantom{1}16.5 &  56$\pm$22\phantom{1}           &  77$\pm$42            &617$\pm$20           &  \phantom{1}112$\pm$\phantom{1}45           &  1.41$\pm$0.64  \\ 
UDF8 & 11.2$\pm$0.15  & 0.98$\pm$0.02 & 1.6 & \phantom{1}35.8 & 149$\pm$90\phantom{1}           &  94$\pm$37             & 73$\pm$\phantom{1}5 &  \phantom{1}152$\pm$\phantom{1}92           &  0.94$\pm$0.66  \\ 
UDF9 & 10.0$\pm$0.10  & 0.06$\pm$0.01 & 0.9 & \phantom{11}0.5 &  23$\pm$25\phantom{1}           &   5$\pm$\phantom{1}2   & 5$\pm$\phantom{1}1  &  \phantom{1}383$\pm$421           &  2.30$\pm$2.56  \\ 
UDF10 & 10.2$\pm$0.15  & 1.14$\pm$0.10 & 1.5 & \phantom{1}37.0 &  45$\pm$22\phantom{1}           &  34$\pm$\phantom{1}7  & $<$35\phantom{444,} &  \phantom{11}39$\pm$\phantom{1}20 &  2.84$\pm$1.71  \\ 
UDF11 & 10.8$\pm$0.10  & 6.29$\pm$0.20 & 1.4 &           162.8 & 162$\pm$94\phantom{1}           & 232$\pm$10            & 172$\pm$14          &  \phantom{11}26$\pm$\phantom{1}15 &  2.57$\pm$1.60  \\ 
UDF12 & \phantom{1}9.6$\pm$0.15  & 1.55$\pm$0.10 & 0.2 & \phantom{11}2.6 &  37$\pm$14\phantom{1} &  21$\pm$\phantom{1}7  & $<$100\phantom{444,}&  \phantom{11}24$\pm$\phantom{1}10 &  9.29$\pm$4.80  \\ 
UDF13 & 10.8$\pm$0.10  & 0.95$\pm$0.05 & 1.3 & \phantom{1}18.0 &  68$\pm$18\phantom{1}           &  60$\pm$19            & 142$\pm$17          &  \phantom{11}72$\pm$\phantom{1}19 &  1.08$\pm$0.38  \\ 
UDF14 & \phantom{1}9.7$\pm$0.10  & 0.05$\pm$0.01 & 1.3 & \phantom{11}1.0 &  44$\pm$17\phantom{1} &   3$\pm$\phantom{1}2  & $<$4\phantom{444,}  &  \phantom{1}880$\pm$383           &  8.78$\pm$3.96  \\ 
UDF15 & \phantom{1}9.9$\pm$0.15  & 1.14$\pm$0.02 & 1.1 & \phantom{1}15.5 &  38$\pm$27\phantom{1} &  25$\pm$\phantom{1}8  & $<$20\phantom{444,} &  \phantom{11}33$\pm$\phantom{1}24 &  4.78$\pm$3.79  \\ 
UDF16 & 10.9$\pm$0.10  & 0.10$\pm$0.05 & 0.6 & \phantom{11}0.5 &  40$\pm$18\phantom{1}           &  25$\pm$\phantom{1}4  & 38$\pm$\phantom{1}3 &  \phantom{1}400$\pm$269           &  0.50$\pm$0.26  \\ 
\hline
\end{tabular}
\end{center}
\end{normalsize}
\end{table*}

\section{Source Properties}
\label{sec:redshifts}

The galaxies revealed by our ALMA imaging have several interesting properties.
First, as is clear from the colour ($i$+$Y$+$H$) postage-stamp images presented
in Fig.\,4, the vast majority are noticeably red.
Indeed, certainly the $z > 1.5$ ALMA sources can essentially be spotted by eye as the
reddest objects in this particular colour representation of the HUDF.
Second, because these objects are actually quite bright (15/16 have $H_{160} < 25$\,mag),
and because of the wealth of supporting spectroscopy and photometry in the HUDF, we have
complete, high-quality redshift information for essentially the whole sample (see Table\,2, and
associated caption). Third, armed with this redshift information and multi-wavelength
photometry (e.g. see Appendix A, Fig.\,A1) we can derive relatively robust stellar masses and star-formation
rates for these galaxies, as we now discuss (see Table\,4).

\subsection{Redshift Distribution}
As explained in Section 2.2.5, and tabulated in Table\,2, the wealth of deep
spectroscopy in the HUDF field results in 13 out of our 16 ALMA-detected galaxies
having optical--near-infrared spectroscopic redshifts. To complete the redshift content of the sample,
we estimated the redshifts
of the remaining sources by adopting the median value from five alternative
determinations of $z_{\rm phot}$ based on different SED fitting codes. The range of
values returned by these codes, coupled with tests of $z_{\rm phot}$ v $z_{\rm spec}$ (for the
13 sources in Table\,2 with spectroscopic redshifts), indicates that the three photometric
redshifts listed in Table\,2 carry a typical rms uncertainty of $\delta z \simeq 0.1$.

The final redshift distribution of the ALMA-detected HUDF galaxy sample is shown in Fig.\,5.
Although the present study probes an order-of-magnitude deeper (in terms of dust-enshrouded SFR)
than the brighter sub-mm/mm selected samples produced by SCUBA, MAMBO, LABOCA, AzTEC, and SCUBA-2 over the last
$15-20$ years, the redshift distribution of detected sources is very little changed. The 
mean redshift of the ALMA HUDF sources is $\langle z \rangle = 2.15$, with 13/16 sources
($\simeq 80$\%) in the redshift range $1.5~<~z~<~3$. This is very similar to the redshift distribution
displayed by, for example, the much brighter AzTEC sources uncovered in the SHADES fields, which have a median
redshift of $z \simeq 2.2$, with $\simeq 75$\% of sources lying in the redshift
range $1.5 < z < 3.5$ (Micha{\l}owski et al. 2012a; see also, e.g., Chapman et al. 2005;
Aretxaga et al. 2007; Chapin et al. 2009), although there is some evidence that the most extreme sources
are confined to somewhat higher redshifts (e.g. Ivison et al. 2007;
Smolcic et al. 2012; Koprowski et al. 2014; Micha{\l}owski et al. 2016). 

In Fig.\,6 we plot the galaxies in the HUDF on the $M_{1500}$-redshift plane, highlighting the
locations of the ALMA-detected galaxies. This shows that the ALMA-detected galaxies span 
a wide range of (observed) UV luminosities.

However, a different picture emerges when
stellar mass is plotted versus redshift, as shown in Fig.\,7. Here it can be seen that essentially
all of the ALMA-detected galaxies have high stellar masses. Indeed, from Fig.\,7 it can then be seen that
the most obvious physical reason for the lack of very high redshift galaxies in our sample
(i.e. only one detection beyond $z = 3.1$) is the absence of high-mass galaxies at these redshifts
within the relatively small cosmological volumes sampled by the 4.5\,arcmin$^2$ field. This is
discussed further in the next subsection. By contrast, the decline in
ALMA detections at $z < 1.5$ is driven by the quenching of star-formation activity in
high-mass galaxies, as is again evident from Fig.\,7.

\begin{figure}
\begin{center}
\includegraphics[scale=0.55]{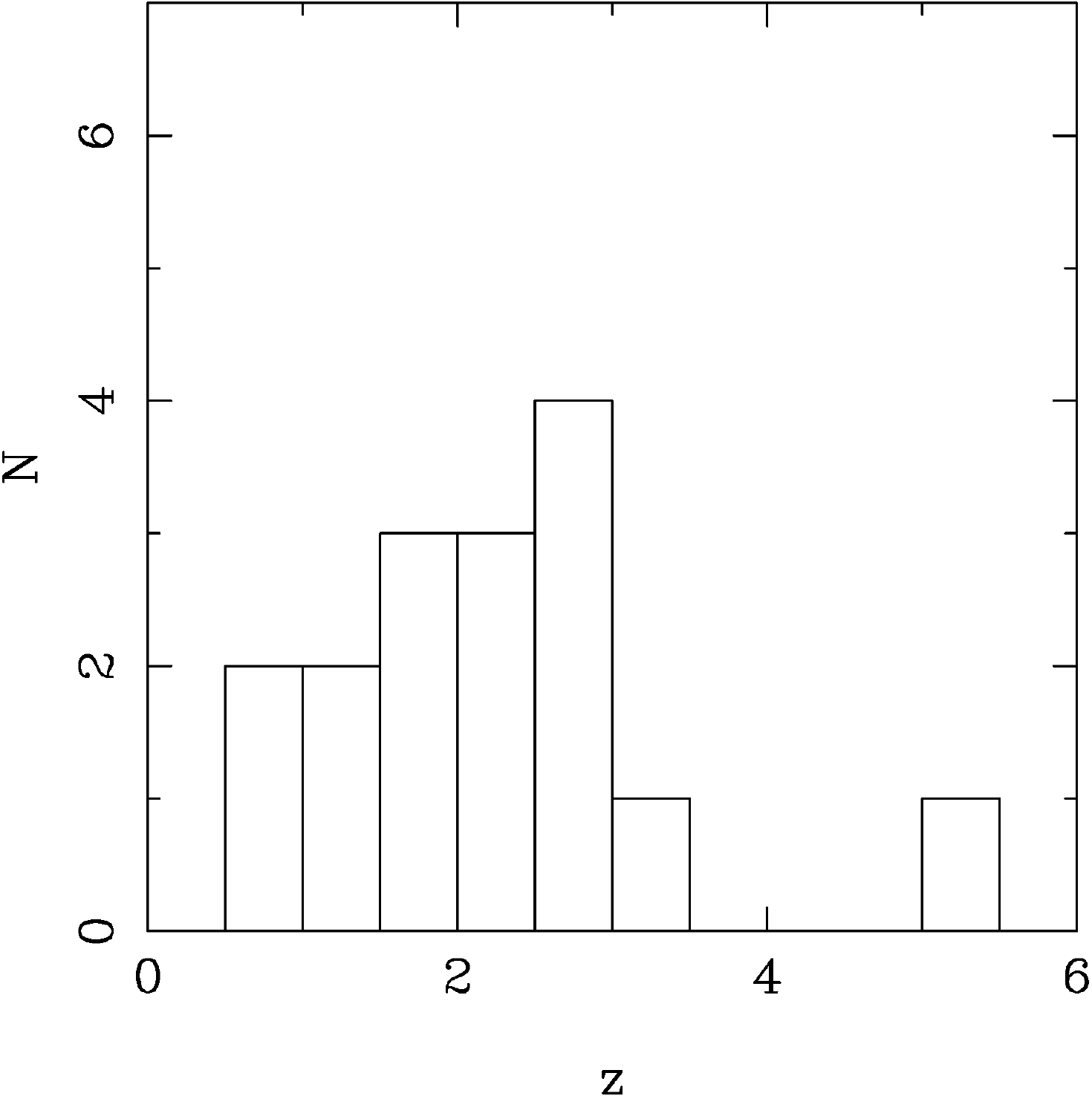}
\end{center}
\caption{The redshift distribution of the 16 ALMA-detected galaxies in the HUDF. The redshift
  information consists of 13 spectroscopic redshifts, and three accurate ($\delta z \simeq 0.1$)
photometric redshifts, as listed in Table\,2.}
\label{fig:redshifts}
\end{figure}

\subsection{Stellar Masses}
To derive the stellar masses of the galaxies, we fitted a range of single-component, and then also
two-component Bruzual \& Charlot (2003) evolutionary synthesis models
to the optical-infrared photometry of the sources (at the redshifts listed in Table\,2).
For the single-component models, the minimum age was set to 50\,Myr, with $\tau$ allowed
to vary from 300\,Myr up to an essentially constant star-formation rate (see Wuyts et al. 2009).
We applied reddening assuming the dust attenuation law of Calzetti et al. (2000), with extinction
allowed to vary up to $A_V \simeq 6$ (see Dunlop, Cirasuolo \& McLure 2007), and the impact of
IGM absorption was modelled following Madau (1995).

\begin{figure*}
\begin{center}
\includegraphics[scale=0.8]{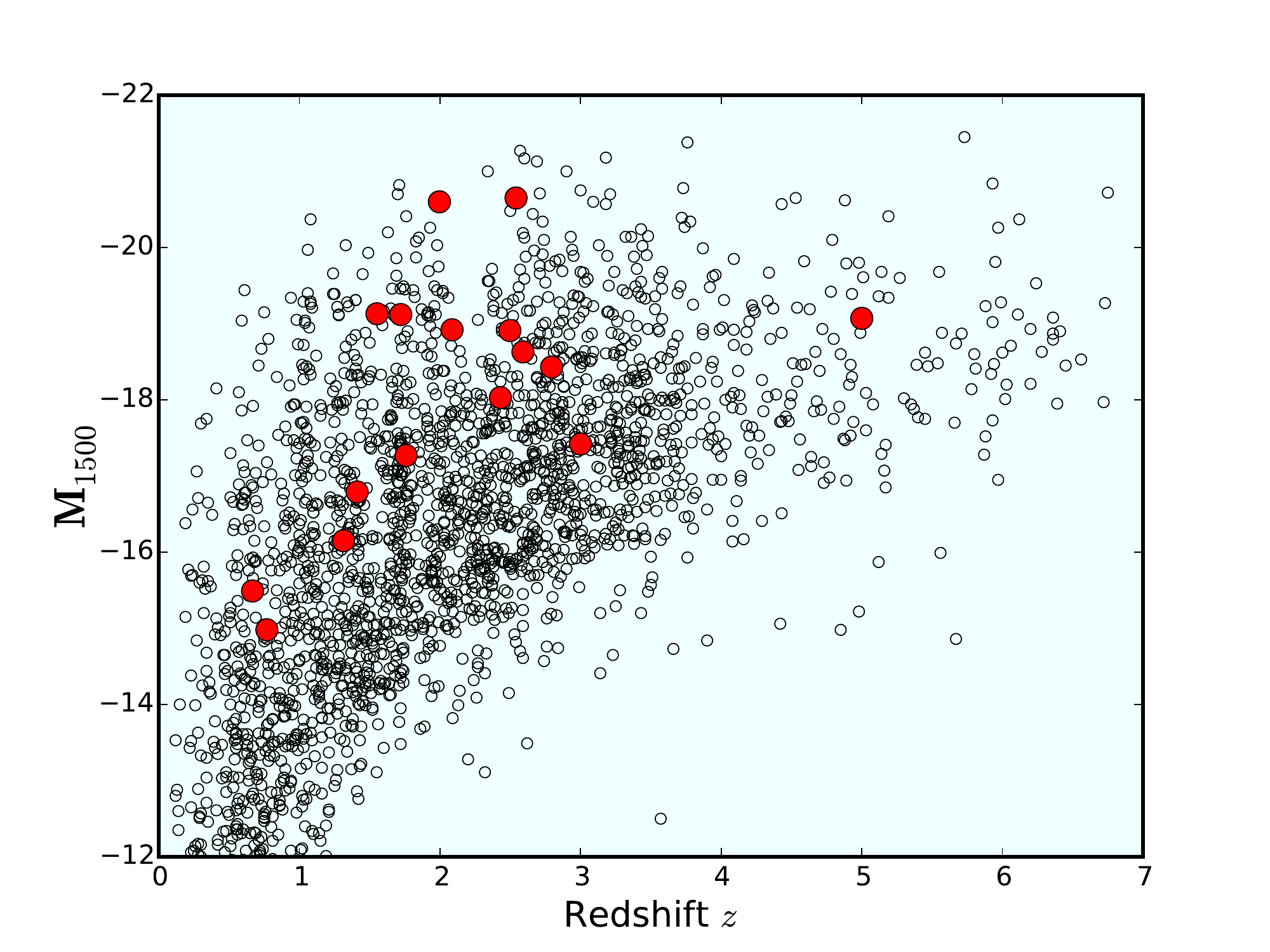}
\caption{UV absolute magnitude versus redshift
  for the galaxies in the HUDF, highlighting (in red) those detected in our ALMA 1.3-mm
  image. It can be seen that the ALMA-detected galaxies span a wide range of raw unobscured
  UV luminosity, and appear unexceptional on the $M_{1500}$--$z$ plane. However, if corrected for
dust obscuration, they would be the brightest galaxies in the field (see Section 6).} 
\end{center}
\end{figure*}

\begin{figure*}
  \begin{center}
\includegraphics[scale=0.8]{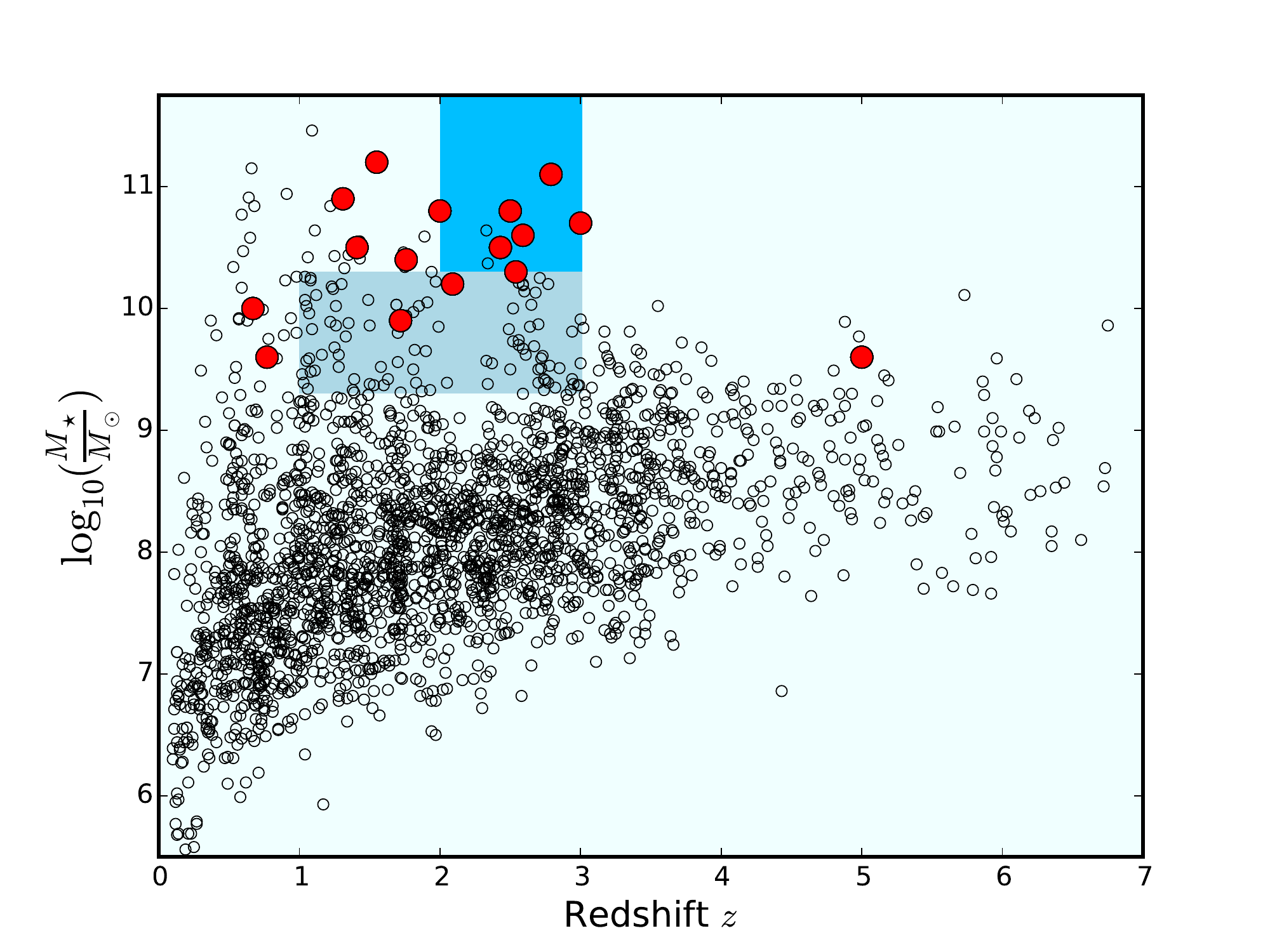}
\caption{Stellar mass versus redshift
  for the galaxies in the HUDF, highlighting (in red) those detected in our ALMA 1.3-mm
  image. The connection between dust-enshrouded SFR and $M_*$ is clear, and indeed, as emphasized by 
  the bright-blue box at the top of the plot, we detect 7 of the 9 galaxies (i.e. $\simeq 80$\%)
  with $M_* \geq 2 \times 10^{10}\,{\rm M_{\odot}}$ at $z > 2$. Also apparent 
  is the emergence of a significant population of quenched high-mass galaxies at $1 < z < 2$, where the ALMA detection
  rate for galaxies with $M_* \geq 2 \times 10^{10}\,{\rm M_{\odot}}$ drops to 5/19 (i.e. $\simeq 25$\%).
  It can be also seen that the lack of ALMA detections beyond $z > 3$ in our
  sample simply reflects the dearth
  of comparably-massive galaxies in the HUDF at these redshifts (due to the evolution of the galaxy stellar mass function).
  The blue-grey rectangle indicates
  the sample of $\simeq 90$ galaxies in the redshift range $1 < z < 3$ and
  mass range $9.3 < \log_{10}(M_*/{\rm M_{\odot}}) < 10.3$
  used to produce the stacked ALMA image shown in Fig.\,8, and discussed in
  Section 7.} 
\end{center}
\end{figure*}

The derived galaxy masses, given in Table\,4, proved to be (perhaps surprisingly) robust to the range of star-formation
histories which provided acceptable fits to the photometry (possibly as a result of the high signal:noise
of the available optical-infrared data in the HUDF). The uncertainties in stellar mass given in Table\,4
reflect the range in stellar masses delivered by acceptable star-formation histories.

We assumed the IMF of Chabrier (2003), and that the stellar masses given in Table\,2 are
$M_*$ `total', which means the mass of living stars plus stellar remnants. One advantage of this choice
is that the conversion from $M_*$ assuming Chabrier (2003) to $M_*$ with a Salpeter (1955) IMF is relatively
immune to age, involving multiplication by a factor $\simeq 1.65$. This $M_*$-total is typically
$\simeq 0.05$ dex larger than $M_*$-living, and $\simeq 0.15$ dex smaller than $M_{\rm gal}$, which includes
recycled gas, although these conversions are functions of galaxy age and star-formation history.

It can be seen that the stellar masses are high, with 13 out of the 16 sources having
$M_* > 10^{10}\,{\rm M_{\odot}}$ (with the adopted Chabrier IMF). Such objects are
rare at these epochs in the relatively small volumes probed by the HUDF. In particular,
from Fig.\,7 (which shows mass versus redshift for all galaxies in the HUDF), it can be seen that
the HUDF contains only 9 galaxies with $M_* > 2 \times 10^{10}\,{\rm M_{\odot}}$ at $z \geq 2$, and that
our ALMA observations have detected 7 of them (i.e. $\simeq 80$\%). This provides compelling
evidence that stellar mass is the best predictor of star-formation rate at these epochs
(rather than, for example rest-frame UV luminosity),
as expected from the `main sequence' of star-forming
galaxies (e.g. Noeske et al. 2007; Daddi et al. 2007; Elbaz et al. 2010; Micha{\l}owski et al. 2012b;
Roseboom et al. 2013; Speagle et al. 2014; Koprowski et al. 2014, 2016;
Renzini \& Peng 2015; Schreiber et al. 2015). The location of our ALMA-detected galaxies
relative to the `main-sequence' is discussed further below in Section\,8.1.

As is also apparent from Fig.\,7, at $1 < z < 2$ the fraction of high-mass
($M_* > 2 \times 10^{10}\,{\rm M_{\odot}}$) galaxies that we detect with ALMA
drops to 5/19 (i.e. $\simeq 25$\%) reflecting the emergence of a significant
population of quenched high-mass galaxies over this redshift range. 

At redshifts $z > 3.5$ we have detected only one object, UDF12, which lies at
$z = 5.0$. Among our ALMA detections, this is in fact the galaxy with the lowest estimated stellar
mass, but from Fig.\,7 it can be seen that it is one of the most massive
galaxies in the HUDF for its redshift (i.e. it is one of the very few galaxies in this field
at $z > 4$ which has $M_* > 10^{9.5}\,{\rm M_{\odot}}$). We can therefore speculate that
this detection may reflect a modest increase in typical specific star-formation rate
between $z \simeq 2$ and $z \simeq 5$ (e.g. Steinhardt et al. 2014; Marmol-Queralto et al. 2016), combined
with the sensitivity of 1.3-mm observations to extreme redshift dusty star-forming galaxies,
and that moderately deeper sub-mm/mm observations of the field may yield
significantly more detections at $z > 3$.

\subsection{Star-formation rates}

The completeness and quality of the redshift information, and the available high-quality
multi-wavelength photometry allows us to derive various alternative estimates
of star-formation rates (SFR) in the 16 ALMA-detected galaxies, which we present
in Table\,4.

First, we convert the rest-frame UV ($\lambda = 1500$\AA) absolute magnitude,
$M_{1500}$ of each source into an estimate of the unobscured SFR. The adopted
calibration means than an absolute magnitude of $M_{1500} = -18.75$ corresponds
to a SFR of 1\,${\rm M_{\odot} yr^{-1}}$, consistent with the conversion
given by Kennicutt \& Evans (2012). The resulting values are given in column 3
of Table\,4, and are unspectacular (with SFR~$< 1\,{\rm M_{\odot}~yr^{-1}}$
for the majority of the sources).

Second, we use the SED fitting, as described above (to derive the stellar masses), to
estimate the extinction-corrected SFR from the UV--near-infrared SED. These estimates
are given in column 5 of Table\,4, with the corresponding best-fitting values of $A_V$
given in column 4 (extinction was allowed to range up to $A_V = 6$\,mag. in the SED fitting,
but the best-fitting values lie in the range 0.2 to 3.1\,mag.).

Third, we estimate dust-enshrouded SFR from the long-wavelength photometry, utilising the
ALMA data, the deconfused {\it Herschel} photometry, and testing the impact of including
or excluding {\it Spitzer} 24-$\mu$m photometry. Given the redshifts of the ALMA-detected
galaxies, the key ALMA 1.3-mm datapoint samples the rest-frame SEDs of the sources significantly
longward of the peak of any feasible far-infrared SED. This is good for the estimation of
robust dust masses, but means that the inferred SFR obviously depends
on the form of the adopted far-infrared SED template.

\begin{figure*}
\begin{center}
\includegraphics[scale=0.7]{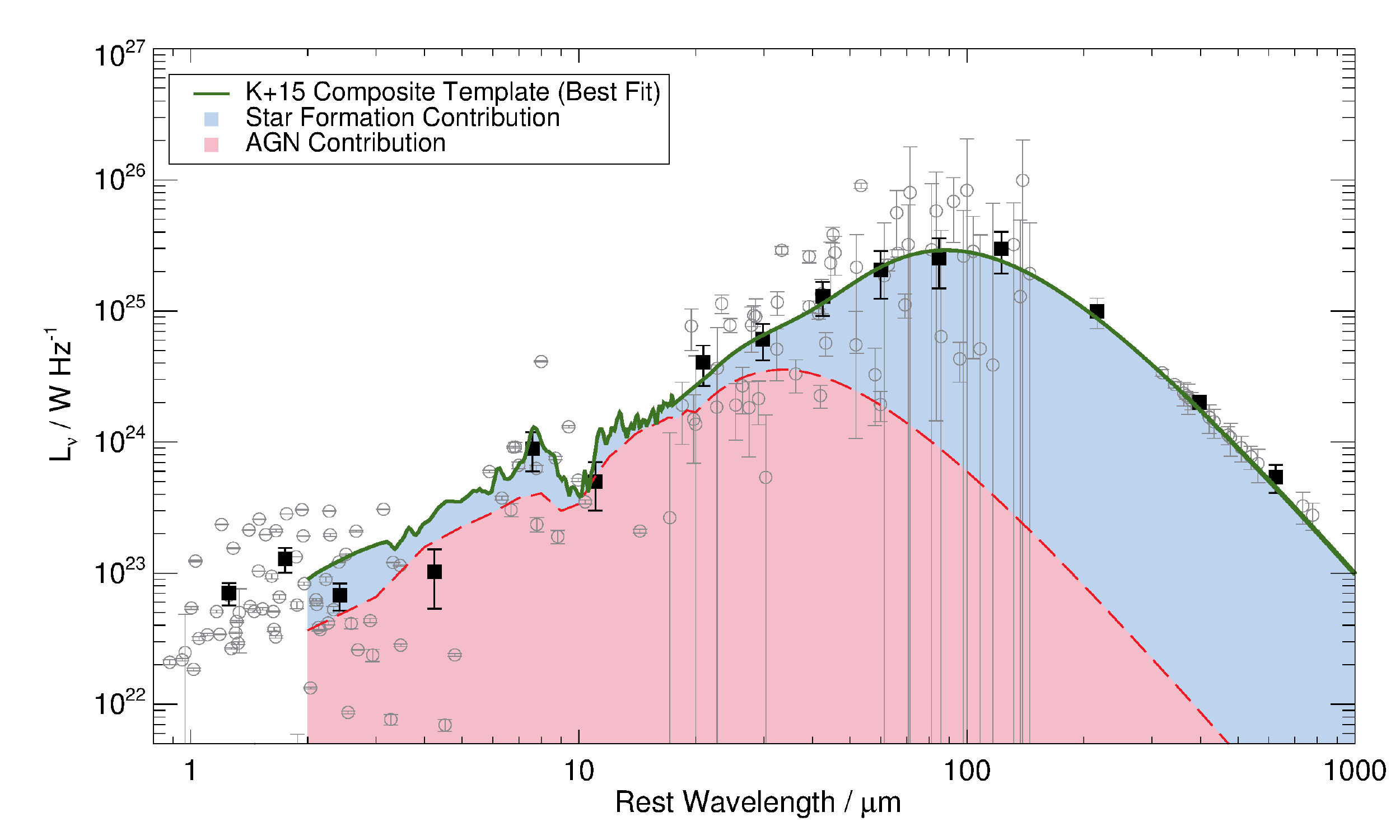}
\end{center}
\caption{The combined {\it Spitzer}+{\it Herschel}+ALMA photometry of the 16 ALMA sources,
  (after de-redshifting and scaling to the same rest-frame 1.3-mm luminosity), fitted by the
  composite star-forming+AGN template of Kirkpatrick et al. (2015). The solid black squares
  indicate the weighted mean of the scaled multi-source photometry within a given wavelength bin.
  The accuracy of the redshift
  information results in the 8-$\mu$m feature being clearly visible in the observed
  combined rest-frame SED. The best-fitting composite template
  has an AGN component which contributes 
  only $\simeq 20$\% to the bolometric rest-frame far-infrared ($8-1000$\,$\mu$m) luminosity, and
  is completely negligible at the wavelengths sampled by the ALMA imaging.}
\label{fig:pope}
\end{figure*}

We therefore investigated the long-wavelength SEDs of the individually detected sources, and also
amalgamated the 16-source photometry into a combined SED (after de-redshifting and normalizing
at $\lambda_{rest} = 1.3$\,mm) which we show in Fig.\,8. This combined SED (which, due to the accuracy of the
available redshift information, displays the rest-frame 8-$\mu$m feature) can be used to establish
the best template SED to use for the estimating of SFR for each source. The best-fitting model SED
shown in Fig.\,8 is the AGN+star-forming composite model of Kirkpatrick et al. (2015), and yields a
best fit with a 20\% AGN contribution to bolometric $8-1000$\,$\mu$m infrared luminosity.
We have therefore used this template to estimate the far-infrared luminosity of each of the ALMA sources,
exploring the impact of both including and excluding the 24-$\mu$m photometry to yield realistic
uncertainties, and then converting to SFR using the conversion of Murphy et al. (2011) with a final
minor scaling applied to convert from a Kroupa to a Chabrier IMF. The resulting values are tabulated in
column 6 of Table\,4, and are adopted hereafter as our best estimates of dust-enshrouded SFR.
We note that, although this calculation has been performed using the composite template
shown in Fig.\,8, and then reducing the far-infrared lumionosity by 20\% to compensate for the
typical AGN contribution, in practice near identical values are obtained by simply fitting
the star-forming component to the ALMA data-point. The star-forming component is essentially
the sub-mm galaxy template of Pope et al. (2008)\footnote{http://www.alexandrapope.com/\#\!downloads/t0u6h},
and at the redshifts of interest here ($z \simeq 2$)
produces a conversion between observed 1.3-mm flux density and SFR that can be approximated by
SFR (in ${\rm M_{\odot} yr^{-1}})$ $\simeq 0.30 \times S_{1.3}$\,(in $\mu$Jy).
With this template, and adopting a Chabrier (2003) IMF, the 
flux-density limit of the current survey thus corresponds to a limiting SFR sensitivity
of $\simeq 40\,{\rm M_{\odot} yr^{-1}}$.

For completeness, we also explored the impact of attempting to determine the best-fitting far-infrared SED
template for each individual source, rather than adopting a single template for all sources. This approach
makes more use of the deconfused {\it Herschel} photometry and limits, but the decision between
alternative SEDs is inevitably rather uncertain on a source-by-source basis. We fitted each source
with an M82, Arp220, or sub-mm galaxy template (Silva et al. 1998; Micha{\l}owksi et al. 2010), and again explored
the impact of including and excluding the 24\,$\mu$m data. The results are given in column 7 of Table\,4;
we give the average of the derived SFR with and without including the 24\,$\mu$m data, with the adopted error
being the larger of the statistical error in the fitting or the range of results dictated by the impact
of the 24\,$\mu$m data. The most important information to be gleaned from these results
is that the derived SFR is generally reassuringly similar to, or slightly lower than the values estimated
from the single template (shown in Fig.\,8) as described above. The somewhat lower values for some sources
simply reflect the fact that, when a source is better fit with the Micha{\l}owski et al. (2010) sub-mm galaxy SED
than either M82 or Arp220, the bolometric luminosity of the Pope et al. (2008) sub-mm galaxy template
is $\simeq 1.4$ larger than that of the Micha{\l}owski et al. (2010) sub-mm template when normalized at rest-frame
wavelengths $\lambda_{rest} \simeq 500$\,$\mu$m (when normalised at $\lambda_{rest} \simeq 150$\,$\mu$m,
the ratio is only $\simeq 1.1$).

Finally, we provide an estimate of SFR based on the new JVLA 6\,GHz photometry of the ALMA sources (Table\,3).
Given the potential for AGN contamination at radio wavelengths, uncertainty in the precise
radio-SFR calibration, and the need to adopt a consistent estimator for SFR for all 16 ALMA sources,
we do not make further use of the radio-based estimates of SFR in this paper.
Nevertheless, the the radio-inferred estimates of SFR given in column 8 of Table\,4 provide
reassurance that our far-infrared derived values are not seriously over-estimated.

In this context, we note from Tables 3 and 4 that there appear to be two radio-loud AGN in our sample, UDF3 and UDF7.
There are also two obvious X-ray AGN, UDF1 and UDF8. As listed in Table\,3, the catalogue produced
by Xue et al. (2011) also yields detections of three more galaxies in our sample, but these detections are $>20$ times fainter
than the obvious X-ray AGN, and, depending on the adopted extinction correction, may in fact be explained
by the star formation activity in these galaxies (although two of these weaker X-ray detections do correspond
to the two radio-loud AGN).

\begin{figure}
  \begin{center}
    \includegraphics[scale=0.39]{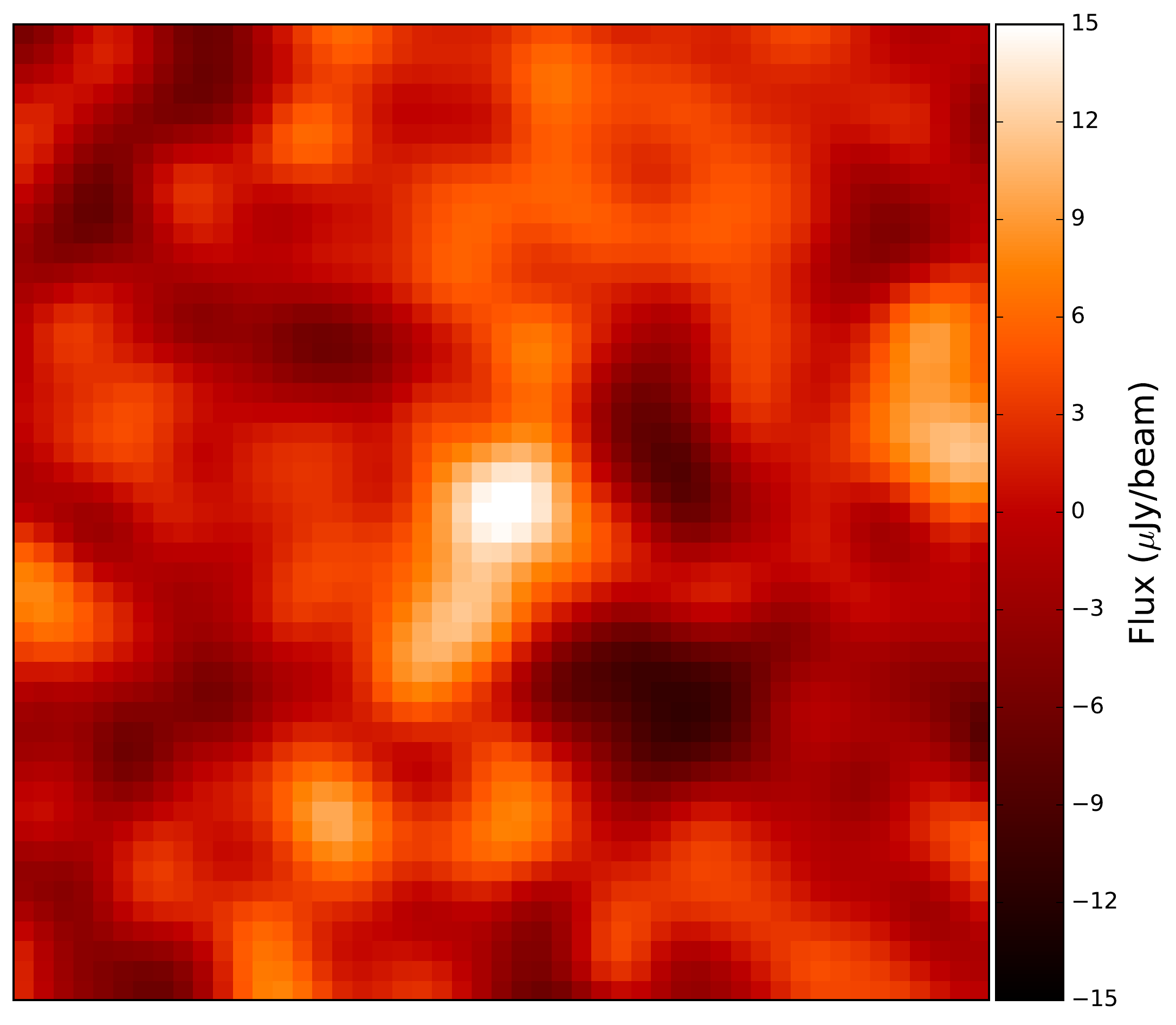}
  \end{center}
  \caption{The result of stacking the ALMA imaging
    on the positions of galaxies in the redshift range
    $1~<~z~<~3$ and the stellar mass range $9.3~<~\log_{10}(M_*/{\rm M_{\odot}})~<~10.3$,
    (excluding the sources already detected and listed in Table\,4).
    The image shown is 5.9 $\times$ 5.9
    arcsec. The stacked `source' includes 89 galaxies, and
    has a mean flux density (point-source + 25\%) of $S_{1.3} = 20.1 \pm 4.6\,{\rm \mu Jy}$, corresponding to a mean
    SFR of $6.0 \pm 1.4\,{\rm M_{\odot} yr^{-1}}$. This means that galaxies in this
    mass and redshift range contribute a total dust-enshrouded SFR~$\simeq~530~\pm~130\,{\rm M_{\odot}~yr^{-1}}$.
    The same sources have an average UV luminosity corresponding to an
    absolute magnitude of $M_{\rm UV} = -19.38$,
    and hence contribute a total unobscured (raw UV) SFR $\simeq 160\,{\rm M_{\odot} yr^{-1}}$.
    These results imply an average ratio of obscured:unobscured
    SFR of $\simeq 3.3$ (or total-SFR/unobscured-SFR $\simeq 4.3$), and a sSFR $\simeq 1.95$\,Gyr$^{-1}$
    at a median mass of $\log_{10}(M_*/{\rm M_{\odot}})=9.6$ and median redshift $z = 1.92$.}
  \label{fig:pope}
\end{figure}

\begin{figure}
\begin{center}
\includegraphics[scale=0.53]{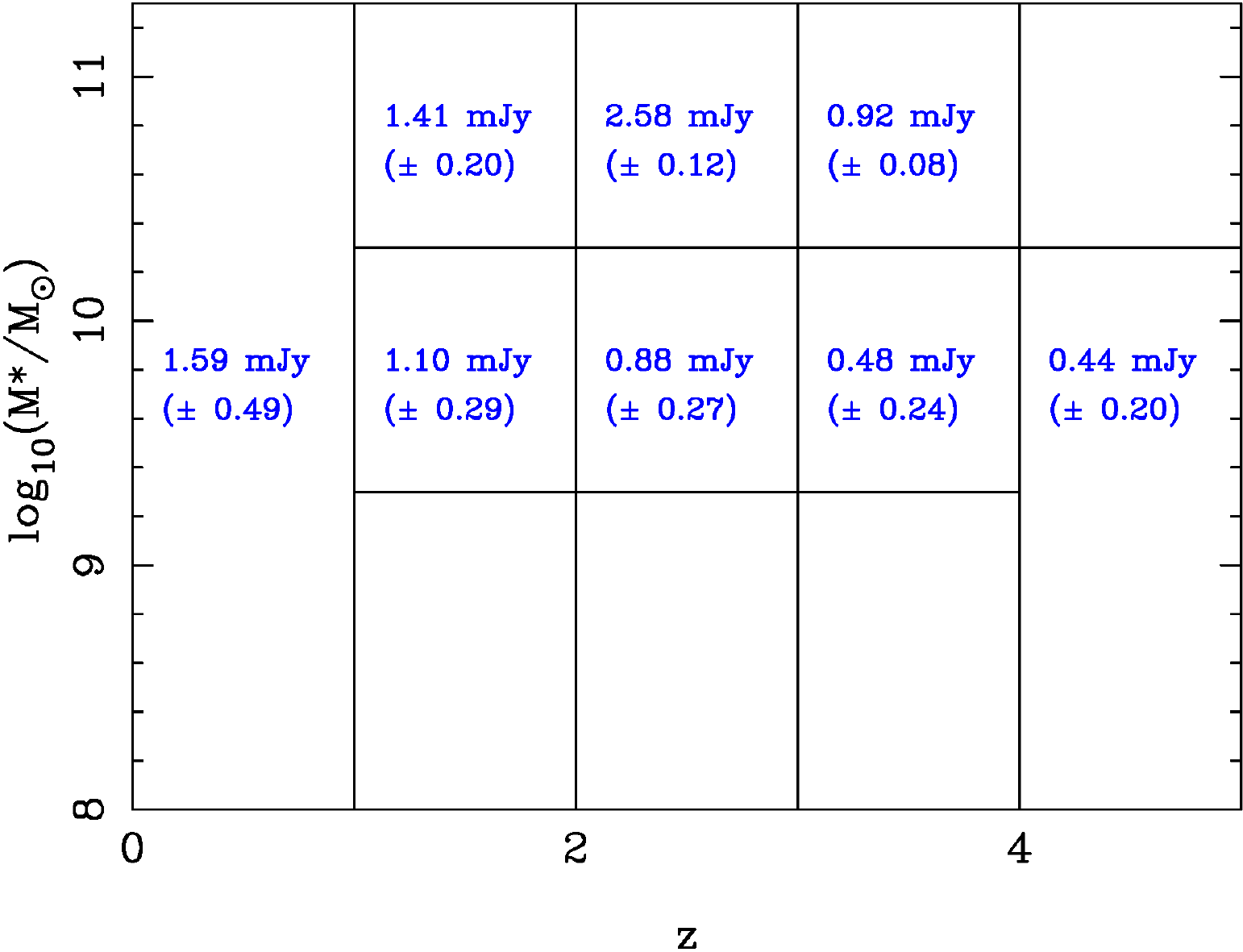}
\end{center}
\caption{The distribution of 1.3-mm flux density found within the HUDF across the galaxy redshift--star-mass
  plane. The figures given here combine the flux densities of the detected sources and the results of stacking in the
  appropriate redshift-mass bins (excluding detected sources to avoid double counting). Results are only given
  for bins within which stacking yielded better than a 2-$\sigma$ detection, typically corresponding to an average
  flux-density $S_{1.3} \ge 0.010$\,mJy. At $z > 4$ the HUDF contains no galaxies more massive than
  $\log_{10}(M_*/{\rm M_{\odot}}) = 10.3$, and so the right-hand cell in the top row is genuinely blank.
  The cells below $\log_{10}(M_*/{\rm M_{\odot}}) = 9.3$ at $1 < z < 3$ may contain additional 1.3-mm flux density,
  but not at a level that could be securely detected via stacking in the current image. At $z < 1$ and $z > 4$,
  low sample size precludes bin subdivision by mass, and so we simply give the total flux detected via stacking
  of all sources in the $z < 1$ and $z > 4$ redshift bins. Addition of all the numbers in this figure, combined
  with an estimated contribution of $\simeq 1$\,mJy per HUDF from sources brighter than $S_{1.3} \simeq 1$\,mJy (equivalent to
  0.8\,Jy deg$^{-2}$) gives an estimate of $10.6 \pm 1.0$\,mJy for the 1.3-mm flux density in the HUDF down to $S_{1.3} \simeq 0.01$\,mJy
  (equivalent to $8.5 \pm 0.9$\,Jy\,deg$^{-2}$; see Fig.\,10). Of this total estimated background, $\simeq 70$\% is
  provided by sources with $\log_{10}(M_*/{\rm M_{\odot}}) > 9.3$ in the redshift range $1 < z < 3$. Within our background
  estimate, $\simeq 45$\% of the flux density is contributed by the 16 individually-detected
  sources in our ALMA image of the HUDF, with only $\simeq 35$\% added
  by our attempts to extend flux retrieval down to $\simeq 0.01$\,mJy sources through stacking.}
\label{fig:flux_mass}
\end{figure}

\section{Stacking, and the mm background}

A key advantage of the HUDF, not shared by many other deep ALMA pointings, is the quality, depth and
completeness of the galaxy catalogue in the field. This enables stacking of chosen
galaxy sub-populations within the ALMA image, allowing us to explore source properties to significantly fainter
flux densities than achieved in the selection of robustly-detected individual sources.
Thus, not only can we attempt to derive an estimate of the total 1.3-mm flux-density present in the
field, but to the extent allowed by population statistics, we can explore how this (and hence dust-obscured
star-formation activity) is distributed as a function of redshift, galaxy
stellar mass, and/or UV luminosity.

We therefore explored stacking of galaxy sub-populations selected from a range of bins
defined by redshift $z$, stellar mass $M_*$, and UV absolute magnitude $M_{UV}$.
Perhaps unsurprisingly, given the clear link between stellar mass and ALMA flux-density revealed
by the location of the ALMA-detected sources in Fig.\,7, the most significant stack detections
were achieved in the next mass bin, which we defined as $9.3 < \log_{10}(M_*/{\rm M_{\odot}}) < 10.3$.
The 1.3-mm image stack of all galaxies in this mass range, and in the redshift range $1 < z < 3$, is shown
in Fig.\,9, and (excluding the two detected galaxies in this bin) produces a 4.4-$\sigma$ detection,
with a mean flux density (point source $+ 25$\%) $S_{1.3} = 20.1 \pm 4.6\,{\rm \mu Jy}$ (corresponding to a mean
SFR of $6.0 \pm 1.4\,{\rm M_{\odot} yr^{-1}}$; reassuringly a stack of the galaxies in this same bin into the JVLA 6-GHz image
yielded a detection with mean flux density $S_{\rm 6GHz} = 262 \pm 24\,{\rm nJy}$,
corresponding to a mean SFR of $6.4 \pm 1.0\,{\rm M_{\odot} yr^{-1}}$).

With 89 galaxies contributing to the stacked detection shown in Fig.\,9, the resulting
inferred total 1.3-mm flux-density in this bin is $1788~\pm~410\,{\rm \mu Jy}$.
We subdivided this bin by redshift (at $z = 2$), and proceeded in an analogous way to seek significant
($>~2$~-~$\sigma$) stacked 1.3-mm detections in other regions of the mass-redshift plane. In practice,
given the available number of galaxies, and the depth of the ALMA imaging, such detections
correspond to bins that yield an average flux-density $S_{1.3} > 10\,{\rm \mu Jy}$.
In Fig.\,10 we show the final result of this process, where, within each bin, we have also added back the
contribution of the individually detected sources. The coverage of the redshift-mass plane is limited by
the fact that i) there are no galaxies in the HUDF with $\log_{10}(M_*/{\rm M_{\odot}}) >  10.3$ to stack
at $z > 4$ (see Fig.\,7); ii) no significant stacked detections were achieved in mass bins confined to 
$\log_{10}(M_*/{\rm M_{\odot}}) <  9.3$, and iii) the limited number of galaxies in the field at $z < 1$ and $z > 4$
prohibited significant stacked detections sub-divided by mass at these redshifts. Nevertheless,
given the evidence for the steep dependence of dust-obscured star formation on stellar mass
explored further below, the sum of the figures given in Fig.\,10 can be expected to yield a reasonably
complete estimate of the 1.3-mm background, and we believe this represents the best estimate
to date of the distribution of this background as a function of redshift and galaxy stellar mass.

\begin{figure*}
  \begin{center}
    \includegraphics[scale=0.5]{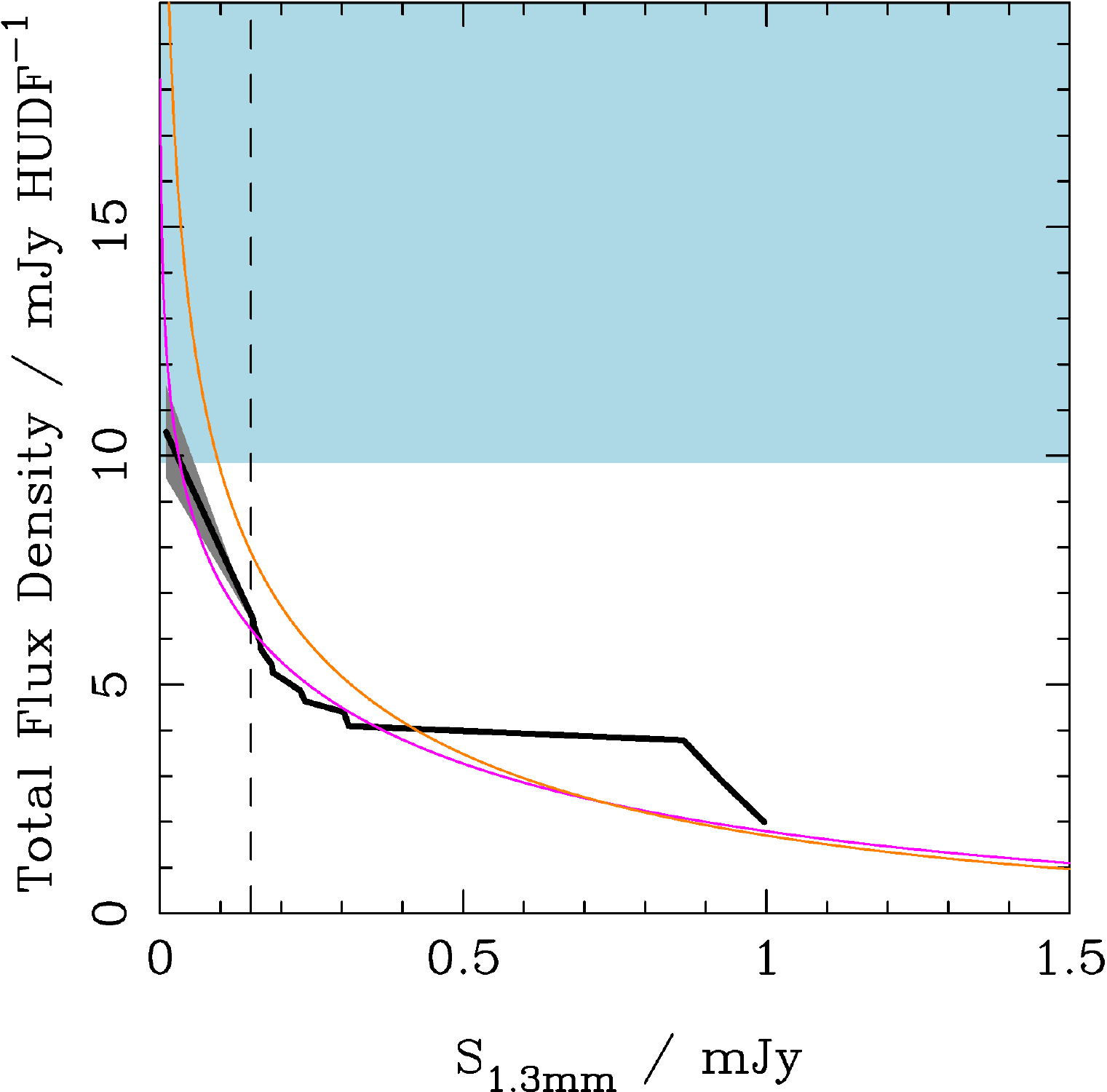}
    \hspace*{2cm}
    \includegraphics[scale=0.5]{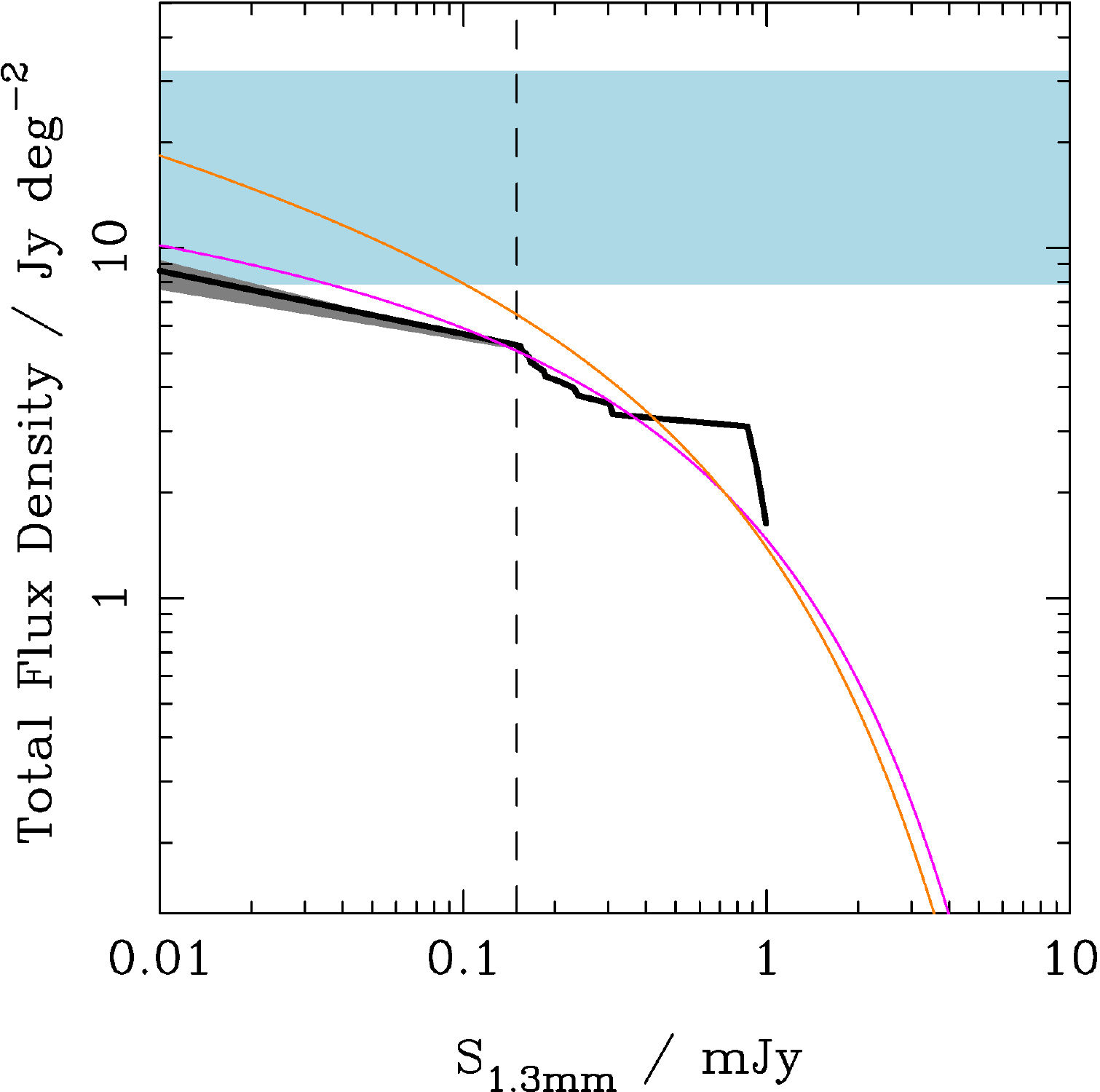}
  \end{center}
  \caption{Total integrated flux density at $\lambda = 1.3$\,mm. The left-hand plot
    shows results per HUDF (i.e. per 4.5\,arcmin$^2$) in linear units, while
    the right-hand plot shows the more standard logarithmic representation in units
    of Jy\,deg$^{-2}$. The solid black line in each plot shows simply the running
    sum of the flux densities of the 16 ALMA HUDF sources, starting at the total
    contributed by sources brighter that $S_{1.3} = 1$\,mJy (i.e. $\simeq 0.8$\,Jy\,deg$^{-2}$;
    Scott et al. 2012) and summing down to our (effective) detection threshold of $S_{1.3} = 0.15$\,mJy (marked
    by the vertical dashed line in both plots). Below that flux density, we extrapolate
    the black line (with grey shading to indicate the uncertainty)
    to account for the contribution estimated from our stacking analysis,
    which samples the fainter population down to $S_{1.3} \simeq 0.01$\,mJy (and adds
    an additional $4.1 \pm 1.0$\,mJy / HUDF). The blue-grey
    shaded region in both panels indicates the (highly-uncertain) estimate of the 1.3-mm
    background as measured by COBE (i.e. 17$^{+16}_{-9}$\,Jy\,deg$^{-2}$; Puget et al. 1996; Fixsen et al. 1998).
    It can be seen that our HUDF derived flux-density total is (just) consistent with the 1-$\sigma$ lower bound
    on the background estimate. The magenta and orange curves give the predicted
    background as a function of 1.3-mm flux-density for the scaled 
    Schechter number-count models of Hatsukade et al. (2016)
    and Fujimoto et al. (2016) already utilised in Fig.\,3,
    and discussed in Section 5.2. Our results can be plausibly
    reconciled with those of Hatsukade et al. (2016), but to achieve the much higher estimated
    background reported by Fujimoto et al. (2016) down to $S_{1.3} = 0.01$\,Jy
    we would require an approximate doubling of the flux density we have actually been able to uncover
    in the HUDF through source detection and stacking.}
\end{figure*}

We note that, at $z > 1$, the numbers given in Fig.\,10 can be reasonably converted into estimates
of total dust-enshrouded SFR per bin, by multiplying by $\simeq 300$ (see Section 6.3). However, at
$z < 1$, this will yield a serious over-estimate of SFR, because of the lack of a negative K-correction 
in this regime (i.e. a significant amount of the flux density in the $z < 1$ bin is contributed
by relatively low-redshift, but intrinsically not very luminous sources). For this reason, and because
at $z \simeq 1$, observations at 1.3-mm sample the far-infrared SEDs of the sources rather far from the
peak of emission, we do not use our stacked results at $z < 1$ in the discussion of star-formation rates
in the remainder of this paper. However, the contribution of sources at $z < 1$ is still of interest
when considering the implications for faint source counts, and the 1.3-mm background.

\begin{figure*}
  \begin{center}
\includegraphics[scale=0.9]{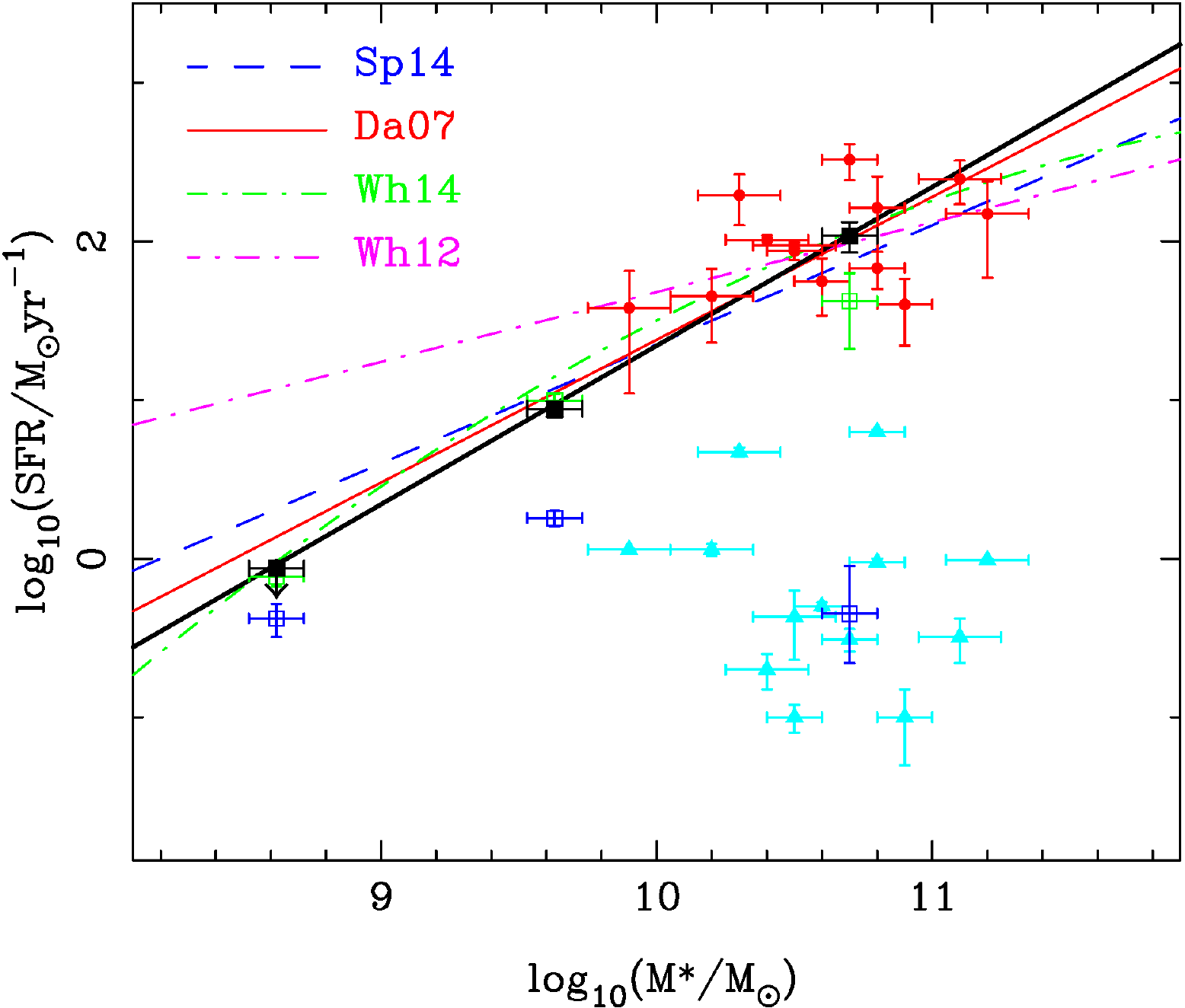}
\caption{Star-formation rate versus galaxy stellar mass at $z \simeq 2$. The 13 ALMA sources
  that lie in the redshift range $1~<z~<3$ (median redshift $z = 2.09$) are shown twice, first
  adopting the total star-formation rate (i.e. UV+FIR SFR; red points) and then adopting only the raw UV
  SFR (cyan triangles). Also shown is the average value in 3 stellar-mass bins at $1~<z~ \leq3$
  (derived from stacking as described in the text)
  for total UV+FIR SFR (black solid squares), raw UV SFR (navy-blue open squares), and dust-corrected
  SFR (from the optical--near-infrared SED fitting; green open squares). The binned points are
  plotted at the median stellar mass of each bin.
  Reassuringly, the green
  and black values agree well in the two lower-mass bins, but the SED $A_V$-corrected values
  fall about a factor of 2 short of the true ALMA-derived average in the highest mass bin.
  The solid black line shows a simple relation of the form SFR~$\propto~M_*^{1.0}$, with sSFR\,=\,2.2\,Gyr$^{-1}$.
  The other curves are proposed fits to the `main-sequence' of star-forming galaxies, at $z \simeq 2$, published by
  Daddi et al. (2007; solid red line, after conversion to a Chabrier IMF), Whitaker et al. (2012;
  dot-dashed magenta line), Whitaker et al. (2014; dot-dashed green curve), and Speagle et al. (2014; dashed blue line).}  
\end{center}
\end{figure*}

In Fig.\,11 we explore our resulting estimate of 
integrated flux density at $\lambda = 1.3$\,mm, as a function of source flux-density, utilising
our detections down to $S_{1.3} \simeq 150\,{\rm \mu Jy}$, and the contribution produced by
the sum of stacked fluxes (excluding the individual source contributions from Fig.\,10) reaching down
to $S_{1.3} \simeq 10\,{\rm \mu Jy}$. The left-hand panel of Fig.\,11
shows results per HUDF (i.e. per 4.5\,arcmin$^2$) in linear units, while
the right-hand panel shows the more standard logarithmic representation in units
of Jy\,deg$^{-2}$. The solid black line in each plot shows simply the running
sum of the flux densities of the 16 ALMA HUDF sources, starting at the total
contributed by sources brighter that $S_{1.3} = 1$\,mJy (i.e. $\simeq 0.8$\,Jy\,deg$^{-2}$;
Scott et al. 2012) down to our (effective) detection threshold of $S_{1.3} = 0.15$\,mJy.
Below this flux-density we extrapolate the line down to $S_{1.3} \simeq 10\,{\rm \mu Jy}$ 
to account for the contribution estimated from our stacking analysis,
(which adds an additional $4.1 \pm 1.0$\,mJy / HUDF).
Our HUDF-derived estimate of total flux density can be seen to be (just)
consistent with the 1-$\sigma$ lower bound
on the estimated background as measured by COBE (i.e. 17$^{+16}_{-9}$\,Jy\,deg$^{-2}$;
Puget et al. 1996; Fixsen et al. 1998).
Our results can be plausibly
reconciled with those of Hatsukade et al. (2016), but to achieve the much higher estimated
backgrounds reported by Fujimoto et al. (2016) or Carniani et al. (2015) down to $S_{1.3} = 0.01$\,mJy
we would require to approximately double the flux-density we have actually been able to uncover
in the HUDF through source detection and stacking.

\section{Discussion}

\label{sec:discussion}

\subsection{The star-forming main sequence}
Armed with the physical knowledge of the properties of the individual
ALMA-detected galaxies (Table\,4), the stacking results discussed in the previous section,
and our knowledge of the redshifts, rest-frame UV luminosities and stellar masses of all galaxies in the HUDF, we now
investigate the implications for the relationship between star formation and stellar mass at $z \simeq 2$.

We confine our attention to the 13 ALMA-detected sources with $1.0 < z \leq 3.0$, derive the average
properties of this sample, and consider also the average results from stacking in the same redshift range
within the two stellar mass bins defined by $9.3 < \log_{10}(M_*/{\rm M_{\odot}}) < 10.3$ and
$8.3 < \log_{10}(M_*/{\rm M_{\odot}}) < 9.3$. The median redshift of the detected sample is $z = 2.086$,
while for the two stacks the median redshifts are $z \simeq 1.92$ and $z = 2.09$. The corresponding median
stellar masses are $\log_{10}(M_*/{\rm M_{\odot}}) = 10.70$, 9.63 and 8.62.

The results are plotted in Fig.\,12. This shows (as solid red circles) the positions on the SFR--$M_*$ plane of the
13 ALMA-detected sources, with associated uncertainties (see Table\,4), with the average (median) values of total
(FIR+UV) SFR plotted for the three mass bins as solid black squares (with standard errors). For the lowest of the
stellar mass bins, no detection was achieved in the ALMA stack of the 391 galaxies in the redshift range $1.0 < z \leq 3.0$,
and so we plot an upper limit derived from the rms of $S_{1.3} = 1.5\,{\rm \mu Jy}$ achieved in this stack (corresponding
to a 1-$\sigma$ limit on average obscured SFR $< 0.45\,{\rm M_{\odot} yr^{-1}}$, and hence a limit of
total SFR $< 0.87\,{\rm M_{\odot} yr^{-1}}$).

The solid black line in Fig.\,12 shows a simple relation of the form SFR~$\propto~M_*^{1.0}$, with sSFR\,=\,2.2\,Gyr$^{-1}$. This 
is clearly an excellent fit to our data. Also shown is the original `main-sequence' (MS) of star-forming galaxies at $z \simeq 2$
derived by Daddi et al. (2007) (SFR~$\propto~M_*^{0.9}$, adjusted in mass normalization to account for the change from
Salpeter IMF to Chabrier IMF),
the shallower $z \simeq 2$ MS presented by Whitaker et al. (2012)
(SFR~$\propto~M_*^{0.6}$), the revised steeper polynomial form presented by Whitaker et al. (2014),
and the result of the meta analysis undertaken by Speagle et al. (2014) (calculated
at $z \simeq 2$). All of these (and many more) published relations are perfectly
consistent with our data at $\log_{10}(M_*/{\rm M_{\odot}}) \simeq 10.7$,
proving beyond doubt that the ALMA-detected galaxies lie on the MS at $z \simeq 2$.
However, over the dynamic range probed here, none fit any better than (or indeed as well as) the simply constant sSFR
relation plotted in black (although both the original Daddi et al. and the revised Whitaker et al. relations
are also clearly acceptable).

In Fig.\,12 we also plot the corresponding results for the raw UV SFR for the individual sources (cyan triangles), and
median values in each stellar mass bin (open navy-blue squares). Finally, in each mass bin we also plot the median
values of dust-corrected UV SFR derived from the optical--near-infrared SED fitting (open green squares),
as would be obtained in the absence of any direct mm/far-IR information (i.e. based on the values given in
column 5 of Table\,4, and analogous results for all galaxies in the two lower-mass bins). In the two lower-mass
bins the positions of these green points are reassuringly close to the black points, indicating that
dust-corrected SFR from optical--near-infrared SED fitting works well for moderately-obscured lower mass galaxies,
and also confirming that the steepness of the MS at low masses is not an artefact of the ALMA stacking procedure.
However, in the highest mass bin, the SED-estimated median SFR falls short of the true ALMA-derived values
by a factor $\simeq 2$. This is perhaps not surprising, as the ratio of median total SFR to median raw
UV-estimated SFR at $\log_{10}(M_*/{\rm M_{\odot}}) = 10.70$ can be seen to be $\simeq 300$. The ALMA-derived results
argue against any flattening of the MS, at least up to stellar masses
$\log_{10}(M_*/{\rm M_{\odot}}) \simeq 11$ (provided quenched galaxies are excluded, although the lack of
quenched high-mass galaxies in the field makes this distinction academic at $z > 2$).

\begin{figure}
\begin{center}
\includegraphics[scale=0.58]{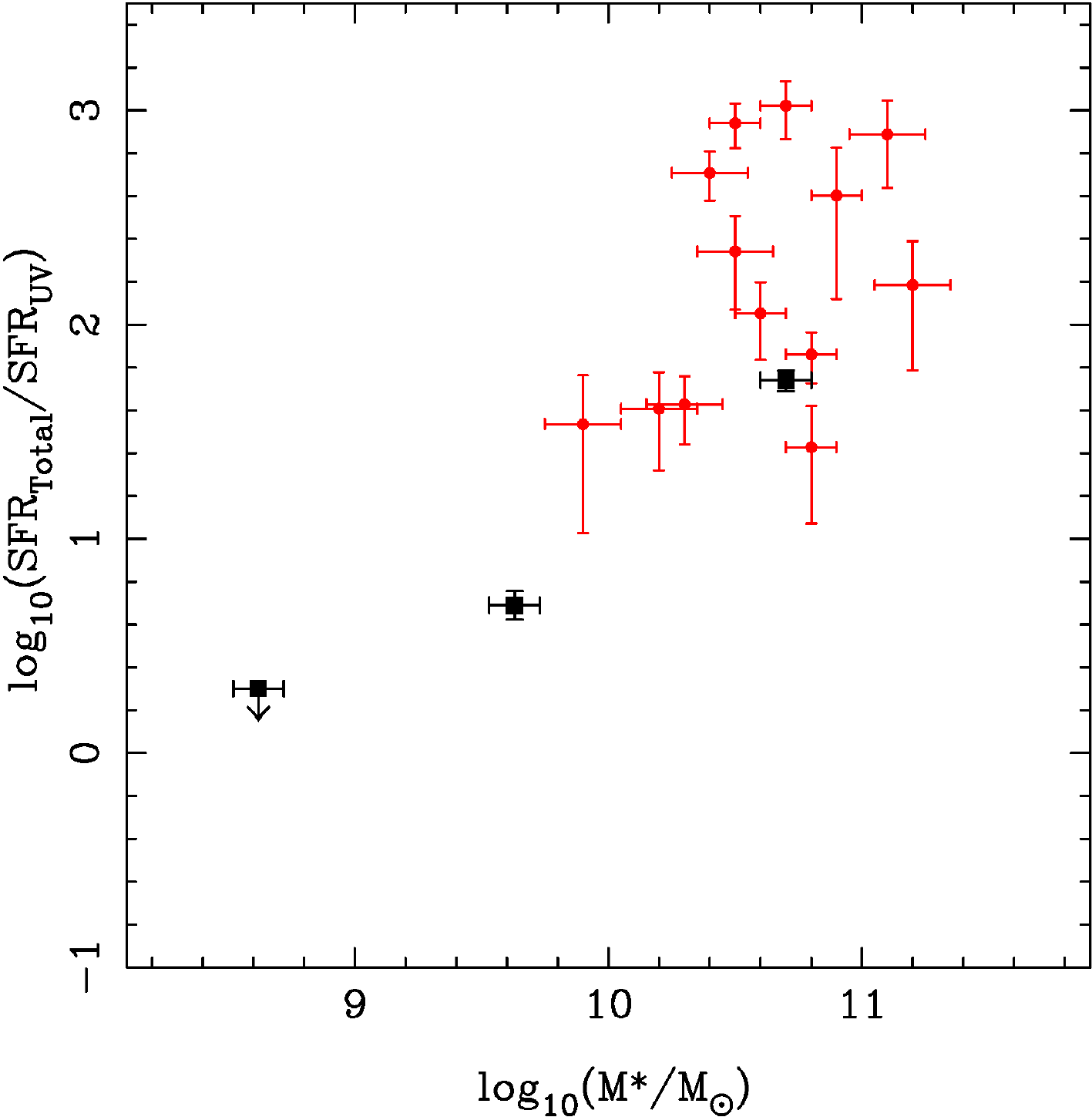}
\end{center}
\caption{The ratio of total SFR (i.e. UV+FIR) to UV-visible SFR as a function of galaxy stellar mass at $z \simeq 2$.
The individual ratios for the 13 ALMA-detected sources at $1 < z \leq 3.0$ are indicated by
the red points, and the median value at $\log_{10}(M_*/{\rm M_{\odot}}) > 10.3$ is $\simeq 200$ (see column 9 in Table\,4).
The black squares indicate the ratios for each of our three mass bins, summing all SFR
(detected-FIR + stacked-FIR + total-UV) in each bin, and dividing by all the raw UV SFR in that mass bin.
The stacked (true total ratio) value at median mass $\log_{10}(M_*/{\rm M_{\odot}}) = 10.7$ is $55 \pm 6$ (significantly
lower than the median ratio of ALMA-detected sources). The average ratio at $\log_{10}(M_*/{\rm M_{\odot}}) = 9.6$ is $4.9 \pm 0.8$,
while at $\log_{10}(M_*/{\rm M_{\odot}}) = 8.6$ our analysis yields a limit on this ratio of $< 2.5$ (see Section 8.2).}

\label{fig:obs_ratio}
\end{figure}

\subsection{Mass dependence of obscuration}

The results discussed in the previous sub-section, and presented in Fig.\,12, clearly imply
a strong mass dependence for the ratio of obscured to unobscured star formation.
We explore this explicitly, again focussing on $z \simeq 2$ (i.e. $1 < z \leq 3$), in Fig.\,13.
Here we plot the ratio of total SFR (i.e. UV+FIR) to UV-visible SFR as a function of mass.
Again we plot the individual ratios for the 13 ALMA detected sources (see column 9 in Table\,4),
and we also plot the ratios for each of our three mass bins (for the lowest mass bin, an upper limit, as in
Fig.\,12), summing all SFR (detected-FIR + stacked-FIR + total-UV)
in each mass bin, and dividing by all the raw UV SFR in that bin.

It can be seen that the median value of this ratio for the detected sources is 152, or in fact 218 if the
two detected galaxies in this redshift range with $\log_{10}(M_*/{\rm M_{\odot}}) < 10.3$ are disregarded (for consistency
with the high mass bin for which $10.3 < \log_{10}(M_*/{\rm M_{\odot}}) < 11.3$). However, the stacked
(true total ratio) value at median mass $\log_{10}(M_*/{\rm M_{\odot}}) = 10.7$ is significantly lower, at $55 \pm 6$. 
The average ratio at $\log_{10}(M_*/{\rm M_{\odot}}) = 9.6$ is $4.9 \pm 0.8$, while at $\log_{10}(M_*/{\rm M_{\odot}}) = 8.6$
our analysis yields $< 2.5$.

Clearly, the ratio of total SFR to UV-visible SFR is a very steep function of mass above $\log_{10}(M_*/{\rm M_{\odot}}) \simeq 9.5$;
over the next decade in mass, this ratio also increases by a factor $\simeq 10$, indicating that essentially 
all the increase in SFR with stellar mass on the main sequence at higher masses is delivered in a dust-obscured form.

This clarifies why sub-mm observations are so effective at detecting high-mass galaxies at $z \ge 2$.
Because the galaxies lie on the star-forming main sequence, a galaxy with $\log_{10}(M_*/{\rm M_{\odot}}) = 10.7$
will have ten times the intrinsic SFR of a galaxy of stellar-mass $\log_{10}(M_*/{\rm M_{\odot}}) = 9.7$, {\it and} the
obscured:unobscured ratio is also typically ten times greater.

\begin{figure*}
\begin{center}
\includegraphics[scale=0.53]{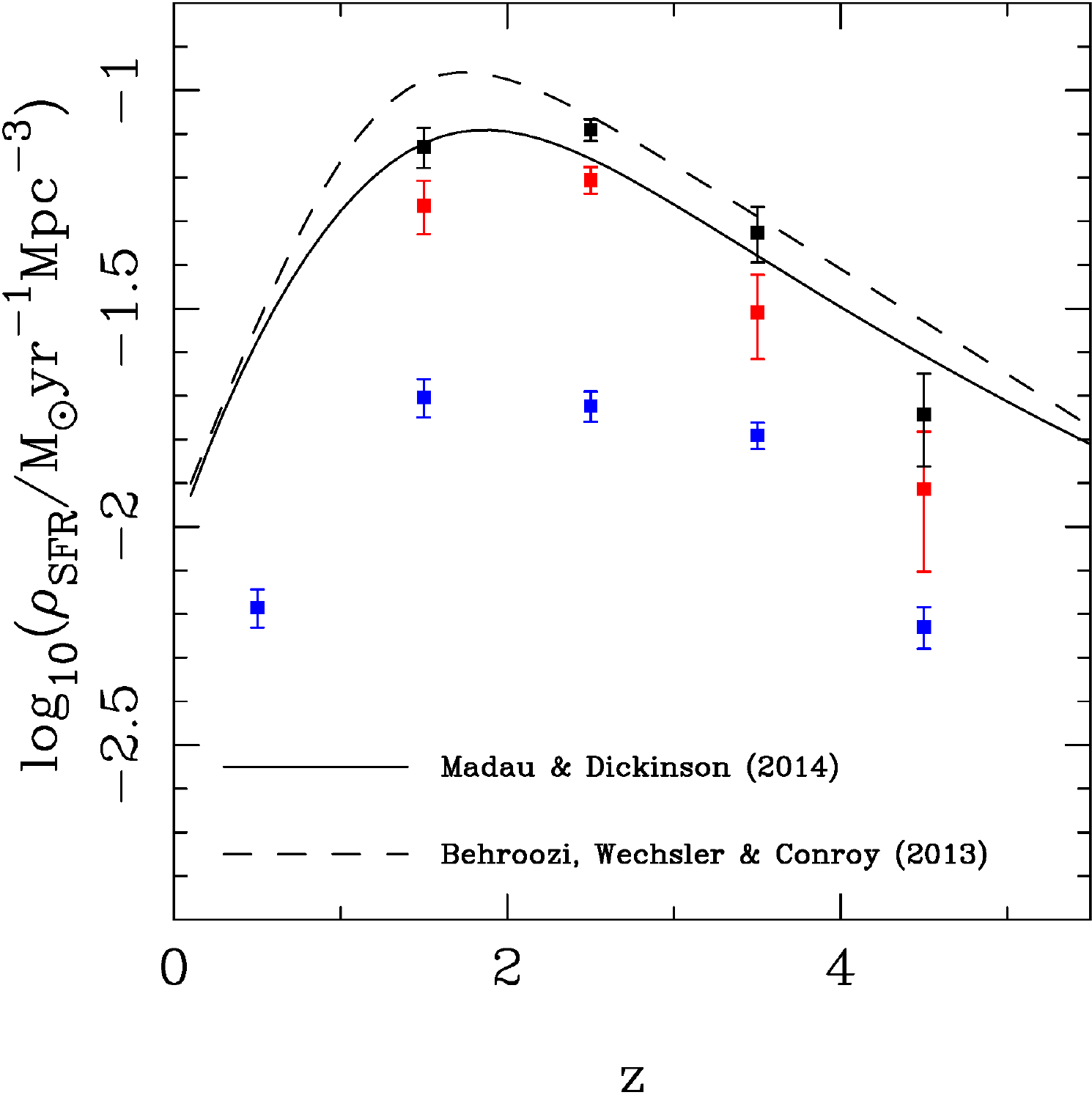}
\hspace*{1.2cm}
\includegraphics[scale=0.53]{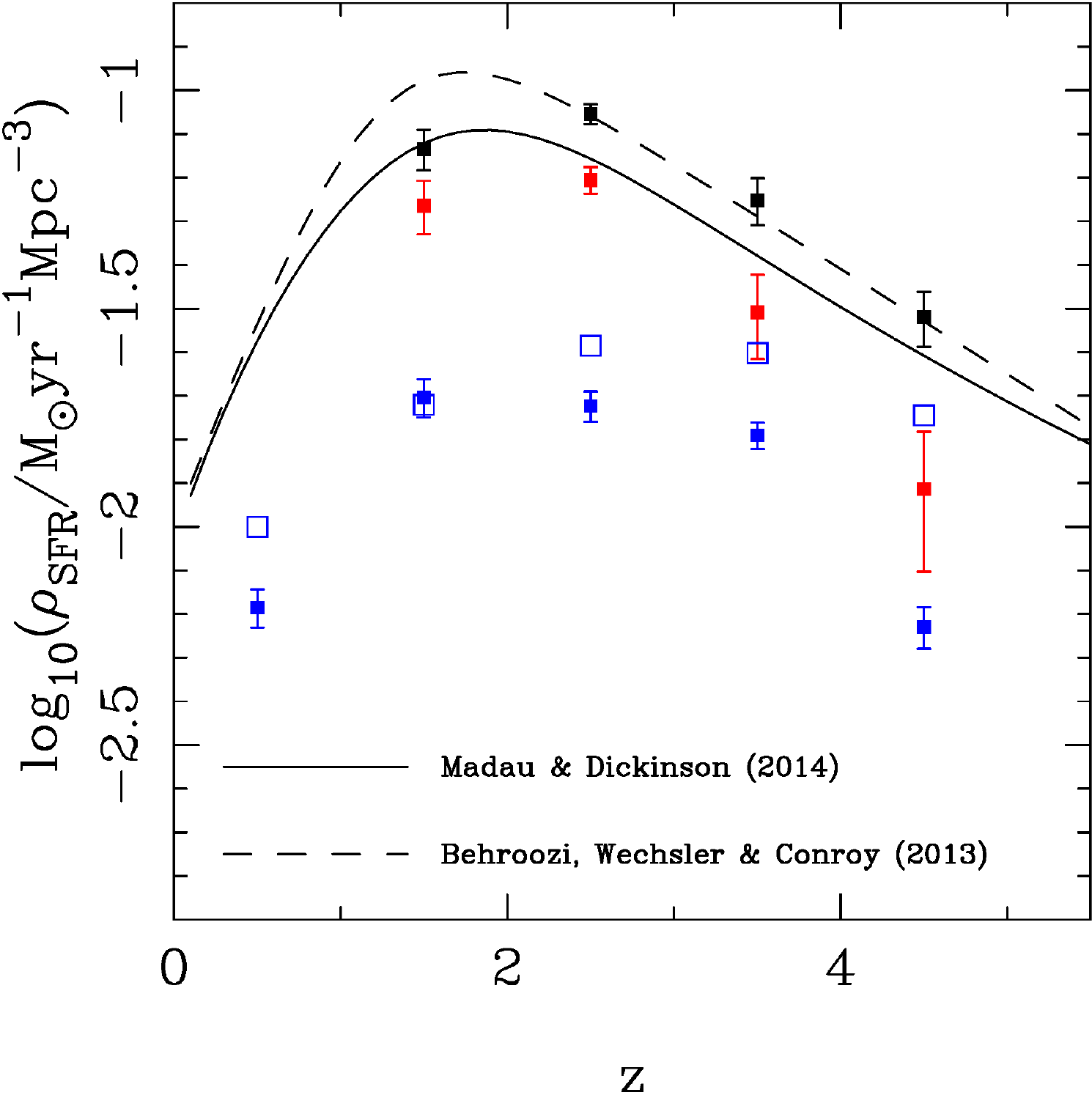}
\end{center}
\caption{The cosmic evolution of comoving star-formation rate density, $\rho_{SFR}$. In the {\bf left-hand panel}
  we plot our new ALMA-derived estimates of $\rho_{SFR}$(obscured) (red points), estimated values of
  $\rho_{\rm SFR}$(visible) produced by summing the raw UV-derived star-formation rates of the galaxies
  in the HUDF (blue points), and the values of $\rho_{\rm SFR}$(total) that result from adding
  these two contributions at each redshift (black points).
  The solid and dashed curves show two recent parametric descriptions of the evolution of $\rho_{\rm SFR}$
  as derived from the literature by Madau \& Dickinson (2014) and Behroozi, Wechsler \& Conroy (2013)
  respectively (scaled to the IMF of Chabrier 2003). In the {\bf right-hand panel} we show the impact of including
  a more complete census of $\rho_{\rm SFR}$(visible), based on the luminosity-weighted integral
  of the evolving UV galaxy luminosity function, down to $M_{\rm UV} = -15$ (Parsa et al. 2016; Bouwens
  et al. 2015). In this panel the values of $\rho_{SFR}$(obscured) are unchanged, but the black points
  indicating $\rho_{\rm SFR}$(total) now include the more complete (and, at high redshift, higher) census 
  of $\rho_{\rm SFR}$(visible), as shown by the open blue squares. At $z \simeq 2$ the agreement with the parametric fit provided by Behroozi et al.
  (2013) is excellent, and our results indicate that there there is a transition from
  $\rho_{\rm SFR}$(visible)/
  $\rho_{\rm SFR}$(obscured) $>$ 1, to  $\rho_{\rm SFR}$(visible)/
  $\rho_{\rm SFR}$(obscured) $<$ 1 at $z \simeq 4$.}
\end{figure*}

\subsection{Cosmic star-formation history}

Finally, we use our new ALMA results,
combined with our existing knowledge of the rest-frame
UV properties of the galaxies in the HUDF, to explore the evolution of cosmic star-formation rate density,
$\rho_{\rm SFR}$. In Fig.\,14 we plot the dust-obscured values of $\rho_{SFR}$ in unit redshift bins from $z = 1$ to $z = 5$,
combining our ALMA detections (Table.\,4) and the results of the stacking summarized in Section 7. In essence,
these results (indicated by the red data-points in both panels of Fig.\,14) are the result of collapsing
the 1.3-mm flux distribution shown in Fig.\,10 along the stellar-mass axis, converting
to $\rho_{\rm SFR}$(obscured) using the SED plotted in Fig.\,8 (see Section 6.3) and dividing by the comoving cosmological
volume sampled by the 4.5\,arcmin$^2$ of the HUDF in each redshift bin. This thus represents a
direct sum of the observed dust-obscured star-formation activity in the field, and does not
rely on, for example, assumptions regarding the poorly constrained faint-end slope of the
far-infrared luminosity function at these redshifts.

In a similar manner, we have summed up all the individual values of raw UV SFR for all the $\simeq 2000$
galaxies uncovered through the {\it HST} imaging of the HUDF, to construct the evolution
of $\rho_{\rm SFR}$(visible) over the same redshift range (indicated by the solid blue points
in both panels of Fig.\,14). These two values are then simply added to produce an estimate
of $\rho_{\rm SFR}$(total), which is plotted as the black points in the left-hand panel of Fig.\,14,
and compared with two recently published parametric fits to the evolution of $\rho_{\rm SFR}$ as derived from
reviews of the existing literature results (Behroozi, Wechsler \& Conroy 2013; Madau \& Dickinson 2014).

From the left-hand panel of Fig.\,14 it can be seen that the results of this simple calculation are remarkably
consistent with the published curves at $z \simeq 1 - 3$ (which, where required, have been scaled
to the Chabrier IMF), but our derived value appears somewhat low at $z \simeq 4.5$. However, by these redshifts
the census of UV SFR produced by this simple summing of galaxy contributions
is inevitably incomplete at the faint end, with the median value of $M_{\rm UV}$ changing from
$-$16.8 at $z \simeq 1.5$ to
$-$18.1 at $z \simeq 4.5$.

Therefore, to enable a proper and consistent accounting of the UV contribution, we integrated
the evolving galaxy UV luminosity function (weighted by UV luminosity) down to a consistent
luminosity limit corresponding to $M_{\rm UV}~=~-15$. To do this we used the latest UV luminosity functions
produced by Parsa et al. (2016) and Bouwens et al. (2015), and the results are shown by the open
blue squares in the right-hand panel of Fig.\,14, where they are also then added to the (unchanged) values of
$\rho_{\rm SFR}$(visible) to produce a revised estimate for  $\rho_{\rm SFR}$(total). Reassuringly, it
can be seen that the two alternative values of $\rho_{\rm SFR}$(visible) are indistinguishable
at $z \simeq 1.5$, where the HUDF data are deep enough to sample the UV luminosity function
to $M_{\rm UV} < -15$ (Parsa et al. 2016). However, at $z < 1$ the HUDF is too small
to properly reflect the contribution made to UV luminosity density by the brighter galaxies,
while at $z > 2$ the direct census becomes increasingly incomplete, and the integration
of the UV luminosity function yields a systematically increasing upward correction.

At $z \simeq 1 - 3$ the overwhelming dominance of dust-obscured star formation means
that completing the UV census in this way makes little impact on the estimated value of
$\rho_{\rm SFR}$(total). However, at $z > 3$ the impact is more pronounced, and by $z \simeq 4 -5$
the effect is dramatic enough to result in $\rho_{\rm SFR}$(visible) being larger than
$\rho_{\rm SFR}$(obscured), and $\rho_{\rm SFR}$(total) being lifted up to values that are
consistently higher than predicted by the Madau \& Dickinson (2014) fit, but in excellent
agreement with (at least the high-redshift end of) the parametric fit obtained by Behroozi et al. (2013).

We stress that this is the first time a direct census of $\rho_{\rm SFR}$(obscured) has been performed
at these redshifts, and that the integration of the UV LF has not always been performed
to a consistently deep luminosity limit at all redshifts. It is therefore reassuring
to see such good agreement with the results of existing literature reviews, but also
interesting to note how the balance of power shifts from unobscured to obscured star formation
with cosmic time. It appears that, at redshift $z > 4$, most of the star formation
in the Universe is unobscured, and relatively modest corrections are therefore
required to infer $\rho_{\rm SFR}$(total) from rest-frame UV observations. By contrast, at $z < 4$
the obscured mode of star formation becomes increasingly dominant until it is primarily
responsible for producing the peak in
$\rho_{\rm SFR}$(total) at $z \simeq 2.5$. Both Behroozi et al. (2013) and Madau \& Dickinson (2014)
favoured a peak in $\rho_{\rm SFR}$(total) at $z \leq 2$, and it may be the case
that the very brightest sub-mm sources not sampled by the small volume of the HUDF
contribute sufficiently around this redshift to both boost $\rho_{\rm SFR}$ by a few percent,
and perhaps shift the peak to slightly lower redshifts. However, recent studies utilising {\it Herschel}
and SCUBA-2 data to probe the dust-obscured contribution out to $z \simeq 3$ also appear to
favour a peak redshift in the range $z \simeq 2-3$ (e.g. Burgarella et al. 2013; Bourne et al. 2016).

While the average metallicity of the Universe must obviously increase with cosmic time,
the apparent transition at $z \simeq 4$ from predominantly visible star formation
at higher redshifts, to primarily
dust-obscured star formation at $z < 4$ is not necessarily driven by an increase in the prevalence of dust
in galaxies of a given mass. Rather it appears to be largely due to 
the rapid growth in the number density of the high-mass galaxies that contain most of the
dust-obscured star formation at $z \simeq 1 -3$ (see Fig.\,7 and Fig.\,13); indeed there
is at most only weak evidence for any evolution in the ratio of obscured:unobscured SFR with
redshift for galaxies selected at constant stellar mass $M_*$ (Bourne et al. 2016).
As is clear from Fig.\,7, deeper sub-mm imaging with ALMA has the potential to resolve this issue, by charting the evolution
of dust-obscured star-formation activity at constant stellar mass from $z \simeq 7$ to the present day. 

\section{Conclusions}
\label{sec:conclusions}

We have constructed the first deep, contiguous and homogeneous ALMA image of the Hubble Ultra
Deep Field, using a mosaic of 45 ALMA pointings at $\lambda = 1.3$\,mm to map the
full $\simeq 4.5$\,arcmin$^2$ area previously imaged with WFC3/IR on the
{\it HST}. The resulting image reaches an rms sensitivity
$\sigma_{1.3} \simeq 35$\,${\rm \mu Jy}$, at a resolution of
$\simeq 0.7$\,arcsec.
A search for sources in this image yielded an initial list of 
$\simeq 50$ $>3.5\sigma$ peaks, but an analogous analysis of the negative image showed
that, as expected from the size and noise level of the map, $30-35$ of these peaks were likely
to be spurious. We then exploited the unparalled optical/near-infrared data in the field
to isolate the real sources, via the identification of robust galaxy counterparts
within a search radius of $\simeq 0.5$\,arcsec (in the process uncovering the need for a
$\simeq 0.25$\,arcsec shift in the {\it HST} co-ordinate system).

The result is a final sample of 16 ALMA
sources with point-source flux densities $S_{1.3} > 120$\,${\rm \mu Jy}$.
The brightest three of these sources were clearly resolved, and so we measured
their total flux densities from image fitting. For the fainter sources
we estimated total flux densities by applying a 25\% boost to their point-source
flux densities (a correction based on a stack of the brightest five sources).

All of the ALMA sources have secure galaxy counterparts with accurate redshifts
(13 spectroscopic, 3 photometric), yielding a mean redshift
$\langle z \rangle = 2.15$. Within our sample, 12 galaxies
are also detected at 6\,GHz in new ultra-deep JVLA imaging.
Due to the wealth of supporting data in this unique field,
the physical properties of the ALMA-detected galaxies
are well constrained, including their stellar masses and UV-visible
star-formation rates. To estimate the dust-obscured star-formation
rates for the sources, we established a template far-infrared SED by fitting
their combined ALMA and (deconfused) {\it Spitzer}+{\it Herschel} photometry. 

Our results confirm previous indications that stellar mass is the best predictor of
star formation rate in the high-redshift Universe, with our ALMA sample containing
7 of the 9 galaxies in the HUDF with $M_*~\geq~2~\times~10^{10}\,{\rm M_{\odot}}$ at
$z \geq 2$. We detect only one galaxy at $z > 3.5$, and show that the lack of
high-redshift detections simply reflects the rapid drop-off of high-mass galaxies
in the field at $z > 3$.

The detected sources, coupled with results of stacking in bins of redshift and mass,
allow us to probe the redshift/mass distribution of the 1.3-mm background down to
$S_{1.3} \simeq 10\,{\rm \mu Jy}$, and we find that our estimate of the total
1.3-mm background provided by detected and stacked sources is (just) consistent
with the background measurement made by COBE.

We find strong evidence for a steep `main sequence' for star-forming galaxies at $z \simeq 2$,  
with SFR $\propto M_*$ and a mean specific SFR $\simeq 2.2$\,Gyr$^{-1}$.
Moreover, we find that $\simeq 85$\% of total star formation at $z = 1 -3$ is enshrouded in dust,
with $\simeq 65$\% of all star formation at this epoch occurring in
high-mass galaxies ($M_* > 2 \times 10^{10}\,{\rm M_{\odot}}$), for which the
average obscured:unobscured SF ratio is $\simeq 200$. Averaged over cosmic volume we find that,
at $z \simeq 2$, the ratio of obscured to unobscured star-formation activity
rises roughly proportional to stellar mass, from a factor $\simeq 5$
at  $M_*\simeq~5~\times~10^{9}\,{\rm M_{\odot}}$, to a factor $\simeq 50$ at  $M_*\simeq~5~\times~10^{10}\,{\rm M_{\odot}}$.

Finally, we combine our new ALMA results with the existing {\it HST} data
to attempt a complete census of obscured and visible star formation in
the field, and hence revisit the cosmic evolution of star-formation
rate density ($\rho_{SFR}$). We find reassuringly good agreement
with recent estimates of the evolution of $\rho_{SFR}$ with redshift,  
and our results indicate that, while most star formation in the young
Universe is visible at rest-frame UV wavelengths, dust-obscured
star formation becomes dominant at $z < 4$, due primarily
to the rise in the number density of high-mass star-forming galaxies.

\section*{Acknowledgments}

JSD acknowledges the support of the European Research Council via the award of an 
Advanced Grant (PI J. Dunlop), and the contribution of the EC FP7 SPACE project 
ASTRODEEP (Ref.No: 312725). RJM acknowledges the support of the European Research Council via the award of a 
Consolidator Grant (PI R. McLure). JEG thanks the Royal Society for support.
MJM acknowledges the support of the UK Science and Technology
Facilities Council, and the FWO Pegasus Marie Curie Fellowship.
RJI acknowledges support from the European Research Council through the 
Advanced Grant COSMICISM 321302. WR acknowledges support from JSPS KAKENHI Grant Number JP15K17604 and
Chulalongkorn University's CUniverse (CUAASC).
RSE acknowledges support from the European Research Council through the
Advanced Grant 669253. JAP acknowledges support from the European Research Council through the 
Advanced Grant 670193. PJ would like to thank NRAO for assistance in the form of a Reber
fellowship.

This paper makes use of the following ALMA data: ADS/JAO.ALMA\#2012.1.00173.S.
ALMA is a partnership of ESO (representing its member states), NSF (USA) and NINS (Japan), together
with NRC (Canada), NSC and ASIAA (Taiwan), and KASI (Republic of Korea), in cooperation with the Republic
of Chile. The Joint ALMA Observatory is operated by ESO, AUI/NRAO and NAOJ.

The National Radio Astronomy
Observatory  is  a  facility  of  the  National  Science  Foundation  operated  under
cooperative  agreement  by  Associated  Universities,  Inc;  VLA  data  are  from  project
ID  VLA/14A-360.

This work is based in part on observations made with the NASA/ESA {\it Hubble Space Telescope}, which is operated by the Association 
of Universities for Research in Astronomy, Inc, under NASA contract NAS5-26555.
This work is also based in part on observations made with the {\it Spitzer Space Telescope}, which is operated by the Jet Propulsion Laboratory, 
California Institute of Technology under NASA contract 1407.
{\it Herschel} is an ESA space observatory with science instruments provided by European-led
Principal Investigator consortia and with important participation
from NASA.

{}

\appendix

\section{Mid--Far-infrared images}

In this Appendix we provide postage-stamp 
{\it Spitzer} MIPS (24\,$\mu$m),
  {\it Herschel} PACS (70, 100 and 160\,$\mu$m), and 
  {\it Herschel} SPIRE (250, 350 and 500\,$\mu$m) images
  centred on each of the 16 ALMA sources listed in Tables\,2 and 4 (for
  which colour {\it HST} images are provided in Fig.\,4). The contrast
  in imaging resolution between ALMA and {\it Herschel} is readily apparent,
  particularly at the longer wavelengths accessed with the SPIRE instrument,
  but useful photometry was nevertheless obtained for many of the sources
  using the de-confusion techniques described in Section\,2.2.2.
  
\begin{figure*}
\begin{center}
\includegraphics[scale=0.75]{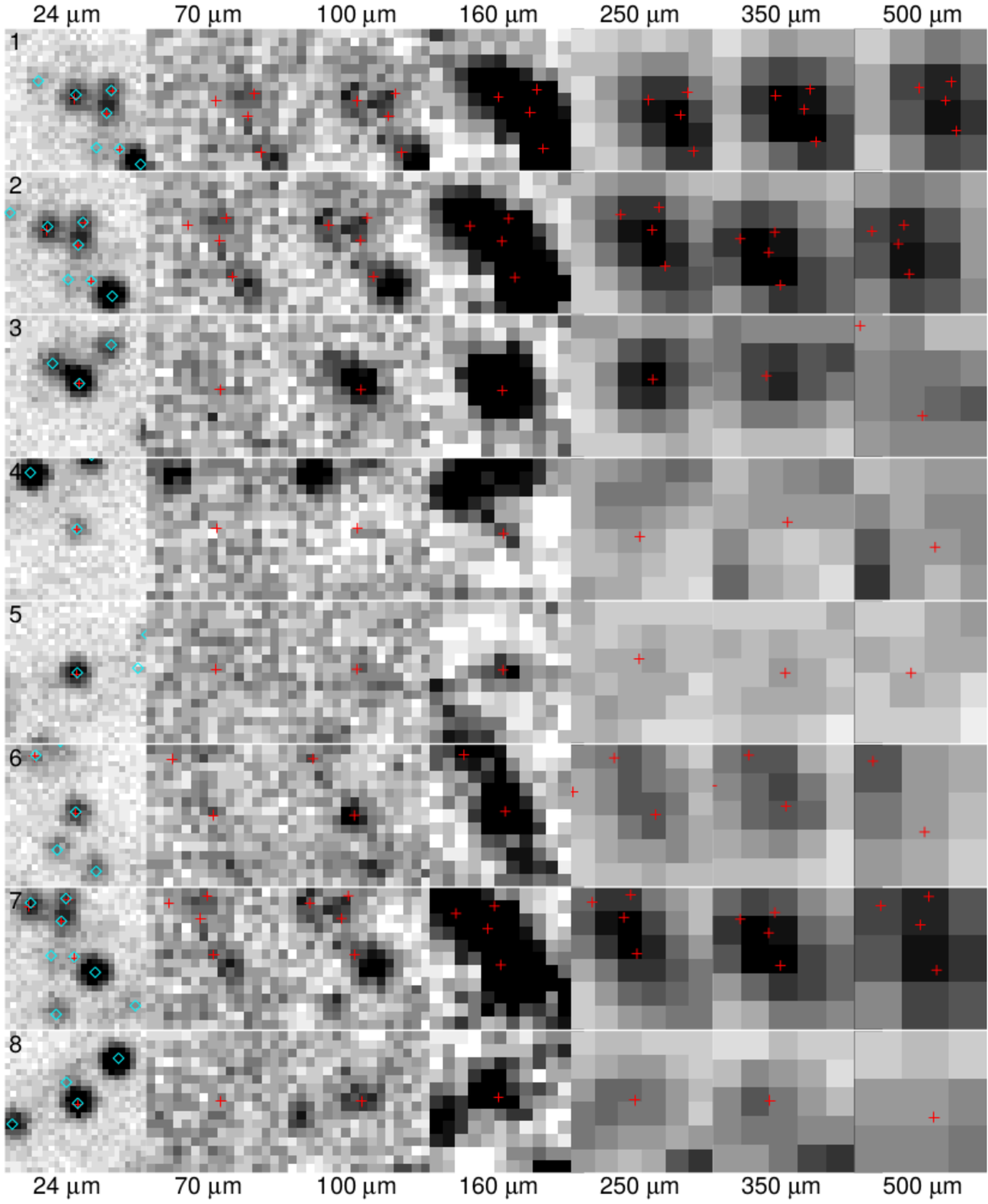}
\end{center}
\caption{Postage-stamp greyscale plots showing 30 $\times$ 30\,arcsec
  {\it Spitzer} MIPS (24\,$\mu$m),
  {\it Herschel} PACS (70, 100 and 160\,$\mu$m), and 
  {\it Herschel} SPIRE (250, 350 and 500\,$\mu$m) images
  centred on each of the 16 ALMA sources. The ALMA positions are marked
  by the red crosses, while the blue diamonds mark 24\,$\mu$m catalogue
  positions.}
\label{fig:spitzer_herschel1}
\end{figure*}

\addtocounter{figure}{-1}
\begin{figure*}
\begin{center}
\includegraphics[scale=0.75]{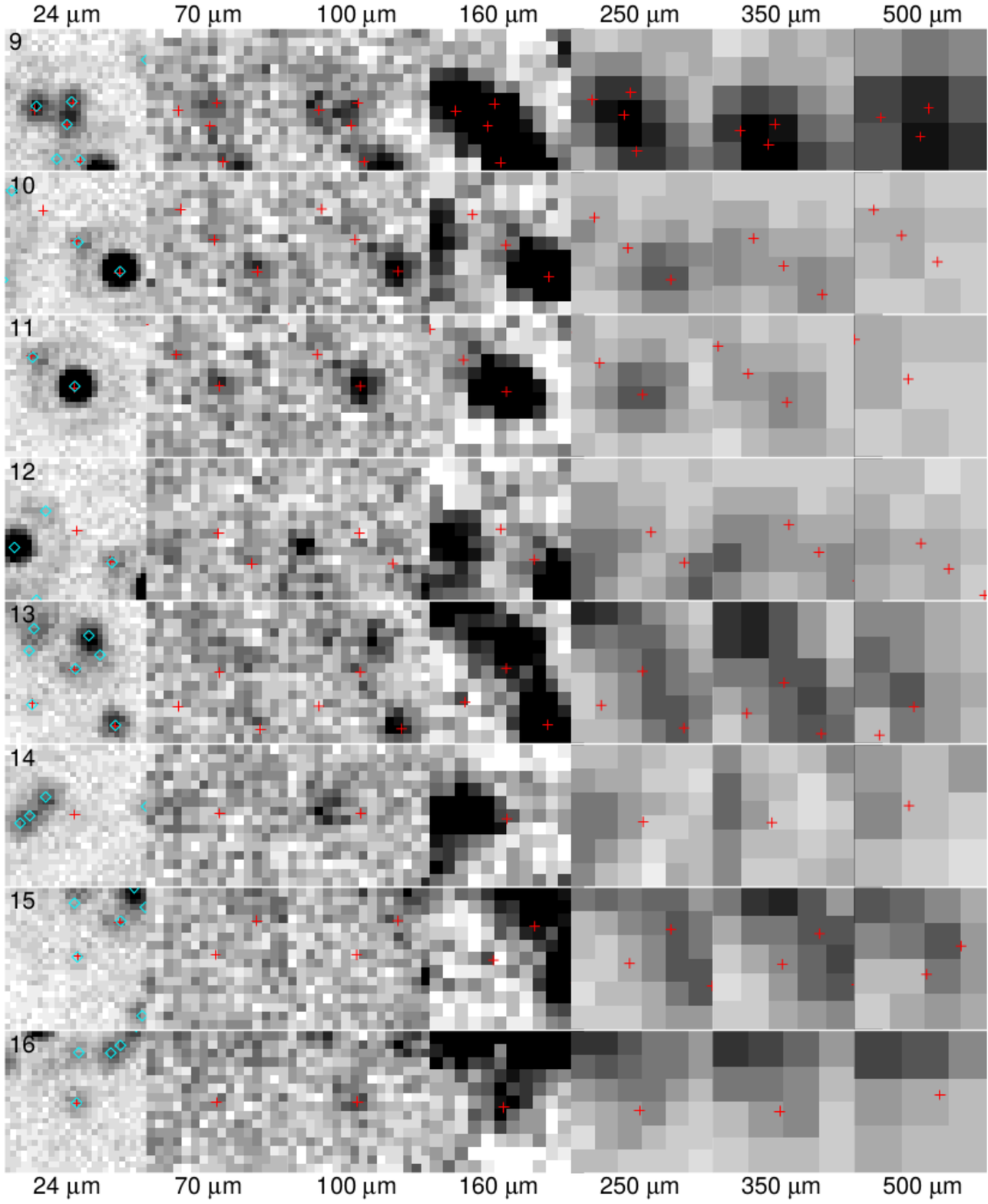}
\end{center}
\caption{(continued).}
\label{fig:spitzer_herschel2}
\end{figure*}

\end{document}